\documentclass[12pt,a4paper]{article}
\usepackage{amsmath,amssymb,mathrsfs,framed,esint,slashed,upgreek}
\usepackage[colorlinks]{hyperref}
\usepackage{xcolor}
\usepackage[all]{xy}

\newtheorem{theorem}{Theorem}[section]

\newtheorem{remark}[theorem]{Remark}
\newtheorem{proposition}[theorem]{Proposition}
\newtheorem{corollary}[theorem]{Corollary}

\newcommand{\rem}[1]{}

\newcommand{\de}{{\rm d}}

\newcommand{\bs}{{\boldsymbol{s}}}
\newcommand{\bH}{{\boldsymbol{H}}}

\newcommand{\bn}{{\boldsymbol{n}}}

\newcommand{\bsigma}{{\boldsymbol{\sigma}}}
\newcommand{\bSigma}{{\boldsymbol{\Sigma}}}

\newcommand{\bxi}{{\boldsymbol{\xi}}}

\newcommand{\beq}{\begin{equation}}
\newcommand{\eeq}{\end{equation}}
\newcommand{\ben}{\begin{eqnarray}}
\newcommand{\een}{\end{eqnarray}}

\renewcommand{\contentsname}{}

\begin{document}

\title{\vspace{-.9cm}Koopman wavefunctions and classical states\\in  hybrid quantum--classical dynamics\vspace{-.2cm}}
\author{Fran\c{c}ois Gay-Balmaz$^1$, Cesare Tronci$^{2,3}$  
\\ 
\footnotesize
\it $^1$CNRS and \'Ecole Normale Sup\'erieure, Laboratoire de M\'et\'eorologie Dynamique, Paris, France
\\
\footnotesize
\it $^2$Department of Mathematics, University of Surrey, Guildford, United Kingdom
\\
\footnotesize
\it %Numerical Methods in Plasma Physics Division, 
$^3$Department of Physics and Engineering Physics, Tulane University, New Orleans LA, United States}
\date{\small\sf For Anthony Bloch, on the occasion of his 65th birthday\vspace{-.3cm}}

\maketitle

\begin{abstract}%\footnotesize
We deal with the reversible  dynamics of coupled quantum and classical systems. Based on a recent proposal by the authors, we exploit the theory of hybrid quantum--classical wavefunctions to devise a closure model for the coupled dynamics in which both the quantum density matrix and the classical Liouville distribution retain their initial positive sign. 
In this way, the evolution allows identifying a classical and a quantum state in interaction at all times, thereby addressing a series of stringent consistency requirements.
After combining Koopman's  Hilbert-space method in classical mechanics with van Hove's unitary representations in prequantum theory,
the  closure model  is made available by  the  variational structure underlying a suitable wavefunction factorization.
Also, we use Poisson reduction by symmetry to show that the  hybrid  model  possesses a noncanonical Poisson structure that does not seem to have appeared before. As an  example, this structure is specialized to the case of quantum two-level systems.
%Combining Koopman's  Hilbert-space method in classical mechanics with van Hove's unitary representations in prequantum theory, we recently proposed a hybrid quantum--classical wave equation for the interaction dynamics of quantum and classical degrees of freedom. Based on Koopman wavefunctions, this construction ensures a positive-definite density matrix of the quantum subsystem, while a similar result is currently unavailable for the sign of the classical distribution. 
%Unless one accepts that an unsigned classical distribution may appear in the presence of quantum--classical coupling
%Then, the question emerges whether it is at all possible to formulate a deterministic hybrid theory also ensuring a positive Liouville density at all times. Here, we provide a positive answer to this question by enforcing a constraint in the above-mentioned   theory of Koopman wavefunctions. 
%We apply the exact factorization method to the Koopman wavefunctions of hybrid quantum--classical dynamics. Then, we combine its underlying Hamiltonian and variational approaches to provide a geometric closure preserving positivity and  allowing for trajectory-based computational models. 
\end{abstract}

% Marco Castrillon
% Joshua Burby
% Phil Morrison
% Cornelia Vizman
% David Martin de Diego
% Hiroaki Yoshimura

\vspace{-1.25cm}
{
\contentsname
\footnotesize
\tableofcontents
}
\addtocontents{toc}{\protect\setcounter{tocdepth}{2}}

\section{Introduction}

This work deals with the interaction dynamics of quantum and classical degrees of freedom.
Going back to the long-standing  problem of quantum measurement, the coupling of quantum and classical systems lies at the heart of several technological and scientific processes. In quantum computing, teleportation and dense coding both involve the interplay of classical and quantum  components. In fully quantum systems, hybrid quantum--classical methods are often employed to mitigate the computational challenges arising from the curse of dimensionality. This approach is well-known in chemical physics, where nuclei are treated as classical while electrons are left fully quantum \cite{Tully}. In solid state physics, spin-orbit coupling may be regarded as the coupling of classical (orbital) degrees of freedom  and quantum spins \cite{Manfredi}. More generally, whenever a scale separation identifies the slow motion of some coordinates, hybrid methods tend to emerge as an attractive strategy to tackle the curse of dimensions arising in many-particle quantum systems. In some cases, hybrid models are formulated in terms of density operators \cite{Aleksandrov,boucher,Gerasimenko,Kapral,PrKi}, while in some other cases these are based on the use of wavefunctions \cite{Barcelo,Hall,Sudarshan,Sugny,WiSr92}.

However,   most of current quantum--classical models fail to satisfy a series of basic self-consistency criteria \cite{AgCi07}, such as the following:
 \begin{enumerate} 
\item the classical system is identified by a phase-space probability density at all times;
 \item the quantum system is identified by a positive-semidefinite density operator $\hat\rho$ at all times;
 \item the model is equivariant under both quantum unitary transformations and classical canonical transformations (Hamiltonian diffeomorphisms);
 \item  in the absence of an interaction potential, the model reduces to uncoupled quantum and classical dynamics;
 \item in the presence of an interaction potential, the {\it quantum purity} $\|\hat\rho\|^2$ is not a constant of motion (decoherence property).
 \end{enumerate}
Notice that this list may not be exhaustive and further consistency requirements may be found in the literature \cite{boucher}. So far, however, current models have been struggling to capture all of the above properties and in some cases no-go arguments have been proposed \cite{PeTe,Terno,Salcedo}. The  model in \cite{Aleksandrov,Gerasimenko}, for example, fails to capture the first two properties, while the model in \cite{Hall} does not satisfy property 4. In addition, the correct prediction of decoherence  remains an outstanding task. To the authors' knowledge, the only model currently available that satisfies all the above requirements is  the {\it Ehrenfest model}:
\beq\label{Ehrenfest1}
i\hbar\frac{\partial \mathcal{P}}{\partial t}+ i\hbar\operatorname{div}(\mathcal{P}\langle X_{\widehat{H}}\rangle)=\big[\widehat{H},\mathcal{P}\big].
\eeq
Here, $\mathcal{P}(q,p)$ is a phase-space density taking values in the space of quantum density operators and, likewise, $\widehat{H}(q,p)$ is a phase-space function taking values in the space of Hermitian operators on the quantum state space. Also, $X_{\widehat{H}}=(\partial_p \widehat{H}, -\partial_q\widehat{H})$ and we have made use of the expectation value notation $\langle\widehat{A}\rangle=\operatorname{Tr}(\mathcal{P}\widehat{A})/\operatorname{Tr}(\mathcal{P})$. The classical density is given by $\rho_c=\operatorname{Tr}{\cal P}$, while the quantum density matrix reads $\hat\rho=\int{\cal P}\,\de q\de p$.   However, while serving as a basis for several numerical methods in chemistry,  equation \eqref{Ehrenfest1} fails to provide a reliable description of decoherence  in terms of  quantum purity evolution \cite{AkLoPr14}. Thus, one is led to look for alternative consistent models beyond Ehrenfest dynamics. 

Given the difficulties in ensuring consistency, sometimes one is tempted to consider alternative approaches such as density functional theory (DFT). Even in this context, mixed quantum--classical DFT is receiving increasing attention \cite{Tavernelli}. 
%In several cases, the classical system can be considered as a heat bath at equilibrium so that its dynamics may be averaged out and one is left with a Lindblad equation for the  quantum system, which is then treated as an open system. Even in this case, the full construction of a Lindblad equation necessitates 
As these alternatives are not immune from important limitations, a consistent model of coupled quantum-classical dynamics will likely open the door to a new generation of numerical simulation approaches. Indeed, while we wait for  quantum computing to get us to the next level of computational power, consistent quantum-classical methods may be able to tackle some of the challenges of many-particle simulations with the currently available resources. Given the ubiquity of many-particle quantum systems in a variety of physical contexts, it is not a moment too soon for us to explore  this important direction.

Compound systems comprising both quantum and classical degrees of freedom are generally referred to as \emph{hybrid quantum--classical systems}, while ``mixed quantum--classical dynamics'' is more common nomenclature in the chemical physics literature \cite{Tully}.
Following earlier ideas by Sudarshan \cite{Marmo,Sudarshan}, we recently obtained a hybrid quantum--classical model  \cite{BoGBTr19,GBTr20,GBTr21b}  by combining Koopman's  Hilbert-space approach to classical mechanics \cite{Koopman} with van Hove's unitary representations in prequantization \cite{VanHove}.  This  hybrid theory  is written on the tensor-product Hilbert space of Schr\"odinger and Koopman wavefunctions and its associated  dynamics is energy-conserving and Hamiltonian. While this essentially implies deterministic dynamics,  we remark that this property may or may not be enforced depending on whether one invokes the occurrence of certain irreversible processes in quantum--classical interaction \cite{Diosi}. Here, we shall restrict to consider reversible dynamics, while leaving the possibility of adding entropy sources  as an interesting future direction. 

%The theory of Koopman wavefunctions is only in its infancy and much of their mathematical structure still needs to be uncovered, along with their physical interpretation. 
While the Koopman theory \cite{BoGBTr19} of hybrid  systems ensures properties 2-5 by construction, a relevant aspect in this approach  is the fact that the expression of the classical Liouville distribution is not sign-definite. Thus one is led to ask if an  initial positive sign is preserved by the evolution, thereby identifying a classical state at all times. This was shown to be true for certain classes of hybrid systems \cite{GBTr20}, although a general statement is still missing and it is not clear whether property 1 holds in the general case. While this property was listed as desirable in \cite{boucher}, the same reference also discusses how the development of negative values in the classical distribution could still be justified by exploiting  analogies to Wigner distributions, which are known to become unsigned in the general case. Following Feynman \cite{Feynman}, analogous arguments were also presented in \cite{BoGBTr19}. The detection of  negative values in the classical density would mean that the classical subsystem may evolve to occupy certain non-classical states that are made available by the quantum--classical coupling. Despite the interesting foundational considerations arising from this point, 
it is  natural to ask whether 
%the possibility of an unsigned Liouville distribution is required by quantum--classical coupling or rather 
it is at all possible to formulate a deterministic hybrid theory beyond the Ehrenfest model  that still ensures a positive Liouville density at all times. In that case, the theory would allow the identification of both a quantum and a classical state during the entire evolution. In this paper, we provide a positive answer to this question by simply modifying the dynamical theory of quantum--classical wavefunctions in \cite{BoGBTr19,GBTr20} to enforce a positive Liouville density.  In order to obtain this variant of the theory, we combine the variational principle underlying the canonical structure of hybrid  wavefunctions with a factorization method recently rediscovered in  chemical physics  and now known as \emph{exact factorization} \cite{AbediEtAl2012}.
%Several open questions in the Koopman theory of hybrid QC dynamics still remain. One of them is whether the  phase-space density for the classical subsystem remains positive in time. The development of negative values could be justified by exploiting  analogies to Wigner distributions \cite{Feynman}, since the  latter are known to become unsigned in the presence of essentially quantum features arising from nonlinear terms in the  Hamiltonian operator. 
%Thus, the question remains whether a classical system may be described by a potentially unsigned phase-space distribution in the presence of interaction with a quantum system. 
%The detection of such negative values in the classical density of hybrid QC systems would raise new interesting foundational questions. Instead of pursuing this direction, here we simply show how the mathematical theory of hybrid Koopman wavefunctions may be modified to ensure a positive-definite classical density at all times. In order to perform this modification in the theory, we combine the variational principle underlying the canonical structure of hybrid Koopman wavefunctions with a factorization method recently rediscovered in the context of chemical physics  and now known as \emph{exact factorization} \cite{AbediEtAl2012}.
%The next few sections review the general background material.
% starting with the prequantum geometry underlying the \emph{Koopman-van Hove equation} of classical mechanics.
As we will see, the sequence of steps leading to the final model amounts to applying a gauge principle that makes the theory invariant under local phase transformations. Physically, this corresponds to the general principle that classical phases are not observable and thus represent a gauge freedom, much in analogy to global phases in standard quantum mechanics.

\paragraph{Plan of the paper.}
Section \ref{sec:KvH} reviews the Koopman-van Hove (KvH) formulation of classical mechanics. In particular, Section \ref{sec:KvHgeom} reviews its underlying prequantum geometry in terms of van Hove representations and their associated momentum map, while Section \ref{sec:Madelung1} discusses the Madelung transform. The classical KvH theory is extended in Section \ref{sec:QChybrids} to include  coupling to a quantum system. The quantum--classical wave equation is presented in Section \ref{sec:QCWE} along with its  variational principle reflecting the underlying canonical structure. In Section \ref{sec:hybden}, the hybrid wavefunction are shown to determine a quantum--classical density operator that can be used to compute both the classical Liouville density and the quantum density matrix. The variational structure of the quantum--classical wave equation is studied in Section \ref{sec:EF} upon using the exact factorization of the hybrid wavefunction. After introducing the general variational setting in Section \ref{sec:VS}, we perform a change of frame expressing the quantum evolution in the phase-space frame associated to classical motion; see Section \ref{sec:classframe}.  Section \ref{sec:closure1} is devoted to formulating the closure model ensuring that the classical density retains its initial sign. As shown in Section \ref{sec:closure}, this property is achieved by enforcing a particular constraint relating the Berry connection  to the canonical one-form on the classical cotangent bundle. Within the same section, we explain how this constraint corresponds to applying a gauge principle. Section \ref{sec:HamStr}  discusses the Hamiltonian structure of the final model, along with Casimir invariants. In particular, a new noncanonical Poisson bracket  is obtained by standard Poisson reduction in Section \ref{sec:HamStr2}. Furthermore,  Section \ref{sec:HamStr3} shows how the quantum-classical Hamiltonian structure is related to a new class of Poisson brackets governing different representations of the classical  Liouville equation. Section \ref{sec:clasquantdens} presents a candidate hybrid density operator and its equivariance properties, along with the dynamics of the quantum and classical states.
Then, Section \ref{sec:twolevel} specializes the closure model to the case of quantum two-level systems. A conclusive discussion is presented in Section \ref{conclusions}.  Section \ref{sec:relevcases} shows how the closure model specializes to previous hybrid schemes, Section \ref{sec:outlook} presents a list of open questions in the form of a plan of action for future work in this direction.

\section{Koopman-van Hove classical mechanics\label{sec:KvH}}

\subsection{The Koopman-van Hove equation of classical mechanics}

Defined on the classical phase-space $T^*{Q}$, Koopman wavefunctions $\Psi(t,q,p)$ were introduced in classical mechanics so that the relation  $\rho=|\Psi|^2$  provides a Hilbert-space representation of the classical Liouville density $ \rho  (t,q,p)$ \cite{Koopman}. Since $|\Psi(t,q,p)|^2$  is phase-invariant, its direct replacement in the classical Liouville equation $\partial_t\rho+\{\rho,H\}=0$ leads to essentially the same equation for $\Psi(t,q,p)$, namely
$i\hbar\partial_t\Psi=\{i\hbar H, \Psi \}$, which  carries the name of {\it Koopman-von Neumann equation} (KvN). The main feature here is that the Liouvillian operator $\widehat{L}_H=\{i\hbar H,\ \}$ is Hermitian so that the KvN equation $i\hbar\partial_t\Psi=\widehat{L}_H\Psi$ identifies a unitary flow on the classical Hilbert space $\mathscr{H}_{\scriptscriptstyle C} = L^2(T^*{Q})$, thereby resembling the usual quantum Schr\"odinger construction. Over the years, this approach has been rediscovered by several eminent scholars  \cite{Berry,Wiener,tHooft}, who were apparently unaware of Koopman's work.

While the Koopman approach  is currently being revived in both quantum and classical settings \cite{Bondar,Mezic,Joseph20,Giannakis}, the KvN equation involves certain unclear aspects discussed in \cite{BoGBTr19} and recently overcome only by resorting to noncanonical methods \cite{TrJo21}. Following Kostant and Souriau \cite{Kostant,Ko1970,Souriau}, in \cite{BoGBTr19,GBTr20} the authors proposed to combine Koopman's construction with van Hove's prequantization theory of unitary representations in classical mechanics. This results in adding to the operator $\widehat{L}_H$ a phase term corresponding to the Lagrangian function $\mathscr{L}:T^*Q\rightarrow \mathbb{R} $ on phase-space, that is  $\mathscr{L}=\mathcal{A}\cdot X_H- H$.
Here, the canonical one-form $ \mathcal{A} \in \Omega  ^1(T^*Q)$ satisfies $\omega=-\de\mathcal{A}$, where  $ \omega $ is the canonical symplectic form on $T^*Q$ and $X_H \in \mathfrak{X} (T^*Q)$ is the Hamiltonian vector field associated to $H$. Upon denoting by $\mathbf{i} _{X}{\cal T}$ the insertion of a vector field $X$ into a covariant tensor $\cal T$, one has $ \mathbf{i} _{X_H} \omega = {\rm d} H$. For example, if $Q=\Bbb{R}$, then one can set ${\cal A}=p\de q$, so that $\omega=\de q\wedge\de p$ and $X_H=(\partial_p H,-\partial_q H)$. The extension to arbitrary dimensions is straightforward. 
%Here, $\mathfrak{X} (T^*Q)$ denotes the space of vector fields on $T^*Q$, while $ \mathbf{i} _{X}\cdot$ stands for the insertion of a vector field $X$ into a differential $n$-form. 
The {\it Koopman-van Hove equation} (KvH) for classical mechanics reads
\beq\label{KvH_eq}
i\hbar\partial_t\Psi=\widehat{L}_H\Psi -\mathscr{L} \Psi  =: \widehat{ \mathcal{L} }_H \Psi \,,
\eeq
where $\mathscr{L}$ acts as standard multiplication and  we defined the {\it prequantum operator} on the right-hand side. Upon introducing  $\widehat{\lambda}=-i\hbar {\rm d} $, this operator can be written as
\beq\label{preqop}
\widehat{\mathcal{L}}_H=\widehat{L}_H  -\mathscr{L}=H+X_H\cdot(\widehat{\lambda}-\mathcal{A})\,.
\eeq
Here, the dot product stands for the standard insertion of a vector field into a one form.
While the name `prequantum operator' is customary in geometric quantization, this operator was also called {\it covariant Liouvillian} in \cite{BoGBTr19}. 
In this construction, there is a one-to-one correspondence $F\leftrightarrow\widehat{\mathcal{L}}_F$ between phase-space functions $F(q,p)$ identifying physical observables and prequantum operators $\widehat{\mathcal{L}}_F$. This is somewhat analogous to Dirac's canonical quantization prescription for quantum observables, which instead reads $F(q,p)\leftrightarrow F(\hat{\sf q},\hat{\sf p})$ with $[\hat{\sf q},\hat{\sf p}]=i\hbar$.  In the classical case, upon introducing the multiplicative operators $(\hat{q},\hat{p})\Psi=({q}\Psi,{p}\Psi)=:\widehat{z}\,\Psi$, we notice that the relation $[\widehat{z}_j,\widehat{\lambda}_k]=i\hbar\delta_{jk}$  takes the KvH equation \eqref{KvH_eq} to a formally equivalent quantum system with the two sets of canonical observables
$(q,\lambda_{q})$ and $(p,\lambda_{p})$.
More importantly, we recall that prequantum operators comprise the Lie algebra structure $[\widehat{\mathcal{L}}_F,\widehat{\mathcal{L}}_K]=i\hbar\widehat{\mathcal{L}}_{\{F,K\}}$, thereby leading to the canonical commutation relation
$\big[\widehat{\mathcal{L}}_{q},\widehat{\mathcal{L}}_{p}\big]=i\hbar$. This last relation was used in \cite{Bondar} to rewrite Schr\"odinger's quantum mechanics in terms of wavefunctions on phase-space.

%\begin{remark}[Quantum Schr\"odinger equation on phase-space]
%In \cite{Bondar}, similar commutation relations were used to express standard quantum mechanics in phase-space, so that the quantum Schr\"odinger equation reads $i\hbar\partial_t\psi=\mathcal{H}(\widehat{\mathcal{L}}_q,\widehat{\mathcal{L}}_p)\psi+\mathcal{F}(\,\overline{\!\mathcal{L}}_q,\,\overline{\!\mathcal{L}}_p)\psi$. Here, $\mathcal{H}$ is the Hamiltonian functional and we have defined the conjugate prequantum operator $\,\overline{\!\mathcal{L}}_F=F-X_F\cdot(\widehat{\Lambda}+\mathcal{A})$ so that $[\widehat{\mathcal{L}}_z,\,\overline{\!\mathcal{L}}_z]=0$, where $z=(q,p)$. Since $[\mathcal{K}(\widehat{\mathcal{L}}_q,\widehat{\mathcal{L}}_p),\mathcal{F}(\,\overline{\!\mathcal{L}}_q,\,\overline{\!\mathcal{L}}_p)]=0$ for any two functions $\mathcal{K}$ and $\mathcal{F}$, the function $\mathcal{F}$ in the phase-space Schr\"odinger equation does not affect the Heisenberg evolution of quantum observables and thus can be left completely arbitrary. In the case $\mathcal{F}=0$, a similar quantum Schr\"odinger equation on phase-space also appeared in \cite{deGosson}, following previous developments in chemical physics \cite{}.
%\end{remark}

The main novelty introduced by \cite{BoGBTr19} in KvH theory consists of the relation between the Koopman wavefunction and the classical Liouville density. This relation is given by a modification of the KvN original prescription $\rho=|\Psi|^2$ recalled above. This modification reads
\begin{equation}\label{KvHmomap}
\rho=|\Psi|^2 -\operatorname{div}( \Bbb{J}\mathcal{A} |\Psi|^2)  + {\rm i} \hbar\{\Psi,\,\overline{\!\Psi\!}\,\}
\,,
\end{equation}
where $\Bbb{J}:T^*(T^*{Q})\rightarrow T(T^*{Q})$ is the vector bundle map associated to the Poisson structure, defined by $\{F,H\}=  {\rm d}F\cdot \Bbb{J}({\rm d}H)$.
In local coordinates, $\boldsymbol{z}=(q,p)$ and one can write $\Bbb{J}$ as the Poisson tensor, so that $\Bbb{J}^{jk}=\{z^j,z^k\}$ and $X_H^j=\Bbb{J}^{jk}\partial_{k} H$.
 It can be shown that the expression \eqref{KvHmomap} satisfies the classical Liouville equation if $\Psi$ is a solution of the KvH equation \eqref{KvH_eq}.
While the first term on the right-hand side of \eqref{KvHmomap} coincides with the KvN prescription, the remaining two terms  vanish under integration in the absence of  singularities. In addition, the last two terms make the expression \eqref{KvHmomap} sign-indefinite; this should not be seen as an issue because the characteristic nature of the classical Liouville equation ensures that the sign of $\rho$ is preserved in time. As shown in \cite{GBTr20}, equation \eqref{KvHmomap} has a deep geometric meaning as it identifies a  momentum map structure. However, in order to make this point clear, we will need to revise the general geometric construction underlying KvH theory. This is the subject of the next section.

\subsection{Geometry of KvH classical mechanics\label{sec:KvHgeom}}
In KvH theory, the operator $-i\hbar^{-1}\widehat{\mathcal{L}}_H$ acts on the classical wavefunction as the infinitesimal action of the group of connection-preserving automorphisms of the prequantum circle bundle $T^*{Q}\times S^1 \rightarrow T^*Q$, where the latter is endowed with the principal connection one-form $\mathcal{A}$. Then, the relation \eqref{KvHmomap} emerges as the momentum map associated to this action. We refer the reader to \cite{GBTr20} for more details, while here we shall only provide a quick summary of the geometry underlying KvH theory.

\paragraph{Connection-preserving automorphisms.} Since the prequantum  bundle is  trivial, its associated group of connection preserving automorphisms is expressed as
\begin{equation}\label{stricts}
\operatorname{Aut}_{\cal A}(T^*{Q}\times S^1):=
\left\{(\eta,e^{\rm i\varphi})\in\operatorname{Diff}(T^*{Q})\,\circledS\, \mathcal{F}(T^*{Q}, S^1)\ \Big|\ \eta^*\mathcal{A}+\de\varphi=\mathcal{A} \right\},
\end{equation}
where $\eta^*$ denotes pullback and $\operatorname{Diff}(T^*{Q})\,\circledS\, \mathcal{F}(T^*{Q}, S^1)$ denotes the semidirect product of the diffeomorphism group with the space of $S^1-$valued functions on phase-space. Since $\de{\cal A}=-\omega$,
the condition $\mathcal{A}-\eta^*\mathcal{A}=\de\varphi$ in \eqref{stricts} implies  that $\eta$ is a symplectic diffeomorphism with generating function 
\beq\label{varphi}
\varphi(z)=\theta+\int_{z_0}^{z}({\cal A}-\eta^*{\cal A}), \qquad z=(q,p ) \in  T ^*Q,
\eeq
for an arbitrary $\theta \in \mathbb{R}$. 

The Lie algebra of $\operatorname{Aut}_{\cal A}(T^*{Q}\times S^1)$ is given by the space 
\[
\mathfrak{aut}_{\cal A}(T^*Q\times S^1)=\{(X, \nu ) \in \mathfrak{X} (T^*Q) \,\circledS\, \mathcal{F} (T^*Q) \mid \pounds _X \mathcal{A} + {\rm d} \nu =0\},
\]  
which inherits the usual semidirect-product bracket structure 
$[(X_1,\phi_1),(X_2,\phi_2)]=([X_2,X_1],\de\phi_1\cdot X_2- \de\phi_2\cdot X_1)$. Here, $\pounds$ denotes the Lie derivative and $[X_2,X_1]=\pounds_{X_2}X_1$ is the usual Jacobi-Lie bracket on vector fields. Importantly,   $\mathfrak{aut}_{\cal A}(T^*Q\times S^1)$
is isomorphic to the Lie algebra algebra $\mathcal{F} (T^*Q)$ of phase-space functions, endowed with the canonical Poisson bracket $\{\,,\}$. In formulas, one has
\begin{align}\nonumber
 \big(\mathcal{F} (T^*Q),\{\,,\}\big) & \to   \big(\mathfrak{aut}_{\cal A}(T^*Q\times S^1), [(\, ,),(\, ,)]\big)
\\
H & \mapsto (X_H, H- \mathcal{A} \cdot X_H).
\label{LA_isomorphism} 
\end{align}
Here, the inverse Lie algebra isomorphism $\mathfrak{aut}_{\cal A}(T^*Q\times S^1)\to\mathcal{F} (T^*Q)$ is given by $(X,\phi)\mapsto\phi+X\cdot\mathcal{A}$. The fact that \eqref{LA_isomorphism} is indeed a Lie algebra isomorphism follows from the result below, which exploits the properties of the insertion $ \mathbf{i} _X$ and the Lie derivative $ \pounds _X$:
\begin{align*}
&{\rm d} (F- \mathcal{A} \cdot X_F) \cdot X_G-{\rm d} (G- \mathcal{A} \cdot X_G) \cdot X_F\\
&= - \pounds _{X_F} \mathcal{A} \cdot X_G +  \pounds _{X_G} \mathcal{A} \cdot X_F\\
&= \mathbf{i} _{[X_F,X_G]} \mathcal{A} - \pounds _{X_F} \mathbf{i} _{X_G} \mathcal{A} + \mathbf{i} _{X_F} \pounds _{X_G} \mathcal{A} \\
&= \mathbf{i} _{[X_F,X_G]} \mathcal{A} - \mathbf{i} _{X_F} {\rm d} ( \mathbf{i} _{X_G} \mathcal{A} )+ \mathbf{i} _{X_F} \mathbf{i} _{X_G} {\rm d} \mathcal{A} + \mathbf{i} _{X_F} {\rm d} ( \mathbf{i} _{X_G} \mathcal{A} )\\
&= - \mathcal{A} \cdot  X_{\{F,G\}}  + \{F,G\}\,.
\end{align*} 
Here, Cartan's magic formula is used to write ${\rm d} (\mathcal{A} \cdot X_F-F)=\pounds _{X_F} \mathcal{A}$ (and analogously for $X_G$) in the first equality. The second equality follows from the general property $\mathbf{i}_{[X,Y]}\alpha=\pounds_{X}\mathbf{i}_Y\alpha-\mathbf{i}_Y\pounds_X\alpha$, holding for any differential form $\alpha$ and any two vector fields $X$ and $Y$. The third equality applies Cartan's magic formula to the last term in the third line.
Consequently, since $[X_F,X_G]=X_{\{F,G\}}$, the Lie bracket structure on $\mathfrak{aut}_{\cal A}(T^*Q\times S^1)$ yields
\begin{align*} 
[(X_F, F- \mathcal{A} \cdot  X_F), (X_G, G- \mathcal{A} \cdot  X_G)]
%&= \left( -[X_F,X_G], {\rm d} (F- \mathcal{A} \cdot  X_F) \cdot X_G - {\rm d} (G- \mathcal{A} \cdot  X_G) \cdot X_F \right) \\
&= \left( X_{\{F,G\}},\{F,G\} - \mathcal{A} \cdot  X_{\{F,G\}} \right), 
\end{align*} 
thereby showing that \eqref{LA_isomorphism} is indeed a Lie algebra isomorphism. We refer the reader to \cite{GBTr20,GBTr,ILM} for a description of the group $\operatorname{Aut}_{\cal A}(T^*{Q}\times S^1)$ in terms of central extensions.

\paragraph{Van Hove representation on Koopman wavefunctions.}
Having characterized the group structure of \eqref{stricts}, we will now explain its relevance to the  KvH equation \eqref{KvH_eq} for classical dynamics. In particular, we shall see that   the relation \eqref{KvHmomap} comprises a geometric momentum map structure  associated to a specific action of the group \eqref{stricts}, that is the van Hove representation.
Indeed, the group $\operatorname{Aut}_{\cal A}(T^*{Q}\times S^1)$  possesses the following unitary (left) action on the Hilbert space $\mathscr{H}_{\scriptscriptstyle C} = L^2(T^*{Q})$ of Koopman wavefunctions:
\beq\label{sdp-action}
%\Psi\mapsto 
\Psi\mapsto\frac1{\sqrt{\operatorname{Jac}(\eta)}}({e^{- {\rm i} \varphi/\hbar}}\Psi) \circ \eta ^{-1} =:U_{(\eta, e^{{\rm i} \varphi })} \Psi
\,,
\eeq  
where $\operatorname{Jac}({\eta})$ denotes the Jacobian determinant of $\eta$ with respect to the canonical  volume form on $T^*Q$. Notice that the defining relation $\eta^*\mathcal{A}+\de\varphi=\mathcal{A}$ leads to  $\operatorname{Jac}(\eta)=1$, although  we retain the general form \eqref{sdp-action}  for later convenience.
Here, the Hilbert space $\mathscr{H}_{\scriptscriptstyle C} = L^2(T^*{Q})$ carries the standard Hermitian inner product
\[
\langle \Psi_1| \Psi_2\rangle= \int_{T^*{Q}\!} \bar{\Psi}_1(z)\Psi_2(z)\,\Lambda,\qquad
\Psi_1,\Psi_2\in \mathscr{H}_{\scriptscriptstyle C},
\]
defined in terms of the Liouville volume form $\Lambda$ on $T^*Q$.
The corresponding real-valued pairing and symplectic form on $\mathscr{H}_{\scriptscriptstyle C}$ are given by
\begin{equation}\label{inner_symplectic}
\langle \Psi_1, \Psi_2\rangle= \operatorname{Re} \langle{\Psi}_1|\Psi_2\rangle
%\int_{T^*{Q}\!}  \bar{\Psi}_1(z)\Psi_2(z)\,\Lambda
\qquad\text{and}\qquad\Omega(\Psi_1, \Psi_2)= 2\hbar \operatorname{Im}\langle{\Psi}_1|\Psi_2\rangle
% \int_{T^*{Q}\!} \bar{\Psi}_1(z)\Psi_2(z)\,\Lambda
\,.
\end{equation}
%The isomorphism \eqref{groupisomorphism} leads to a unitary action of the group central extension \eqref{scont} on $\mathscr{H}_{\scriptscriptstyle C} = L^2(T^*{Q})$; in practice, the action $\Psi\mapsto U_{(\eta, e^{{\rm i}\theta})}\Psi$ is given by \eqref{sdp-action} with $\eta\in {\operatorname{Diff}}_\omega(T^*{Q})$ and $\varphi$ given in \eqref{varphi}. 
We now focus on the infinitesimal generator of the van Hove representation \eqref{sdp-action}:
%As shown in \cite{GBTr20,GBTr,ILM}, the group structure of the central extension \eqref{scont} leads to an associated Lie algebra structure which can be identified with the space $\mathcal{F}(T^*{Q})$ of smooth functions on phase-space, with the Lie bracket given by the canonical Poisson structure.
 by using the Lie algebra isomorphism \eqref{LA_isomorphism}, the unitary action $\Psi\mapsto U_{(\eta, e^{{\rm i} \varphi })}\Psi$ in \eqref{sdp-action}  leads to the infinitesimal action $\Psi\mapsto- {\rm i} \hbar^{-1}\widehat{\cal L}_H\Psi$, thereby recovering the prequantum operator \eqref{preqop}.

\paragraph{Momentum map for the Van Hove representation.}
At this point, we have provided the background that is necessary to present the explicit momentum map structure underpinning the relation \eqref{KvHmomap} between the KvH construction and the Liouville equation of classical mechanics.
It is easy to see \cite{BoGBTr19,GBTr20} that, upon identifying the dual space ${\cal F}(T^* {Q})^*$ of phase-space functions with the space of densities $\operatorname{Den}(T^*{Q})$, the momentum map associated to the unitary action $\Psi\mapsto U_{(\eta, e^{{\rm i}\theta})}\Psi$ is given by $J(\Psi)=|\Psi|^2 +\operatorname{div}(  \Bbb{J}  \mathcal{A} |\Psi|^2)  + {\rm i} \hbar\{\Psi,\,\overline{\!\Psi\!}\,\}$, thereby recovering the expression \eqref{KvHmomap} for the Liouville density, i.e. $ \rho  =J( \Psi  )$.  This equivariant momentum map is formally a Poisson map with respect to the Poisson structure
\[
\{\!\!\{f,h\}\!\!\}(\Psi)=\frac1{2\hbar}\operatorname{Im}\left\langle\frac{\delta f}{\delta \Psi}\bigg|\frac{\delta h}{\delta \Psi}\right\rangle
%\int_{ T^*{Q}}\ \overline{\!\frac{\delta f}{\delta \Psi}\!}\ \, \frac{\delta h}{\delta \Psi}\,\Lambda
\]
associated to the symplectic form \eqref{inner_symplectic} on $\mathscr{H}_{\scriptscriptstyle C}$ and the following Lie-Poisson structure on the space $\operatorname{Den}(T^*{Q})$:
\beq\label{LiouvBracket}
\{\!\!\{f,h\}\!\!\}(\rho)=\int_{T^*{Q}}
\rho\left\{\frac{\delta f}{\delta \rho},\frac{\delta h}{\delta \rho}\right\} \Lambda\,.
\eeq
The KvH equation \eqref{KvH_eq} is a Hamiltonian system on $\mathscr{H}_{\scriptscriptstyle C}$ with respect to the symplectic form  \eqref{inner_symplectic} and with the Hamiltonian functional $h( \Psi )=\int_{T^*Q}\bar \Psi \widehat{ \mathcal{L} }_H \Psi \Lambda = \int_{T^*Q}  \rho   H\Lambda $, where $ \rho  =J ( \Psi )$. In particular, the Hamiltonian functional depends on $ \Psi $ only through $ J( \Psi )$. Hence, by momentum map collectivization \cite{GuSt80}, if $\Psi(t)$ is a solution of 
the KvH equation, the density \eqref{KvHmomap} solves the classical Liouville equation $\partial_t\rho=\{H,\rho\}$, as in the following diagram: \\
%\begin{figure}[h]

\vspace{-.45cm}
{\small\center\ 
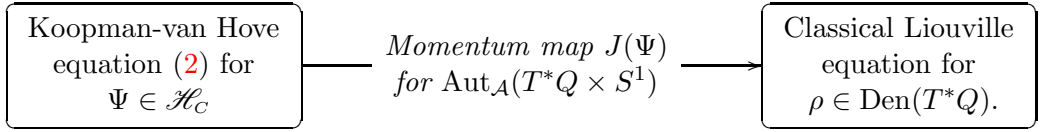
\begin{xy}
\xymatrix{
&*+[F-:<3pt>]{\begin{array}{c}\text{Koopman-van Hove}\\
\text{equation \eqref{KvH_eq} for}\\
\text{ $\Psi \in \mathscr{H}_{\scriptscriptstyle C}$}
\end{array}}\ar[rrrrr]|{\begin{array}{c}
\textit{Momentum map ${J}(\Psi)$}\\
%\text{map $\mathcal{J}$}\\
\textit{for $\operatorname{Aut}_{\cal A}(T^*Q \times S ^1 )$} 
\end{array}
} &\hspace{1cm} & & & &
*+[F-:<3pt>]{
\begin{array}{c}
\text{Classical Liouville}\\
\text{equation %\eqref{Liouville}
for}\\
\text{ $ \rho  \in \operatorname{Den}(T^*Q)$.}
\end{array}
}
}
\end{xy}
}
%\caption{Momentum map relation from classical KvH theory to Liouville dynamics.}
%\end{figure}

\bigskip
\noindent
Thus, while relating the KvH equation \eqref{KvH_eq} to standard classical mechanics,  the momentum map $J(\Psi)$ identified by \eqref{KvHmomap} requires the full geometric construction underlying the van Hove transformations  \eqref{sdp-action}. In order to unfold the information contained in the evolution of the KvH wavefunction,  it is convenient at this stage to look at the dynamics of its phase and amplitude. This is the topic of the next section.

\subsection{Madelung transform of the KvH equation\label{sec:Madelung1}}
As discussed  in \cite{GBTr21a,GBTr20}, the Madelung transform of complex partial differential equations simply corresponds to expressing the equation in terms of amplitude and phase by writing $\Psi$ in polar form. In the present section, this procedure will be applied to the KvH equation.

\paragraph{Madelung equations and Koopman phase.} In the case of the KvH equation \eqref{KvH_eq} $\Psi$ is defined on phase-space and we write
$\Psi(t,z)=R(t,z)e^{{\rm i}S(t,z)/\hbar}$.
This leads to the following equations for the amplitude $R$ and the phase $S$:
\begin{align}\label{KvHMadelung1}
\partial_t S+\{S,H\}=&\ \mathscr{L}
\\
\partial_t R+\{R,H\}=&\ 0
\,,
\label{KvHMadelung2}
\end{align}
where we recall the expression of the Lagrangian $\mathscr{L} = \mathcal{A} \cdot X_H-H \in \mathcal{F}(T^*{Q})$.
Thus, while \eqref{KvHMadelung2} recovers the standard Koopman-von Neumann equation for the amplitude $R=|\Psi|$, the KvH construction comprises also the dynamics \eqref{KvHMadelung1} of the classical phase. Notice that \eqref{KvHMadelung1} is equivalently written as
\begin{equation}\label{classicalphase_evolution}
\frac{\de }{\de t}S(t, \eta(t,z))=\mathscr{L}(\eta(t,z))
\end{equation}
where $\eta(t)$ is the flow of $X_H$.  
We remark that, due to the Lie derivative relation $\pounds_{X_H}\mathcal{A}= \de \mathscr{L}$ and Cartan's magic formula $\pounds_{X_H}={\rm d} \mathbf{i}_{X_H} + \mathbf{i}_{X_H}{\rm d}$, the phase dynamics \eqref{KvHMadelung1} also produces the relation
\begin{equation}\label{dS_theta}
(\partial_t+\pounds_{X_H})({\rm d} S-{\mathcal{A}})=0\,.
\end{equation}
In this paper, we will work with differential forms on manifolds rather than vector calculus on Euclidean spaces, so that in the case $Q=\Bbb{R}_n$ one has $\de S(\boldsymbol{z})=\nabla S(\boldsymbol{z})\cdot\de \boldsymbol{z}$.
Notice that, as a result of \eqref{dS_theta}, the relation ${\rm d} S={\mathcal{A}}$ would be preserved in time thereby recovering the KvN prescription $\rho=|\Psi|^2$ via the momentum map \eqref{KvHmomap}. However, as pointed out in \cite{Joseph20}, this possibility would require introducing topological singularities, which  we prefer to avoid in this work. 

\paragraph{Momentum map for the Madelung transform.}
As the dynamics of the KvH phase and amplitude has been characterized, we now move on to present their associated momentum map structure, which will be useful for later purpose; see Section \ref{sec:VS}.
In the standard treatment of the Madelung transform \cite{Madelung}, it is convenient to introduce the quantities
\[
\sigma= R^2 {\rm d}  S=\hbar\operatorname{Im}(\bar\Psi {\rm d}   \Psi)
\qquad\text{and}\qquad D=R^2=|\Psi|^2
\,.
\]
In the case of quantum wavefunctions on the configuration space, these identify the probability current and density, respectively, thereby taking the Schr\"odinger equation for $\Psi$ into a set of hydrodynamic equations.
However, the situation is different in the KvH context. In the latter case, the equations \eqref{KvHMadelung1}-\eqref{KvHMadelung2} for  phase and amplitude  decouple. 
Nevertheless, similarly to the quantum case \cite{KhMiMo2019}, the quantities $(\sigma,D)$ comprise a momentum map structure that  deserves some discussion. In particular, the momentum map
\begin{equation}\label{Joli}
\mathcal{J} (\Psi)=(\hbar\operatorname{Im}(\bar\Psi {\rm d} \Psi),|\Psi|^2)=:(\sigma,D)
\end{equation} 
  is associated to the representation   of the prequantum bundle automorphisms $\operatorname{Aut}(T^*{Q}\times S^1)\simeq \operatorname{Diff}(T^*{Q})\,\circledS\,\mathcal{F}(T^*{Q},S^1) $ on $\mathscr{H}_{\scriptscriptstyle C}$. 
This representation is formally the same as in \eqref{sdp-action}, except for the fact that in the latter case $\eta$ is enforced to be symplectic and $\varphi$ is prescribed by \eqref{varphi}. Indeed, in \eqref{sdp-action} one restricts to consider the action of the subgroup of connection-preserving automorphisms, that is
\[
\operatorname{Aut}_{\cal A}(T^*{Q}\times S^1)\subset \operatorname{Aut}(T^*{Q}\times S^1)\simeq\operatorname{Diff}(T^*{Q})\,\circledS\,\mathcal{F}(T^*{Q},S^1)
\,.
\]
Thus, the momentum map $\rho  = J(\Psi)$ in \eqref{KvHmomap} for the classical Liouville equation can be related to the momentum map $ ( \sigma , D)= \mathcal{J} (\Psi)$ in \eqref{Joli} via the dual of the Lie algebra inclusion $\iota: \mathcal{F}(T^*{Q}) \hookrightarrow \mathfrak{X}(T^*{Q})\,\circledS\, \mathcal{F}(T^*{Q})$. See further details in \cite{GBTr21a,GBTr20}.
Then, we are left with a relation between the momentum map \eqref{KvHmomap} for the classical Liouville equation and the momentum map \eqref{Joli} associated to the KvH Madelung transform, that is 
\beq\label{LAincl}
J ( \Psi )= \iota ^*\mathcal{J} ( \Psi ), 
\qquad\quad\text{ with }\qquad 
\iota ^* (D, \sigma )= D+\operatorname{div}(\sigma-D{\cal A})
\,,
\eeq
where we emphasize the different notations $J$ and $\cal J$ for the momentum maps \eqref{KvHmomap} and \eqref{Joli}, respectively.

%This set of geometric structures is summarized in the diagram below:
In summary, we have  reviewed how several quantities appearing in KvH classical mechanics are interconnected and their relations are most often given by specific momentum maps associated to  particular automorphism groups of the prequantum bundle, as illustrated in the diagram below:

%\begin{figure}[h!]
{\noindent
\footnotesize
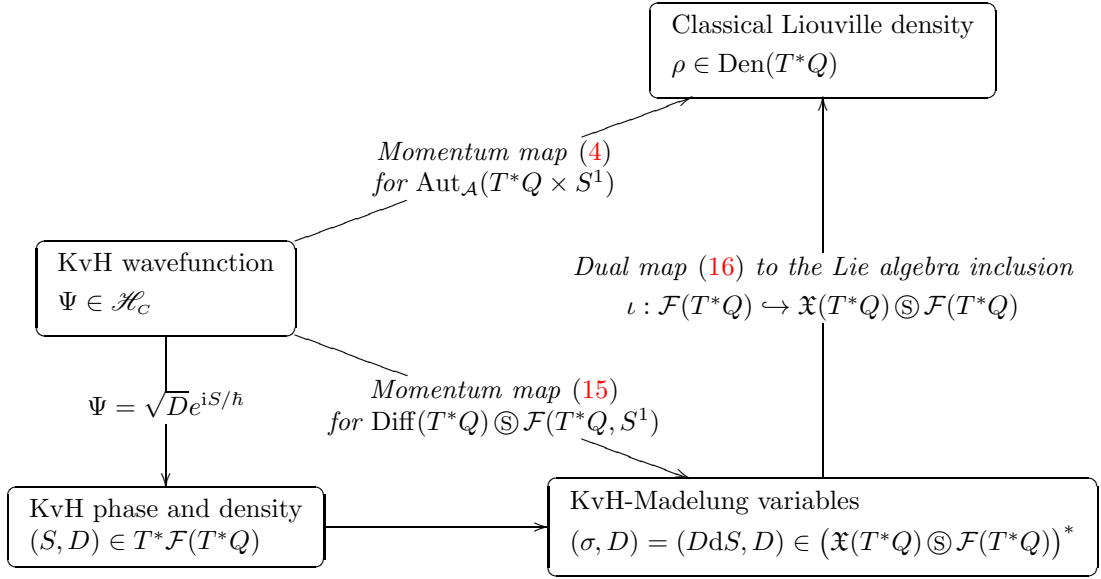
\begin{center}
\hspace{.3cm}
%\noindent
%\begin{framed}
\begin{xy}
\xymatrix{
& & &  &*+[F-:<3pt>]{
\begin{array}{l}
\vspace{0.1cm}\text{Classical Liouville density}\\
\vspace{0.1cm} \rho  \in \operatorname{Den}(T^*Q)%\\
%\vspace{0.1cm}\displaystyle h(\rho) = \int_{T^*Q} H \rho  \Lambda \\
%\vspace{0.1cm} \partial _t \rho  +\{ \rho  , H\}=0
\end{array}
} &\\
& & & & &\\
&
*+[F-:<3pt>]{
\begin{array}{l}
\vspace{0.1cm}\text{KvH wavefunction}
\\
\vspace{0.1cm}\Psi \in \mathscr{H}_{\scriptscriptstyle C}%, \; H \in \mathcal{F}(T^*Q)
\\
%\vspace{0.1cm}\displaystyle h(\Psi)= \int_{T^*Q}\bar\Psi \widehat{\cal L}_H\Psi\,\Lambda\\
%{i}\hbar \partial _t \Psi = \widehat{\cal L}_H\Psi
\end{array}
} \ar[ddrrr]|{\begin{array}{c}\textit{Momentum map \eqref{Joli}}\\
\textit{for $
\operatorname{Diff}(T^*Q) \,\circledS\, \mathcal{F}(T^*Q, S^1)$}\\
\end{array}}
\ar[uurrr]|{\begin{array}{c}\textit{Momentum map \eqref{KvHmomap}}\\
\textit{for 
$\operatorname{Aut}_{\cal A}(T^*{Q}\times S^1)$}\\
\end{array}}
\ar[dd]|{\begin{array}{l}\Psi= \sqrt{D}e^{{\rm i} S/\hbar}\end{array}} & & & \\
& & & & &\\
&*+[F-:<3pt>]{\begin{array}{l}\text{KvH phase and density}\\
\text{$(S,D)\in T^*\mathcal{F}(T^*Q)$}
%\\
%\text{for $(S,D)$}
\end{array}}\ar[rrr] & & &
*+[F-:<3pt>]{
\begin{array}{l}
\vspace{0.1cm}\text{KvH-Madelung variables}\\
\vspace{0.1cm} ( \sigma , D) =(D\de S,D)  \in \big( \mathfrak{X}(T^*Q) \,\circledS\, \mathcal{F}(T^*Q) \big) ^*%\\ 
%\vspace{0.1cm}\displaystyle h ( \sigma , D) = \int_{T^*Q} H \rho  \Lambda \\
%\vspace{0.1cm} \partial _t \rho  +\{ \rho  , H\}=0
\end{array}
} \ar[uuuu]|{\begin{array}{c}\vspace{0.1cm}\textit{Dual map \eqref{LAincl} to the Lie algebra inclusion} \\
\iota:  \mathcal{F}(T^*Q)\hookrightarrow \mathfrak{X}(T^*Q) \,\circledS\, \mathcal{F}(T^*Q)\end{array}} &
}
\end{xy}
%\end{framed}\vspace{-.5cm}
%\it\caption{Schematic description of the various quantities appearing in Koopman-van Hove classical mechanics and some of the mappings between them.}
%\label{figure1}
%\end{figure}
\end{center}
}

\bigskip
\noindent
The next sections are devoted to illustrating the dynamical theory of hybrid quantum--classical wavefunctions.

\section{Hybrid quantum--classical wavefunctions\label{sec:QChybrids}}

Once the theory of classical wavefunctions has been reviewed in the context of the KvH construction, one is tempted to construct a hybrid quantum--classical dynamics on the tensor-product Hilbert space of classical and quantum wavefunction. This idea was first proposed by Sudarshan \cite{Marmo,Sudarshan} and later criticized in \cite{PeTe,Terno} due to apparent interpretative issues. While Sudarshan's construction was based on KvN classical dynamics, the present treatment overcomes the interpretative issues by resorting to KvH theory.

\subsection{The quantum--classical wave equation\label{sec:QCWE}}

As pointed out in \cite{BoGBTr19,GBTr20}, the KvH framework can be exploited to construct a hybrid description of coupled quantum--classical systems. 
Upon starting with a KvH equation for two classical particles with coordinates $z_1=(q_1,p_1)$ and $z_2=(q_2,p_2)$, one applies  a partial quantization procedure on one of them (say, particle 2), so that $\partial_{p_2}\Psi(z_1,z_2)=0$ and $p_2\Psi(z_1,z_2)\to-i\hbar\partial_{x_2}\Psi(z_1,z_2)$. Then, upon dropping subscripts and changing the notation as $q_2\to x$ and $\Psi\to\Upsilon$, one obtains  the following \emph{quantum--classical wave equation} \cite{BoGBTr19,GBTr20,GBTr21b}
\begin{equation}\label{hybrid_KvH}
{\rm i}\hbar\partial_t\Upsilon=\{{\rm i}\hbar \widehat{H},\Upsilon\} + \big(\widehat{H} - {\mathcal{A}}\cdot X_{\widehat{H}}\big)\Upsilon
\,.
\end{equation}
This equation was recently considered in \cite{Andre}, where it was shown to be Galilean-covariant.
Here, the phase-space function $\widehat{H}(z)$   takes values in the space of unbounded Hermitian operators on the quantum Hilbert space $\mathscr{H}_{\scriptscriptstyle Q}:=L^2( M)$, where $M$ is the quantum configuration manifold.  Also, $\Upsilon \in L^2(T^*{Q}\times M)$ is a hybrid wavefunction $\Upsilon(z,x)$ depending on both the classical and the quantum coordinates, denoted by $z\in T^*{Q}$ and $x\in M$, respectively. The operator-valued vector field $X_{\widehat{H}}$ is defined as $\mathbf{i}_{X_{\widehat{H}}}\omega=-\de\widehat{H}$, where $\mathbf{i}$ is the standard vector-field insertion and $\de\widehat{H}$ is an operator-valued one-form. Similarly, one has $\{\widehat{H},\Upsilon\}=-  {\rm d} \Upsilon \cdot  X_{\widehat{H}}$. 
 As usual, we assume that $M$ is endowed with a volume form $\mu$ so that the inner product and symplectic form on $L^2(T^*{Q}\times M)$ are defined by the immediate generalization of the classical definitions \eqref{inner_symplectic}. To keep consistent with the previous notation, here we shall denote the \emph{hybrid quantum--classical Hilbert space} by
\begin{equation}
\mathscr{H}_{\scriptscriptstyle QC}:=L^2(T^*{Q}\times M)
\,.
\end{equation}

Let us now discuss some of the main points of the quantum--classical wave equation \eqref{hybrid_KvH}. First, we see that the \textit{hybrid quantum--classical Liouvillian} 
\beq\label{hybliouv}
\widehat{\cal L}_{\widehat{H}}=\{{\rm i}\hbar \widehat{H},\ \} + \big(\widehat{H} - {\mathcal{A}}\!\cdot\! X_{\widehat{H}}\big)
\eeq
 is an unbounded Hermitian operator on $\mathscr{H}_{\scriptscriptstyle QC}$ so that  \eqref{hybrid_KvH}  takes the compact form ${\rm i}\hbar\partial_t\Upsilon=\widehat{\cal L}_{\widehat{H}}\Upsilon$ and the 
 hybrid wavefunction $\Upsilon$ undergoes unitary dynamics. We notice that, similarly to the injective correspondence $H\mapsto \widehat{\cal L}_{{H}}$ from the purely classical case, the hybrid correspondence $\widehat{H} \rightarrow \widehat{\cal L}_{\widehat{H}}$ is also injective. The algebraic properties of quantum--classical Liouvillians were recently studied in \cite{GBTr20}. 
 
A particularly important point for the later developments in this paper involves the Hamiltonian structure of \eqref{hybrid_KvH}. Upon considering the immediate generalization of the symplectic form in \eqref{inner_symplectic}, we notice that the quantum--classical wave equation \eqref{hybrid_KvH} is  Hamiltonian with the following Hamiltonian functional expressed in terms of the quantum Liouvillian:
\begin{equation}\label{hybHam}
h(\Upsilon)=\int_{T^*{Q}}\!\big\langle\Upsilon\big|\widehat{\cal L}_{\widehat{H}}\Upsilon\big\rangle\,\Lambda
:=
\int_{T^*{Q}}\int_M\big( \,\overline{\!\Upsilon\!}\,\,\widehat{\cal L}_{\widehat{H}}\,\Upsilon\big)\,\Lambda\wedge \mu
\,.
\end{equation}
From now on, we will use $\langle\cdot|\cdot\rangle$ to denote only the Hermitian inner product on $\mathscr{H}_{\scriptscriptstyle Q}$, unless otherwise specified.
In particular, the quantum--classical wave equation \eqref{hybrid_KvH} arises as the critical condition for the following variational principle:
\beq\label{DFVP}
\delta\int_{t_1}^{t_2}\!\int_{T^*{Q}}\!\big\langle\Upsilon,i\hbar\partial_t\Upsilon-\widehat{\cal L}_{\widehat{H}}\Upsilon\big\rangle\,\Lambda\, {\rm d} t=0\,,
\eeq
with arbitrary variations $ \delta \Upsilon $ vanishing at $t=t_1,t_2$. Here, the notation $\langle\cdot,\cdot\rangle=\operatorname{Re}\langle\cdot|\cdot\rangle$ stands for the real-valued pairing on $\mathscr{H}_{\scriptscriptstyle Q}$. 
Notice that \eqref{DFVP} is a standard phase-space  variational principle of the form 
\beq\label{geomDFVP}
\delta\int_{t_1}^{t_2}\! \big( \Theta ( \Upsilon ) \cdot\dot \Upsilon - h( \Upsilon ) \big) \, {\rm d} t=0\,.
\eeq
 Indeed, given the Hilbert space $\mathscr{H}_{\scriptscriptstyle QC}$, its hybrid symplectic form extending the classical version \eqref{inner_symplectic} can be written as $  \Omega =-{\rm d} \Theta$, where $\Theta$  is the canonical one-form. Specifically, we have $\Theta ( \Upsilon ) \cdot \Upsilon' : = - \hbar \operatorname{Im} \int_{T^*Q} \left\langle  \Upsilon (z)| \Upsilon'(z) \right\rangle  \Lambda$
 where $\Upsilon'\in\mathscr{H}_{\scriptscriptstyle QC}$ and the dot product notation $\cdot$ stands for the general vector-covector contraction (in this case, on an infinite-dimensional space).
 %at $\Upsilon$, that is $\Upsilon'\in T_{\Upsilon}\mathscr{H}_{\scriptscriptstyle QC}$.

The canonical Hamiltonian structure identified by \eqref{DFVP} will be crucial in the next sections. Indeed, while so far we have simply reviewed the current status of the hybrid theory, the following sections will exploit the variational structure \eqref{DFVP} to present a new noncanonical variant ensuring that the classical density retains its initial sign over time. However, before we enter the details,  the next section illustrates how the quantum and classical densities are related to the hybrid wavefunction.
%\todo{CT: can we avoid the index notation and use cross products and divergences? Actually, is there an easy way to write the divergence in the KvH momap in a coordinate-free way? In my understanding, that is $\pounds_{J{\mathcal{A}}}|\Psi|^2$, where J is the Poisson tensor contracting with the one-form ${\mathcal{A}}$. Then, the corresponding part of the infinitesimal generator may be written as $-\pounds_{J{\mathcal{A}}}H$. Do you agree? Maybe we want to write this some place?\\
%\color{blue}FGB: I am not sure to see what you mean. By a general formula we have
%\[
%\operatorname{div}(|\Psi|^2 J{\mathcal{A}})=|\Psi|^2\operatorname{div}(J{\mathcal{A}}) + \pounds_{J{\mathcal{A}}}|\Psi|^2.
%\]
%Now,  with respect to the Liouville form we have $\operatorname{div}(J{\mathcal{A}})= \sum_i \partial_{p_i} (-p_i)=-n$, since in local coordinates the vector field is $J{\mathcal{A}} (q,p)= - p_i \partial_{p_i}$. So, we have
%\[
%\operatorname{div}(|\Psi|^2 J{\mathcal{A}})= - n |\Psi|^2 + \pounds_{J{\mathcal{A}}}|\Psi|^2
%\]}

\subsection{The hybrid density operator\label{sec:hybden}}

As shown in \cite{BoGBTr19}, the Hamiltonian structure of the quantum--classical wave equation \eqref{hybrid_KvH} leads to defining a hybrid density operator for the evaluation of expectation values. This is a necessary ingredient for the identification of the quantum and classical densities, which are obtained by suitable projections of the hybrid quantity. Notice that, while in the present construction the entire information is encoded in the hybrid wavefunction, other QC theories \cite{Aleksandrov,boucher,Gerasimenko,Diosi,Kapral} consider the hybrid density operator as the elementary object. As discussed in \cite{BoGBTr19}, these two approaches are essentially different.

In the present context, the hybrid density operator can be identified by rewriting the Hamiltonian functional \eqref{hybHam} by using integration by parts as follows:
\beq\label{Dintro}
h(\Upsilon)=\int_{T^*{Q}}\!\big\langle\Upsilon\big|\widehat{\cal L}_{\widehat{H}}\Upsilon\big\rangle\,\Lambda
=\operatorname{Tr}\int _{T^*{Q}\!}\widehat{H}(z)\widehat{\cal D}(z)\,\Lambda
\,.
\eeq
In analogy to the expression \eqref{KvHmomap} of the classical Liouville density, the   hybrid density operator $\widehat{\cal D}$ is given as
\begin{equation}\label{hybridDenOp}
\widehat{\cal D}=\Upsilon\Upsilon^\dagger - \operatorname{div}\!\big( \Bbb{J}{\mathcal{A}} \Upsilon \Upsilon^\dagger\big) + {\rm i}\hbar\{\Upsilon,\Upsilon^\dagger\}\in\operatorname{Den}(T^*Q)\otimes\operatorname{He}(\mathscr{H}_{\scriptscriptstyle Q})\,,
\end{equation}
so that $\operatorname{Tr}\int_{\scriptscriptstyle T^*\!Q}\widehat{\cal D}\,\Lambda=1$ and we have denoted by $\operatorname{He}(\mathscr{H}_{\scriptscriptstyle Q})$ the space of Hermitian operators on $\mathscr{H}_{\scriptscriptstyle Q}$. Here, we have defined the \emph{quantum adjoint} as
\begin{equation}\label{gigi}
\Upsilon^\dagger(z)\psi:=\langle\Upsilon(z)|\psi\rangle= \int_M \bar{\Upsilon}(z,x)\psi(x)\,\mu
\,,
\end{equation}
for all $ z  \in T^*Q$ and all $\psi\in \mathscr{H}_{\scriptscriptstyle Q}$.
Given the hybrid density  $\widehat{\cal D}$, one computes the quantum density operator 
\beq\label{QuantDensMat}
\hat\rho_q:=\int_{ T^*\!Q\!}\widehat{\cal D}(z)\,\Lambda = \int_{T^*\!Q\!} \Upsilon(z)\Upsilon^\dagger(z)\,\Lambda
\,,
\eeq
%whose kernel is given exactly by \eqref{QDM}, i.e. ${\cal K}_{\hat\rho}(x,x')=\int_{T^*{Q}}{\cal K}_{\widehat{\cal D}}(z;x,x')\Lambda = \int_{T^*{Q}}\Upsilon(z,x)\bar{\Upsilon}(z,x')\Lambda$. 
which is evidently positive semidefinite by construction. 
On the other hand,  the classical density reads
$
\rho_c(z)=\operatorname{Tr}\widehat{\cal D}(z)$.
Here the trace is  computed only with respect to the quantum degrees of freedom, so that
\begin{align}\label{rhoc}
\rho_c(z):=&
\operatorname{Tr}\widehat{\cal D}(z) =\int_M  \Big(|\Upsilon(z,x)|^2-  \operatorname{div}\!\big( \Bbb{J}\mathcal{A}|\Upsilon(z,x)|^2\big) + {\rm i}\hbar \{\Upsilon, \bar\Upsilon\}(z,x)\Big)\,\mu\,.
\end{align}
As anticipated, this expression is not positive--definite and it is still unclear whether negative values may develop during the time evolution. In \cite{GBTr20}, we identified an infinite family of hybrid Hamiltonians $h( \Upsilon )$ preserving the initial sign of $\rho_c$. However, more general results are still lacking. Otherwise, the hybrid model based on the quantum-classical wave equation \eqref{hybrid_KvH} satisfies the properties 2-5 of desirable self-consistency criteria presented in the Introduction. We address the reader to \cite{GBTr20} for details on property 3.

We conclude this section by pointing out that  the quantity $-i\hbar\hat\rho_q$, with $\hat\rho_q$ as in \eqref{QuantDensMat}, and the classical density $\rho_c$ in \eqref{rhoc} are momentum maps for the natural actions on $\mathscr{H}_{\scriptscriptstyle CQ}$ of the unitary group $\mathcal{U}(\mathscr{H}_{\scriptscriptstyle Q})$ and the connection-preserving prequantum bundle automorphisms $\operatorname{Aut}_{\cal A}(T^*{Q}\times S^1)$, respectively. Eventually, all these quantities are related as in the following diagram:

%\begin{figure}[h!]
{\noindent
\scriptsize
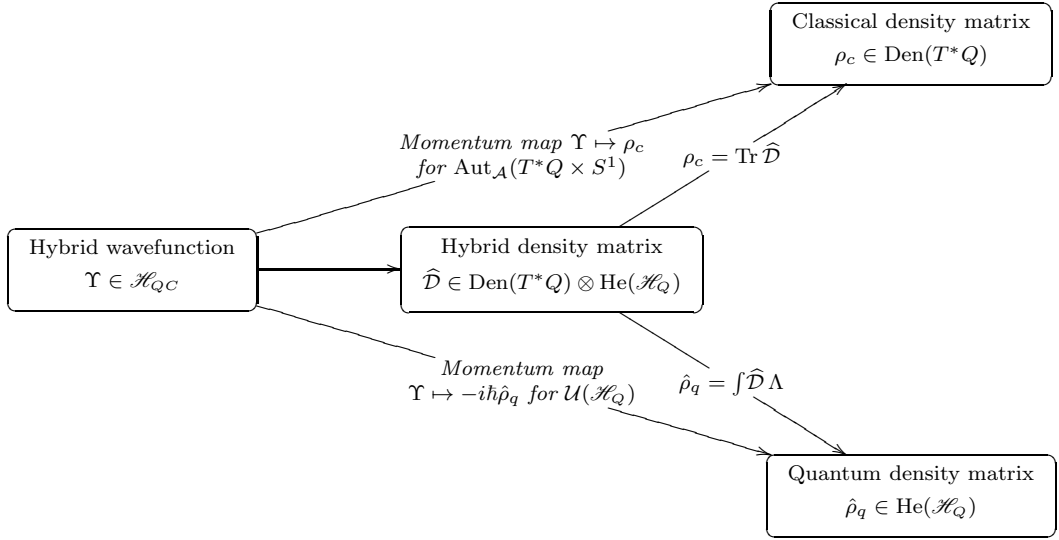
\begin{center}
\begin{xy}
\hspace{.5cm}\xymatrix{
& & & &  *+[F-:<3pt>]{
\begin{array}{c}
\vspace{0.1cm}\text{Classical density matrix}\\
\vspace{0.1cm} \rho  _c\in\operatorname{Den}(T^* Q)
\end{array}
} \\
& & & & \\
& *+[F-:<3pt>]{
\begin{array}{c}
\vspace{0.1cm}\text{Hybrid wavefunction}\\
\vspace{0.1cm} \Upsilon\in \mathscr{H}_{\scriptscriptstyle QC}
\end{array}
}
\ar[rr]
\ar[uurrr]
|{\begin{array}{c}\textit{Momentum map $\Upsilon\mapsto\rho_c$}\\
\textit{for 
$\operatorname{Aut}_{\cal A}(T^*{Q}\times S^1)$}\\
\end{array}}
\ar[ddrrr]|{\begin{array}{c}\textit{Momentum map}\\
\textit{$\Upsilon\mapsto-i\hbar{\hat\rho}_q$ for 
$\mathcal{U}(\mathscr{H}_{\scriptscriptstyle Q})$}\\
\end{array}}
& & *+[F-:<3pt>]{
\begin{array}{c}
\vspace{0.1cm}\text{Hybrid density matrix}\\
\vspace{0.1cm} \widehat{ \mathcal{D}}
\in\operatorname{Den}(T^* Q)\otimes \operatorname{He}(\mathscr{H}_{\scriptscriptstyle Q}) 
\end{array}
}\ar[uur]|{\begin{array}{c}\rho_c=\operatorname{Tr}\widehat{\cal D}
\end{array}}\ar[ddr]|{\begin{array}{c}\hat{\rho}_q=\int \!\widehat{\cal D}\,\Lambda
\end{array}}  \\
& & & & \\
& & & & *+[F-:<3pt>]{
\begin{array}{c}
\vspace{0.1cm}\text{Quantum density matrix}\\
\vspace{0.1cm} \hat{\rho}_q \in \operatorname{He}(\mathscr{H}_{\scriptscriptstyle Q})
\end{array}
} \\
}
\end{xy}
%\end{figure}
\end{center}
}

\noindent
This completes  our review of the present status of the KvH theory of hybrid quantum--classical dynamics. As mentioned previously, the main purpose of this work is to exploit this framework in order to formulate a closure model ensuring that the sign of the classical density is preserved in time. We achieve this goal by combining the hybrid KvH theory with a particular factorization of the quantum--classical wavefunction. These approaches will be combined within the context of the variational structure given by \eqref{DFVP}. This is the topic of the rest of this paper.

\section{Exact factorization of hybrid wavefunctions\label{sec:EF}}

In this section, we apply a constraint to the hybrid quantum--classical model from Section \ref{sec:QChybrids} so that both the quantum and classical densities  are positive at all times. Since the quantum density operator \eqref{QuantDensMat} is already positive-definite by construction, the enforcement of this constraint concerns the classical density \eqref{rhoc} and thus it needs to involve the classical degrees of freedom. However, it appears that the latter  cannot be  isolated in a simple way since the current form of the theory involves a hybrid wavefunction $\Upsilon(z,x)$ in which quantum and classical coordinates, i.e. $x$ and $z=(q,p)$, respectively, are treated on an equal footing.

In order to circumvent this difficulty, we resort to a method from chemical physics \cite{AbediEtAl2012}. Known under the name of \emph{exact factorization}, in our context this method simply consists in rewriting the hybrid wavefunction as follows:
\beq\label{EFDef}
\Upsilon(t,z,x)=\chi(t, z)\psi(t, x;z)\,,
\qquad\text{with}\qquad
\int_M|\psi(t,x;z)|^2\,\mu=1, \quad \forall z \in T^*Q
\,,
\eeq
where the semicolon indicates that the wavefunction $\psi$ depends on the classical coordinates only parametrically. In other words, one has a  Koopman  wavefunction $\chi(t)\in L^2(T^*Q)$ and a parameterized Schr\"odinger  wavefunction $\psi(t)\in{\cal F}(T^*Q,\mathscr{H}_{\scriptscriptstyle Q})$, where ${\cal F}(T^*Q,\mathscr{H}_{\scriptscriptstyle Q})$ denotes the space of mappings $T^*Q\to\mathscr{H}_{\scriptscriptstyle Q}$. While so far we restricted to the infinite-dimensional case $\mathscr{H}_{\scriptscriptstyle Q}:=L^2( M)$, in what follows we shall consider an arbitrary quantum Hilbert space $\mathscr{H}_{\scriptscriptstyle Q}$ with inner product $\langle\cdot|\cdot\rangle$, induced norm $\|\!\cdot\!\|$,  and real-valued pairing $\langle\cdot,\cdot\rangle=\operatorname{Re}\langle\cdot|\cdot\rangle$.
Notice that both  wavefunctions $\chi$ and $\psi$ in the factorization \eqref{EFDef} are only defined up to a phase factor $e^{i\phi(t,z)}$, thereby requiring a gauge fixing that will be eliminated in the following sections. 

The name \emph{exact factorization} arises from the fact that the relation \eqref{EFDef}  generally identifies an  exact solution of \eqref{hybrid_KvH}, as long as $\|\Upsilon\|$ is nowhere vanishing. Indeed, in the latter case, one may always define $\psi:=\Upsilon/\|\Upsilon\|$ and $\chi:=\|\Upsilon\|$. As we refrain from discussing the technicalities associated to more general cases, here we shall simply regard \eqref{EFDef} as an ansatz. Over the years, factorizations of the type \eqref{EFDef}  appeared  in the context of standard quantum mechanics. For example, this is the common approach to the hydrodynamic formulation of the Pauli equation \cite{BBirula}. However, it was only in \cite{AbediEtAl2012} that this type of wavefunction factorization was recognized to have more general validity. The geometric underpinning of \eqref{EFDef} was studied in \cite{FoHoTr19,FoTr,HoRaTr21} in the context of Euler-Poincar\'e reduction \cite{HoMaRa1998}, and the results therein provide the basis for the present work.

\subsection{Variational structure\label{sec:VS}}
This section sets up the framework for our key results  by presenting the variational structure of the evolution equations for the quantities $\chi$ and $\psi$ in \eqref{EFDef}. The treatment follows closely the discussion in \cite{FoHoTr19}. As we will see, the main point about this section is the emergence of a Berry connection on phase-space: while the presence of this connection will allow removing the gauge choice entirely, it will also lead to a natural strategy for devising a noncanonical closure model for  hybrid quantum--classical dynamics.

\paragraph{Direct variational formulation.}
As a first step, we substitute the ansatz \eqref{EFDef} into the variational principle \eqref{DFVP}, thereby obtaining
\beq
\delta\int_{t_1}^{t_2}\!\int_{T^*\!Q\!}\Big(\operatorname{Re}(i\hbar\bar\chi\partial_t\chi)+|\chi|^2\langle\psi,i\hbar\partial_t\psi\rangle
-\langle\psi,\bar\chi \widehat{\mathcal{L} } _{\widehat{H}}(\chi \psi)\rangle\Big)\,\Lambda\,\de t=0
\,,
\label{VP1}
\eeq
where we recall  the hybrid quantum--classical Liouvillian $\widehat{ \mathcal{L} }_{\widehat{H}}$  defined in \eqref{hybliouv}. Since the variations $ \delta \chi $ and $ \delta \psi $ are arbitrary, it may be shown \cite{FoHoTr19} that this variational principle is entirely equivalent to replacing the ansatz \eqref{EFDef} in the original equation \eqref{hybrid_KvH}. Also, notice that here we have not enforced the constraint $
%\int_M|\psi(t,x;z)|^2\,\sigma
\|\psi(z)\|^2
=1$; 
as we will see, this condition is preserved in time by the equation of motion for $\psi$.

At this point, it may be convenient to express $\chi$ in terms of its Koopman density $D$ and phase $S$, following the discussion on the Madelung transform from Section \ref{sec:Madelung1}. Upon using  the polar form $\chi=\sqrt{D}e^{iS/\hbar}$, the variations $ \delta D $ and $ \delta S$ are again arbitrary so that \eqref{VP1} becomes
\beq
\delta\int_{t_1}^{t_2}\bigg[\int\!D\big(\partial_tS-\langle\psi,i\hbar\partial_t\psi\rangle\big)\Lambda
+h(D,S,\psi)\bigg]\,\de t=0
\,.
\label{VP1bis}
\eeq
Here, we have dropped the integration domain ${T^*Q}$ for convenience of notation and we have rewritten the original Hamiltonian functional \eqref{hybHam} as
\begin{align}
h(D,S,\psi)=& \int\langle\psi,\bar\chi \widehat{ \mathcal{L} }_{\widehat{H}}(\chi\psi)\rangle\,\Lambda
=
\int \!D\,\Big(\big\langle\psi,X_{\widehat{H}}\psi\big\rangle\cdot\de S
+\big\langle\psi, \widehat{ \mathcal{L} }_{\widehat{H}}\psi\big\rangle\Big)\,\Lambda
\,.
\label{covHam}
\end{align}
Then, upon introducing the short-hand notation $\langle \widehat{A}\rangle=\langle\psi\big|\widehat{A}\psi\rangle$, the equations of motion for $(S(t),D(t))$ obtained from \eqref{VP1bis} are written as
\begin{align}
&\partial_t D+\operatorname{div}(D\langle X_{\widehat{H}}\rangle)=0,\label{mario}\\
&\partial_t S+\langle X_{\widehat{H}}\rangle\cdot\de S=\big\langle\psi,i\hbar\partial_t\psi-\widehat{ \mathcal{L} }_{\widehat{H}}\psi\big\rangle,\label{S_equ}
\end{align}
and taking the differential of the latter yields
\begin{equation}\label{dS_equ} 
\big(\partial_t +\pounds_{\langle X_{\widehat{H}}\rangle}\big)\de S=\de \big\langle\psi,i\hbar\partial_t\psi-\widehat{ \mathcal{L} }_{\widehat{H}}\psi\big\rangle.
\end{equation} 
It may be useful to notice that, since the exact factorization \eqref{EFDef} is defined only up to a gauge, the Koopman phase $S$ can be chosen at will and may even be set to zero. However, in the present work we prefer to avoid fixing a particular gauge. Also, we  notice that, since $\pounds_{\langle X_{\widehat{H}}\rangle}{\cal A}=\de\langle X_{\widehat{H}}\cdot{\cal A}\rangle-\langle\de\widehat{H}\rangle$,  equation \eqref{dS_equ} can also be cast in the form
\beq\label{giselle}
\big(\partial_t +\pounds_{\langle X_{\widehat{H}}\rangle}\big)(\de S-{\cal A})=\langle\de\widehat{H}\rangle+\de \big\langle\psi,i\hbar\partial_t\psi- (X_{\widehat{H}}\cdot{\cal A}+\widehat{ \mathcal{L} }_{\widehat{H}})\psi\big\rangle
\eeq
which extends \eqref{dS_theta} to the hybrid case.

Other than the equations \eqref{mario}-\eqref{S_equ}, the variational principle \eqref{VP1bis} also produces the equation for  $\psi(t)$, which at this stage we write compactly as \cite{FoHoTr19}
\beq\label{psieq1}
i\hbar\partial_t(D\psi)+i\hbar D\partial_t\psi=\frac{\delta h}{\delta\psi}
\,.
\eeq
By expanding both sides above one verifies that the normalization $\|\psi(z)\|^2=1$ is indeed preserved in time, although this point will be made clear in the next section.

\paragraph{Alternative variational principle and Lagrangian paths.}
The system of equations \eqref{mario}, \eqref{dS_equ}, and \eqref{psieq1} may be given a variational structure partly analogous to a Lie-Poisson variational principle \cite{Cendra} as follows. From the stationarity condition \eqref{mario} associated to $ \delta S$, 
%namely
%\begin{equation}\label{stat_cond} 
%\partial _t D + \operatorname{div}(D {\cal X})=0 \qquad\text{for}\qquad {\cal X}= \left\langle  \psi \big|  X_{\widehat{H}}  \psi \right\rangle,
%\end{equation}
the evolution of $D(t)$ is given as a push-forward $\eta(t)_*D_0$ of the initial condition $D_0$ as follows:
\begin{equation}\label{stat_cond_integrated}
D=\eta_*D_0= (D_0 \circ \eta ^{-1} )\operatorname{Jac}( \eta  ^{-1})\,,\ \qquad\text{with}\qquad \dot \eta \circ \eta ^{-1} = \mathcal{X} \qquad\text{and}\qquad \mathcal{X} =  \langle    X_{\widehat{H}}   \rangle.
\end{equation}
A variational formulation alternative to \eqref{VP1} may be obtained by inserting a Lagrange multiplier $ \sigma $ to impose the last condition in \eqref{stat_cond_integrated} and considering the constrained variations $ \delta D$ and $ \delta \mathcal{X} $ induced by the free variations $ \delta \eta $ in the first two conditions in \eqref{stat_cond_integrated}. This gives the variational principle
\beq
\delta\int^{t_2}_{t_1}\!\int\Big(\sigma \cdot {\cal X} +D\langle\psi,i\hbar \partial _t \psi\rangle- \sigma \cdot 
\big\langle\psi, X_{\widehat{H}}\psi\big\rangle  
-D\,\big\langle\psi,\widehat{ \mathcal{L} }_{\widehat{H}}\psi\big\rangle\Big)\,\Lambda\, {\rm d} t
=0
\,,
\label{VP2}
\eeq
with respect to arbitrary variations $\delta\sigma$ and $\delta\psi$ and constrained Euler-Poincar\'e variations \cite{HoMaRa1998}
\beq
\delta {\cal X}=\partial_t{\cal Y}+{\cal X}\cdot\nabla{\cal Y}-{\cal Y}\cdot\nabla{\cal X}
\qquad\text{and}\qquad \delta D=-\operatorname{div}(D{\cal Y})
\,,
\label{nicola}
\eeq
As usual in the theory of Euler-Poincar\'e variations, ${\cal Y}:=\delta \eta \circ \eta ^{-1} $ is arbitrary and vanishing at the endpoints. The $\sigma-$terms in \eqref{VP2} comprise the structure of a Lie-Poisson variational principle \cite{Cendra} and one can express $(D,\sigma)$ in terms of  the momentum map \eqref{Joli} to write $\sigma=D {\rm d} S$. Upon using the Lie derivative notation, the stationarity conditions associated to \eqref{VP2} yield
\beq\label{fluidmomedens}
\left(\frac{\partial}{\partial  t}  + \pounds _ \mathcal{X} \right)\frac{ \sigma }{D}= {\rm d} \left\langle\psi,i\hbar\frac{\partial \psi}{\partial  t}-\widehat{ \mathcal{L} }_{\widehat{H}}\psi\right\rangle, 
\qquad\qquad \mathcal{X} = \langle X_{\widehat{H}}\rangle
\,,
\eeq
as well as \eqref{psieq1}. 
Therefore the equations \eqref{mario}, \eqref{dS_equ}, and \eqref{psieq1} for  $D$, $ {\rm d} S$, and $ \psi $, respectively, are obtained upon selecting the invariant solution $ \sigma = D{\rm d} S$  of equation \eqref{fluidmomedens}. Notice that one can also write \eqref{fluidmomedens} in the form \eqref{giselle}. In summary, the exact factorization \eqref{EFDef} takes the original variational principle \eqref{DFVP} for the hybrid wavefunction into \eqref{VP1}. Then, the Madelung momentum map \eqref{Joli} takes \eqref{VP1} into the the form \eqref{VP2}. This process is illustrated by the following diagram:

%\begin{figure}[h!]
{\noindent
\footnotesize
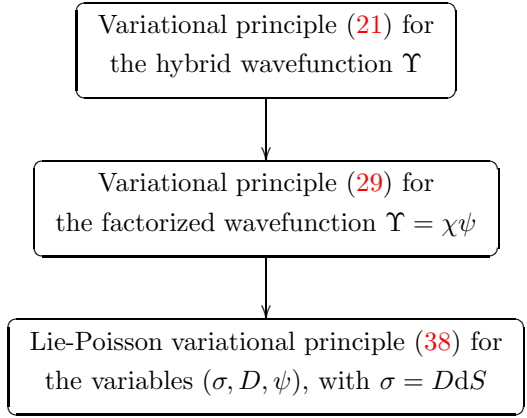
\begin{center}
\hspace{2.5cm}
\begin{xy}
\hspace{-1cm}\xymatrix{
& *+[F-:<3pt>]{
\begin{array}{c}
\vspace{0.1cm}\text{Variational principle \eqref{DFVP} for}\\
\vspace{0.1cm}\text{the hybrid wavefunction $\Upsilon $}
\end{array}
}
\ar[d]
& \\
& *+[F-:<3pt>]{ 
\begin{array}{c}
\vspace{0.1cm}\text{Variational principle \eqref{VP1} for}\\
\vspace{0.1cm}\text{the factorized wavefunction $\Upsilon = \chi \psi $}
\end{array}
}\ar[d] & \\
&*+[F-:<3pt>]{
\begin{array}{c}
\vspace{0.1cm}\text{Lie-Poisson variational principle \eqref{VP2} for}\\
\vspace{0.1cm}\text{the variables $( \sigma , D, \psi )$, with $ \sigma = D {\rm d} S$}
\end{array}
} 
}
\end{xy}
\end{center}
}
%\end{figure}

\smallskip
The flow $\eta$ of the vector field $\cal X$ introduced above will play a central role in the next sections. As it appears from \eqref{stat_cond_integrated},  the diffeomorphism $\eta$ advances the classical quantity $D$ by the push-forward action on densities, thereby leading to an analogy with fluid density transport. This analogy also emerges in equation \eqref{fluidmomedens}, which indeed involves an {\it advective derivative} (in fluid dynamics terminology) along the vector field $\cal X$ on the left-hand side. Then, the flow $\eta$ can be thought of as identifying  the Lagrangian frame of a phase-space fluid with density $D$ and momentum density $\sigma$, and moving with Eulerian velocity $\cal X$. Since $D$ and $\sigma$ are defined by the Madelung transform  of the classical wavefunction $\chi=\sqrt{D} e^{iS/\hbar}$, we may refer to this fluid as the `classical fluid' so that the Lagrangian fluid frame $\eta$ will be called \emph{classical  frame}.
This Lagrangian fluid frame  governs the dynamics of the Poincar\'e integral on phase-space. Indeed, we may use the relation $\pounds_{\langle X_{\widehat{H}}\rangle}{\cal A}=\de\langle X_{\widehat{H}}\cdot{\cal A}\rangle-\langle\de\widehat{H}\rangle$ to write 
\beq
\frac{\de}{\de t}\oint_{\gamma(t)}  \mathcal{A} =-\oint_{\gamma(t)}\langle\de\widehat{H}\rangle
\ \implies\ 
\frac{\de}{\de t}\eta^*\omega=\eta^*\de \langle\de\widehat{H}\rangle
\,,
\label{caterina}
\eeq
where  $\gamma (t)= \eta (t, \gamma _0)$ is an arbitrary loop moving with $\eta(t)$ and  the implication is a direct consequence of Stokes theorem. The purely classical case is recovered when $\widehat{H}=H\operatorname{Id}$, since in that case $\langle\de\widehat{H}\rangle=\de H$.

\paragraph{Densities and Berry connection.}
We notice that under the assumption \eqref{EFDef} the quantum and classical densities in \eqref{QuantDensMat}-\eqref{rhoc} are written as
\beq\label{densities2}
\hat{\rho}_q=\int\! D\psi\psi^\dagger\,\Lambda
\qquad\text{and}\qquad 
\rho_c=D+\operatorname{div}\!\left[\Bbb{J}\left(\sigma+D{\cal A}_B-D{\cal A}\right)\right]\,,
\eeq
where 
\beq
\label{BerryConn}
{\cal A}_B:=\langle\psi,-i\hbar\de\psi\rangle \in  \Omega ^1 (T^*Q)
\eeq
is the \emph{Berry connection on phase-space}. Notice that this quantity can be written in geometric terms as $ \mathcal{A}_B= - \psi ^* \Theta $, where we recall that $\psi:T^*Q\to\mathscr{H}_{\scriptscriptstyle Q}$ and $\Theta$ is the  canonical one-form on $\mathscr{H}_{\scriptscriptstyle Q}$, analogue to that appearing in \eqref{geomDFVP}. Specifically, $ \Theta ( \psi ) \cdot \psi' = - \hbar \operatorname{Im} \left\langle \psi |\psi' \right\rangle = \left\langle \psi , i \hbar \psi' \right\rangle $, so that the corresponding Berry curvature ${\cal B}=\de\mathcal{A} _B$ reads $ {\cal B}=- \psi ^* \de\Theta=:  \psi ^* \Omega $. For the general momentum map structure of the Berry connection, see \cite{Tronci2018}. 

We recognize that, if we could enforce the condition 
\beq\label{constraint1}
D^{-1}\sigma+{\cal A}_B={\cal A},
\eeq
then the classical density $\rho_c$ in \eqref{densities2} would satisfy a pure transport equation retaining the initial sign. Notice that, upon recalling $\sigma=D\de S$, the condition \eqref{constraint1}  enforces a time-independent Berry curvature, that is ${\cal B}(t)=-\omega = {\rm d} \mathcal{A}  $. In order to enforce this condition, it is convenient to make the entire treatment manifestly gauge invariant by expressing the variational principle in term of the projection $\rho=\psi\psi^\dagger$. This is the topic of the next section.

\begin{remark}[von Neumann operators]\label{VNops}
The relation \eqref{constraint1} may be difficult to satisfy. Indeed, this requires finding a particular form of $\psi$ such that ${\cal A}+\langle\psi,i\hbar\de\psi\rangle=\de S$.
  Then, one may think of extending the exact factorization method to allow for a more general type of relation that is solved by a wider class of $\psi$. Following \cite{FoHoTr19,GBTr21b}, this extension may be realized upon resorting to von-Naumann operators, as recently shown in \cite{GBTr22}. 
 \rem{ %%%%%%%%%%%%%%%%%%%%%% 
 In this case,  the quantum--classical wave equation \eqref{hybrid_KvH} is replaced  by $i\hbar\partial_t\widehat{\Theta}=[\widehat{\cal L}_{\widehat{H}},\widehat\Theta]$, where $\widehat{\Theta}$ is a unit-trace Hermitian operator on the quantum--classical Hilbert space $\mathscr{H}_{\scriptscriptstyle QC}$. Upon writing $\widehat{\Theta}=\sum_{a=1}^Nw_a\Upsilon_a\Upsilon_a^\dagger$,  a convenient factorization of $\widehat{\Theta}$ may be realized by mimicking \eqref{EFDef} as $\Upsilon_a=\chi_a\psi$. Notice that, unlike the rest of this paper, here the superscript $\dagger$ denotes the adjoint in the full hybrid space $\mathscr{H}_{\scriptscriptstyle QC}$. The relation $\Upsilon_a=\chi_a\psi$ means that the quantum part  $\psi$ of each hybrid wavefunction $\Upsilon_a$ does not depend on the label $a$, so that \eqref{VP1} is replaced by
\[
\delta\int_{t_1}^{t_2}\!\int\sum_{a}^N\!\Big(\operatorname{Re}(i\hbar\bar\chi_a\partial_t\chi_a)+|\chi_a|^2\langle\psi,i\hbar\partial_t\psi\rangle
-\langle\psi,\bar\chi_a \widehat{\mathcal{L} } _{\widehat{H}}(\chi_a \psi)\rangle\Big)\Lambda\,\de t=0
\,.
\]
 Following  \cite{FoHoTr19},  one shows that this procedure leaves the final equations \eqref{mario},  \eqref{fluidmomedens}, and \eqref{psieq1}  unchanged. Indeed, upon writing $\chi_a=\sqrt{D_a}e^{iS_a/\hbar}$ and denoting $D=\sum_{a=1}^Nw_a D_a$ and $\sigma=\sum_{a=1}^Nw_a D_a\de S_a$, one is left  with the same variational principle as in \eqref{VP2}. However, since now $\de(D^{-1}\sigma)\neq0$, one  allows for a larger class of $\psi$  satisfying the relation \eqref{constraint1}. 
}  %%%%%%%%%%%%%%%%%%%%%%
 \end{remark}

\subsection{Quantum wavefunction in the classical frame\label{sec:classframe}}

So far, the evolution of $\psi$ has been  written as in \eqref{psieq1}. However,  the latter equation does not disclose the geometric  structure underlying the quantum evolution, which is instead the  subject of this section. In particular, we will see how the quantum wavefunction $\psi$ evolves unitarily in the classical fluid frame identified by the Lagrangian flow $\eta$ of the classical phase-space fluid introduced in the previous section. Within the present construction, this is a key step to devise a closure model for hybrid quantum--classical dynamics.

\paragraph{Unitary evolution of the quantum wavefunction.}
In order to unfold the geometric structure of the quantum evolution \eqref{psieq1},  it is convenient to define the functional
\begin{equation}\label{function_f}
f(D,\psi)=\int \!D\,\big\langle\psi, (\widehat{ \mathcal{L} }_{\widehat{H}}-{\cal A}_B\cdot X_{\widehat{H}})\psi\big\rangle\,\Lambda= - \int \!D \,\big\langle\psi,\big(X_{\widehat{H}} \cdot ( \mathcal{A} _B +  i \hbar\, {\rm d} )   + \widehat{\mathscr{L}} \,\;\big)\psi\big\rangle\, \Lambda 
\end{equation}
so that the relation
\begin{equation}\label{h_f}
h(D, \sigma, \psi )=\int \big(\sigma \cdot 
\big\langle\psi, X_{\widehat{H}}\psi\big\rangle  
+ D\,\big\langle\psi,\widehat{ \mathcal{L} }_{\widehat{H}}\psi\big\rangle\big) \Lambda =f(D, \psi )+\int (\sigma+D{\cal A}_B) \cdot \langle X_{\widehat{H}}\rangle\,\Lambda
\end{equation} 
%and
%\[
%-\operatorname{div}(i\hbar D\langle\bX_{\widehat{H}}\rangle)\psi+i\hbar D\partial_t\psi=\frac12\frac{\delta f}{\delta\psi}-\frac12\frac{\delta}{\delta\psi}\int \!D\langle X_{\widehat{H}}\rangle\cdot{\cal A}_B\,\Lambda
%\]
yields equation \eqref{psieq1} in the form
\beq
\label{psieq2}
i\hbar D \big(\partial_t+\langle X_{\widehat{H}}\rangle\cdot {\rm d} \big)\psi=
\frac1{2}\frac{\delta f}{\delta\psi}+(\sigma+D{\cal A}_B)\cdot  X_{\widehat{H}}\psi
\,.
%=\left(\frac{\delta f}{\delta\rho}-D{\cal A}_B\cdot  X_{\widehat{H}}\right)\psi
\eeq
While this form of the  $\psi-$equation is still quite cumbersome, we observe that the functional $f(D,\psi)$ can be written in terms of $D(z)$ and $\rho(z)=\psi(z)\psi(z)^\dagger$. Indeed, we have
\begin{equation}\label{computations1} 
\big\langle\psi, X_{\widehat{H}} \cdot ({\cal A}_B+i\hbar\, {\rm d} \big) \psi\big\rangle
=
\langle\rho,i\hbar \, {\rm d} \rho \cdot X_{\widehat{H}}\rangle
=\frac12
\left\langle\rho,i\hbar\left[X_{\widehat{H}}^\ell,\partial_\ell\rho\right]\right\rangle
=\frac12
\left\langle i\hbar X_{\widehat{H}},\left[\rho, {\rm d} \rho\right]\right\rangle\,,
\end{equation} 
where $\left\langle A  , B \right\rangle = \operatorname{Re} \operatorname{Tr}( A  ^\dagger B )$ and we recall that  $\rho(z)$ and $
\widehat{H}(z)$ are both Hermitian.  
The first equality above is verified as follows:
\begin{align}\nonumber
\operatorname{Re}\big(i\hbar\operatorname{Tr} ( \rho X_{\widehat{H}} \cdot  {\rm d} \rho  )\big) 
=&\ 
\operatorname{Re}\big(
i\hbar\operatorname{Tr} \!\left( \psi\psi^\dagger X_{\widehat{H}} \cdot  ( {\rm d} \psi\psi^\dagger+\psi {\rm d} \psi^\dagger) \right) \big)
\nonumber
\\
=&\ 
\langle\psi,i\hbar X_{\widehat{H}} \cdot  {\rm d} \psi  \rangle
+
 \langle {\rm d} \psi,i\hbar\psi\rangle \cdot \langle X_{\widehat{H}}\rangle
\nonumber
\\
=&\ 
\big\langle\psi , X_{\widehat{H}} \cdot ({\cal A}_B+i\hbar \,{\rm d} \big)\psi\big\rangle
\label{computations2}
\,,
\end{align}
where we used $ \left\langle \psi | \psi \right\rangle =\|\psi\|^2=1$. Hence, we can write the function $f$ in \eqref{function_f} as
\begin{align*}
f(D,\rho)=&\,\frac12\int\! D\big(\!\left\langle  X_{\widehat{H}},i\hbar\left[\rho, {\rm d} \rho\right]\right\rangle-2\big\langle\,\widehat{\!\mathscr{L}\,}\big\rangle\big)\Lambda
\\
=&\, \frac12\int\! D\big(
\big\langle \rho, i\hbar\{\rho,\widehat{H}\}+i\hbar\{{\widehat{H}},\rho\}\big\rangle-2\big\langle\,\widehat{\!\mathscr{L}\,}\big\rangle\big)\,\Lambda
\,.
\end{align*}
Finally, the chain rule $\delta f/\delta\psi=2(\delta f/\delta\rho)\psi$ takes \eqref{psieq2} into the form
\beq
\label{psieq3}
i\hbar  D\big(\partial_t+\langle X_{\widehat{H}}\rangle\cdot {\rm d} \big)\psi= \bigg(\frac{\delta f}{\delta\rho}+(\sigma+D{\cal A}_B)\cdot  X_{\widehat{H}}\bigg)\psi
\,,
\eeq
which unfolds the geometric structure of the original form \eqref{psieq1}: since the parenthesis on the right-hand side is Hermitian, the evolution above indicates a unitary flow  expressed in the classical phase-space frame moving with the velocity vector field $\langle X_{\widehat{H}}\rangle$. For example, we have  $(\partial_t+\langle X_{\widehat{H}}\rangle\cdot {\rm d} )\|\psi\|^2=0$, so that the  normalization condition in \eqref{EFDef} is preserved in time. Indeed, we notice that the left-hand side of \eqref{psieq3} is nothing else than a material/advective derivative.

As a direct consequence of \eqref{psieq3} and following the treatment in \cite{FoHoTr19}, we have the following statement:

\begin{proposition} Denote by ${\cal U}(\mathscr{H}_{\scriptscriptstyle Q})$ the unitary group on $\mathscr{H}_{\scriptscriptstyle Q}$ and by $ \mathfrak{u}(\mathscr{H}_{\scriptscriptstyle Q})$ its Lie algebra.  Let $U(t)\in{\cal F}(T^*Q,{\cal U}(\mathscr{H}_{\scriptscriptstyle Q}))$ be a time-dependent function on $T^*Q$ with values in ${\cal U}(\mathscr{H}_{\scriptscriptstyle Q})$. Then, the solutions of \eqref{psieq1} are written in the form
\begin{equation}\label{sdp_action} 
\psi (t)= (U(t) \psi _0 ) \circ \eta (t) ^{-1} 
\,,
\end{equation} 
%i.e., $\psi(t, z)=U(t,\eta^{-1}(t,z))\psi_0(\eta^{-1}(t, z))$, 
where $\eta (t) \in \operatorname{Diff} (T^*Q)$ is the flow of $ \mathcal{X} $. In addition, equation \eqref{psieq1} is equivalent to
\[
\partial_t\psi+{\cal X}\cdot\de\psi=\xi\psi
\,,
\]
where  ${\cal X}=\langle X_{\widehat{H}}\rangle$ and $\xi(t):=\dot{U}(t)U(t)^{-1}\circ\eta(t)^{-1} \in \mathcal{F} (T^*Q, \mathfrak{u}(\mathscr{H}_{\scriptscriptstyle Q}))$ is given by 
\[
\xi=-\frac{i}{\hbar D}\bigg(\frac{\delta f}{\delta\rho}+(\sigma+D{\cal A}_B)\cdot  X_{\widehat{H}}\bigg)
\,.
\]
\end{proposition}
As we shall see, this step eliminates the gauge choice and leads to a variational principle that is manifestly gauge invariant, that is, it does not depend explicitly on the Berry connection \eqref{BerryConn}.

\begin{remark}[Semidirect product structure] For  later purpose, it is important to note that the evolution \eqref{sdp_action} arises from a left action of the semidirect product group 
\beq\label{sdpgroup}
\operatorname{Diff}(T^*Q) \,\circledS\, \mathcal{F} (T^*Q, \mathcal{U} (\mathscr{H}_{\scriptscriptstyle Q}))
\eeq
 on the space $ \mathcal{F} (T^*Q, \mathscr{H}_{\scriptscriptstyle Q})$. The multiplication rule in this semidirect product is given by $(\eta _1,U_1)(\eta _2, U_2)=(\eta _1 \circ \eta _2, (U_1 \circ \eta _2)U_2)$, from which we verify directly  that $ \psi \mapsto (U \psi ) \circ \eta ^{-1} $ defines a left action of $( \eta , U) \in \operatorname{Diff}(T^*Q) \,\circledS\, \mathcal{F} (T^*Q, \mathcal{U} (\mathscr{H}_{\scriptscriptstyle Q}))$ on $ \psi \in  \mathcal{F} (T^*Q, \mathscr{H}_{\scriptscriptstyle Q})$.
Note also that the expressions of $ \mathcal{X} $ and $ \xi  $ are given by
\begin{equation}\label{sdp_LA} 
(\dot \eta , \dot U)( \eta , U) ^{-1} = ( \dot \eta \circ \eta ^{-1} , (\dot U U ^{-1} ) \circ \eta ^{-1} )=( \mathcal{X} , \xi ) \in \mathfrak{X} (T^*Q) \,\circledS\, \mathcal{F} (T^*Q, \mathfrak{u}(\mathscr{H}_{\scriptscriptstyle Q})),
\end{equation} 
as it appears from the right-trivialization. 
Semidirect-product groups of the type \eqref{sdpgroup}  appeared on several occasions in the geometric mechanics of continuum systems and they were shown to be particularly relevant in the theory of complex fluids \cite{GBRa2009,Holm02,Tronci12}. Within the context of fully quantum systems, a similar structure recently emerged in  nonadiabatic molecular dynamics \cite{FoHoTr19,HoRaTr21}.
\end{remark}
\rem{ %%%%%%%%%%%%%%%%%%%%%%%%%%
\begin{remark}[Purely classical case]
The case of purely classical dynamics is recovered by setting $\widehat{H}(z)=H(z)\widehat{\boldsymbol{1}}$. In this case equation \eqref{psieq2} becomes
\beq
\label{psieq4}
i\hbar  \big(\partial_t+ X_{{H}}\cdot {\rm d} \big)\psi=
H\psi+  X_{{H}}\cdot(D^{-1}\sigma+{\cal A}_B-{\cal A})\psi
\,.
\eeq
Since the left-hand side carries purely phase terms, the $\psi-$evolution relation \eqref{sdp_action} reduces to
\begin{equation}\label{sdp_action2} 
\psi (t)= (e^{-i\phi(t)/\hbar} \psi _0 ) \circ \eta (t) ^{-1} 
,
\end{equation} 
where $\phi(t)\in \mathcal{F} (T^* Q)$ and the semidirect-product group \eqref{sdpgroup} drops to $\operatorname{Diff}(T^*Q) \,\circledS\, \mathcal{F} (T^*Q, \mathcal{U} (1))$. Here, $\mathcal{U} (1)$ denotes the group of complex phase factors. More importantly, the relation \eqref{sdp_action2} yields a Berry connection ${\cal A}_B(t)=-\eta(t)_*({\cal A}_{B}(0)+\de\phi(t))$, where $\eta(t)_*$ denotes the push-froward of the one-form by $\eta(t)$. This implies a Berry curvature ${\cal B}_{jk}=\partial_j{\cal A}_{Bk}-\partial_k{\cal A}_{Bj}=2\hbar\operatorname{Im}\langle \partial_j \psi | \partial_k \psi\rangle$
evolving as ${\cal B}(t)=-\eta_*(t){\cal B}(0)$, which vanishes unless the initial wavefunction $\psi _0 $ in \eqref{sdp_action2}  carries nontrivial holonomy. Here, we recall from Section \ref{sec:VS} that $ \mathcal{B} = \psi ^* \Omega$, where $ \Omega $ the canonical symplectic form on $\mathscr{H}_{\scriptscriptstyle Q}$, so that trivial holonomy amounts to $ \psi ^* \Omega =0$.
% In this case, since $D^{-1}\sigma=\de S$, the relation \eqref{constraint1} becomes $D(t)^{-1}\sigma(t) ={\cal A}+\eta_*(t){\cal A}_{B}(0)+\de(\eta_*(t)\phi(t))$, thereby implying ${\cal B}(t)=\omega={\cal B}(0)$.
\end{remark}
}%%%%%%%%%%%%%%%%%%%%%%%%%%
%\comment{CT: is this an inconsistency because ${\cal B}(t)$ is not generally preserved in time? I am not entirely sure what's going on because we know that the relation \eqref{constraint1} is not compatible with exact solutions. On the other hand, I was not expecting to find such a contradiction. What do you think?}
%\todo{FGB: \textcolor{red}{I don't see a contradiction since \eqref{constraint1} is not preserved by the equations. You were just checking if by chance for the special case $\widehat{H}=H(z) \operatorname{Id}$, then \eqref{constraint1} leads to something that is preserved, and it is not. Is it a problem?}} 
%\todo{FGB: \textcolor{red}{Maybe the case $\widehat{H}=H(z) \operatorname{Id}$ is also interesting to mention for the closure model?}} 

\paragraph{Euler-Poincar\'e variational principle.} From  \eqref{sdp_action}, we observe that the density matrix $ \rho  = \psi \psi ^\dagger$ evolves under the group action
$
\rho(t)  = (U(t) \rho  _0 U(t)^\dagger) \circ \eta(t) ^{-1},
$
with $\rho_0=\psi_0\psi_0^\dagger$. Consequently, one also obtains the equation
\beq\label{rhoeq}
\partial_t \rho+ {\rm d} \rho \cdot {\cal X}=[\xi, \rho]
\,.
\eeq
In addition, upon recalling \eqref{BerryConn}, we have
$\langle\psi,i\hbar \partial _t \psi\rangle=\langle\rho,i\hbar\xi\rangle+ {\cal A}_B \cdot {\cal X}$ 
and thus introducing the variable
\[
m:=\sigma+D{\cal A}_B
\] 
takes the variational principle \eqref{VP2} into the Euler-Poincar\'e form
\beq
\delta\int^{t_2}_{t_1}\!\!\bigg(\int\!\big( m\cdot{\cal X}+\langle\rho,i\hbar D\xi- 
X_{\widehat{H}}
\cdot m\big\rangle
\big)\Lambda-f(D,\rho)\bigg)\,
{\rm d} t=0,
\label{VP5}
\eeq
where the variable $m$ plays the role of a Lagrange multiplier enforcing the condition ${\cal X}=\langle X_{\widehat{H}}\rangle$. In the context of Euler-Poincar\'e reduction \cite{HoMaRa1998}, the relation \eqref{VP5} identifies  an Euler-Poincar\'e variational principle for the group \eqref{sdpgroup} acting on the variables $(D,\rho)\in \operatorname{Den}(T^*Q)\times \mathcal{F} (T^*Q,\operatorname{He}(\mathscr{H}_{\scriptscriptstyle Q}))$ as 
$D \mapsto (D \circ \eta  ^{-1} )\operatorname{Jac}( \eta  ^{-1} )$ and $\rho\mapsto (U \rho   U^\dagger) \circ \eta^{-1}$.
Here, arbitrary variations $ \delta m$ are accompanied by  the constrained variations
\begin{equation}\label{EP_variations}
\begin{aligned} 
\delta {\cal X}=&\ \partial_t{\cal Y}+{\cal X}\cdot\nabla{\cal Y}-{\cal Y}\cdot\nabla{\cal X},\\
\delta \xi =& \ \partial _t \Sigma + [ \Sigma , \xi  ] + {\rm d}  \Sigma  \cdot {\cal X} - {\rm d}   \xi \cdot {\cal Y},\\
\delta \rho =&\  [\Sigma, \rho] - {\rm d}  \rho \cdot {\cal Y},\\
\delta D=&\ -\operatorname{div}(D{\cal Y}),
\end{aligned}
\end{equation} 
where ${\cal Y}= \delta \eta \circ \eta ^{-1} $ and $ \Sigma = \delta U U ^{-1} \circ \eta ^{-1}$ are again arbitrary. The expression for $ \delta \mathcal{X} $ and $ \delta \xi $ follow from $ \mathcal{X} = \dot \eta  \circ \eta ^{-1} $ and $ \xi = \dot U U ^{-1} \circ \eta ^{-1}$, see \eqref{sdp_LA}, and those for $ \delta \rho  $ and $ \delta D$ follow from $\rho  = (U \rho  _0 U^\dagger) \circ \eta ^{-1}$ and $D= (D_0 \circ \eta ^{-1} )\operatorname{Jac}( \eta  ^{-1})$.
For convenience, we illustrate the step leading to the variational principle \eqref{VP5} by the following diagram:
%\todo{\color{magenta} FGB: Previous sentence: strictly speaking, the following diagram illustrates just the last step for the variational principles, while the result above is the summary of the result of the whole section (i.e. from the hybrid wave up to \eqref{VP5}--\eqref{EP_variations}). Agree?}

{\noindent
%\begin{figure}[h!]
\small
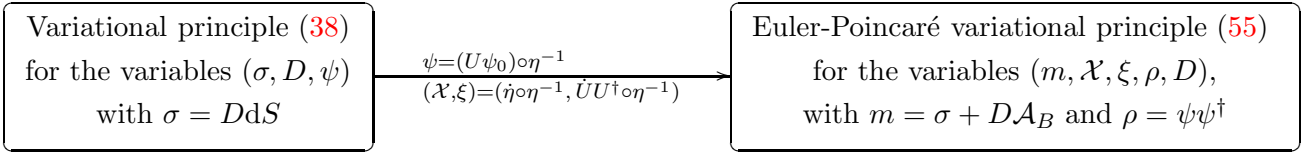
\begin{center}
\hspace{2cm}\begin{xy}
\hspace{-1cm}\xymatrix{
&*+[F-:<3pt>]{
\begin{array}{c}
\vspace{0.1cm}\text{Variational principle \eqref{VP2}}\\
\vspace{0.1cm}\text{for the variables $( \sigma , D, \psi )$}\\
\vspace{0.1cm}\text{with $ \sigma = D {\rm d} S$}
\end{array}
}\ar[rr]
   &  ^{\hspace{-0.3cm}\psi = (U \psi _0 ) \circ \eta  ^{-1} \hspace{-0.3cm}} _{\hspace{-0.3cm} ( \mathcal{X} , \xi ) =(\dot \eta\circ\eta^{-1} , \,\dot UU^\dagger\circ\eta^{-1})\hspace{-0.3cm}} & 
*+[F-:<3pt>]{
\begin{array}{c}
\vspace{0.1cm}\text{Euler-Poincar\'e variational principle \eqref{VP5} }\\
\vspace{0.1cm}\text{for the variables $( m , \mathcal{X} , \xi , \rho  , D )$,}\\
\vspace{0.1cm}\text{with $m=\sigma+D{\cal A}_B$ and $\rho=\psi\psi^\dagger$}
\end{array}
} 
}
\end{xy}
%\end{figure}
\end{center}
}

\medskip
\noindent
Notice that at this point the phase choice has been entirely eliminated since the quantum evolution is now  expressed in terms of the phase-invariant quantity $\rho$. In summary, we have the following statement.

\begin{proposition} Assume that a hybrid wave function, written in the exact factorization form
\[
\Upsilon (t,z)= \chi (t,z) \psi (t,z), \qquad \chi (t,z) \in \mathbb{C} , \quad \psi (t,z) \in \mathscr{H}_{\scriptscriptstyle Q}, \quad \| \psi (t,z)\|^2=1,
\]
is a solution of the quantum--classical wave equation $ i \hbar \partial _t \Upsilon = \widehat{ \mathcal{L} }_{\widehat{H}} \Upsilon $ in  \eqref{hybrid_KvH}.
Define $D \in \operatorname{Den}(T^*Q)$ and $S \in \mathcal{F} (T^*Q)$ such that  $ \chi = \sqrt{D}e^{i S/ \hbar }$, and let 
\begin{equation} \label{gina}
\rho = \psi \psi ^\dagger \in \mathcal{F} (T^*Q, \operatorname{He}( \mathscr{H}_{\scriptscriptstyle Q})) 
\,,\qquad\qquad
m%= D( {\rm d} S + \mathcal{A} _B)
= D( {\rm d} S - \left\langle \psi , i \hbar {\rm d} \psi \right\rangle ) \in \mathfrak{X} (T^*Q) ^*.
%\Omega ^1 (T^*Q)\otimes\operatorname{Den}(T^*Q).
\end{equation} 
Then, the evolution equations for $D(t)$, $ \rho(t)$, and $m(t)$ arise as the critical conditions for the Euler-Poincar\'e variational formulation \eqref{VP5}--\eqref{EP_variations} for the semidirect product group \eqref{sdpgroup}.
In particular, the corresponding solutions $D(t)$ and $ \rho  (t)$ can be written in terms of the action of the  group \eqref{sdpgroup} as
\begin{equation}\label{action_summary} 
D(t)= (D_0 \circ \eta (t) ^{-1} )\operatorname{Jac}( \eta (t) ^{-1} )  \qquad\quad\text{and}\qquad\quad \rho  (t)= (U(t) \rho  _0 U(t)^\dagger) \circ \eta(t) ^{-1}.
\end{equation} 
\end{proposition} 

Instead of going ahead to rewrite the equations \eqref{mario} and \eqref{psieq1} in terms of the new variables, we now exploit the new  variational principle \eqref{VP5} to enforce a constraint so that the sign of the classical density $\rho_c$ in \eqref{densities2} is preserved in time.

\section{Closure model\label{sec:closure1}}

As discussed above, the variational principle \eqref{VP5} has the  advantage of eliminating the gauge fixing involved in the Berry connection, which is now implicitly accounted for by the variable $m$.
While the dynamical system resulting from \eqref{VP5} comprises the three variables $m$, $D$, and $ \rho  $, at this stage we aim to find a closure model preserving the initial sign of the classical density $\rho_c$ in  \eqref{densities2}. Generally speaking, we define a closure as a relation of the type $m=m(D, \rho  )$ so that the variable $m$ can be eliminated. In this section, we will realize this closure by restricting to consider the special case $m=m(D)$.
In particular, since the classical density  is now rewritten as $\rho_c=D+\operatorname{div}\!\left[\Bbb{J}\left(m-D{\cal A}\right)\right]$, we observe that setting 
\beq\label{sara}
m=D{\cal A}
\eeq
would return a pure transport equation for $\rho_c=D>0$. Recently, similar approaches were used in \cite{GBTr21a,GBTr21b} to devise simple closure models based on von Neumann operators in the Koopman setting. Here, we take an alternative route by exploiting the full power of the  exact factorization method. Notice that the present approach does not require a specific type of hybrid Hamiltonian $\widehat{H}(z)$, which indeed is left as completely general.

%\todo{\textcolor{magenta}{FGB: Somewhere we should emphasize that this closure (hence the positivity) can be done for arbitrary hybrid Hamiltonians  $\widehat{H}$ (not only for instance for expressions such as $ - \frac{ \hbar ^2 }{2m } \Delta _x + H_I(q,p,x)$,...)}}

\subsection{Constraint and variational closure\label{sec:closure}}
Importantly, the relation \eqref{constraint1}  does not appear to emerge as an exact solution of the system \eqref{mario}-\eqref{psieq1} and therefore enforcing $m=D{\cal A}$ requires adding a constraint  to the variational principle \eqref{VP5}. Equivalently, here we simply replace $m=D{\cal A}$ in \eqref{VP5}, which then becomes
\begin{equation}\label{VP7}
\delta  \int^{t_2}_{t_1}\!\left[\int\!\big(D{\cal A}\cdot{\cal X}+\langle\mathcal{P},i\hbar\xi\rangle\big)\,\Lambda
-h(D,\mathcal{P}) \right] {\rm d} t
=0,
\end{equation} 
with $\mathcal{P}:=D\rho$ and
\begin{equation}\label{h_D_tilde_rho} 
h(D, \mathcal{P}  )=\int \! \Big\langle\mathcal{P},\widehat{H}+\frac{i\hbar}{2D} \big[{\rm d} \mathcal{P}, X_{\widehat{H}}\big]\Big\rangle \Lambda 
\,.
\end{equation}
This construction is summarized by the following diagram:

{
%\begin{figure}[h!]
\noindent\small
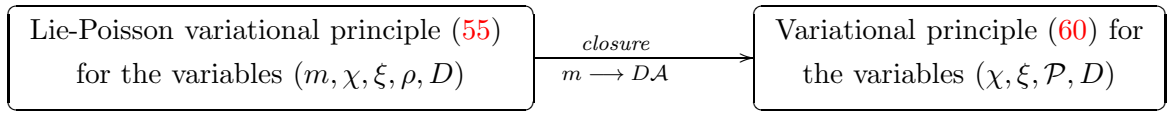
\begin{center}
\hspace{2cm}\begin{xy}
\hspace{-1cm}\xymatrix{
&*+[F-:<3pt>]{
\begin{array}{c}
\vspace{0.1cm}\text{Lie-Poisson variational principle \eqref{VP5} }\\
\vspace{0.1cm}\text{for the variables $( m , \chi , \xi , \rho  , D )$}
\end{array}
}\ar[rr]^{  \textit{closure} }_{   m \;\longrightarrow \; D \mathcal{A} }
& \hspace{1cm} &
*+[F-:<3pt>]{
\begin{array}{c}
\vspace{0.1cm}\text{Variational principle \eqref{VP7} for}\\
\vspace{0.1cm}\text{the variables $(\chi , \xi , \mathcal{P}   , D )$}
\end{array}
} 
}
\end{xy}
%\end{figure}
\end{center}
}

\medskip
\noindent
For later convenience, here we have introduced the density variable $\mathcal{P}$ satisfying $\partial _t\mathcal{P}  + \operatorname{div}(\mathcal{P} {\cal X}) = [ \xi , \mathcal{P}  ]$ and $\delta\mathcal{P}  + \operatorname{div}(\mathcal{P} {\cal Y}) = [ \Sigma , \mathcal{P}  ]$. The other variations to be considered in \eqref{VP7} are given in \eqref{EP_variations}.
Collecting the terms proportional to $ \mathcal{Y} $ in the variations of \eqref{VP7} gives the condition
%\[
%\left(\frac{\partial}{\partial t}+\pounds_{{\cal X}}\right)\left(\frac1D\frac{\delta \ell}{\delta {\cal X}}\right)=\de\frac{\delta \ell}{\delta D}+\frac1D\left\langle\mathcal{P},\de\frac{\delta \ell}{\delta \mathcal{P}}\right\rangle-\frac1D\left\langle\frac{\delta \ell}{\delta \xi},\de \xi\right\rangle
%\,.
%\]
%We have ${\delta \ell}/{\delta {\cal X}}=D\boldsymbol{\cal A}$ and
%\[
%\frac{\delta \ell}{\delta \mathcal{P}}=i\hbar\xi-\frac{\delta h}{\delta \mathcal{P}}
%\,,\qquad\qquad
%\frac{\delta \ell}{\delta \xi}=-i\hbar\mathcal{P}
%\] Thus,
\beq
\pounds_{{\cal X}} \mathcal{A} 
%=\de\left({\cal X}\cdot\boldsymbol{\cal A}-\frac{\delta h}{\delta D}\right)+\frac1D\left\langle\mathcal{P},\de\bigg(i\hbar\xi-\frac{\delta h}{\delta \mathcal{P}}\bigg)\right\rangle+\frac1D\left\langle i\hbar\mathcal{P},\de \xi\right\rangle
=\de \!\left({\cal X}\cdot \mathcal{A} -\frac{\delta h}{\delta D}\right)-\left\langle\de \frac{\delta h}{\delta \mathcal{P}}\right\rangle
\,,
\label{anna}
\eeq
\rem{%%%%%%%%%%%%%%%%%%%%%%%%%%
\begin{framed}
\begin{align*}
(\partial_t+\pounds_{\cal X}) \mathcal{A} 
%=\de\left({\cal X}\cdot\boldsymbol{\cal A}-\frac{\delta h}{\delta D}\right)+\frac1D\left\langle\mathcal{P},\de\bigg(i\hbar\xi-\frac{\delta h}{\delta \mathcal{P}}\bigg)\right\rangle+\frac1D\left\langle i\hbar\mathcal{P},\de \xi\right\rangle
=&\ \de \!\left({\cal X}\cdot \mathcal{A} -\frac{\delta h}{\delta D}\right)+D^{-1}\left\langle \frac{\delta h}{\delta {\psi}},\nabla\psi\right\rangle
\,,
\end{align*}
\begin{align*}
(\partial_t+\pounds_{\cal X})\mathcal{A}_B=&\ \langle\xi\psi,-i\hbar\de\psi\rangle+\langle\psi,-i\hbar\de(\xi\psi)\rangle
\\
=&\ 
 \langle i\hbar\xi\psi,\de\psi\rangle+\de\langle\psi,-i\hbar\xi\psi\rangle-\langle\de\psi,-i\hbar\xi\psi\rangle
 \\
=&\
2 \langle i\hbar\xi\psi,\de\psi\rangle+
\de\langle\psi,-i\hbar\xi\psi\rangle
\\
=&\
 \left\langle \frac{1}D\left(\frac{\delta h}{\delta {\psi}}\psi^\dagger+\psi\frac{\delta h}{\delta {\psi}}^\dagger+2\alpha\psi\psi^\dagger\right)\psi,\de\psi\right\rangle
+
\frac12\de\left\langle\psi,-\frac{1}D\left(\frac{\delta h}{\delta {\psi}}\psi^\dagger+\psi\frac{\delta h}{\delta {\psi}}^\dagger+2\alpha\psi\psi^\dagger\right)\psi\right\rangle
\\
=&\
 \left\langle \frac{1}D\frac{\delta h}{\delta {\psi}},\de\psi\right\rangle
-\de\left(\frac{\alpha}D+\frac{1}D\left\langle\psi,\frac{\delta h}{\delta {\psi}}\right\rangle\right)
\,,
\end{align*}
where we have used
\[
i\hbar D\xi=\frac12\left(\frac{\delta h}{\delta {\psi}}\psi^\dagger+\psi\frac{\delta h}{\delta {\psi}}^\dagger\right)+\alpha\psi\psi^\dagger
\]
for an arbitrary $\alpha(z)$.
Notice that
\[
(\partial_t+\pounds_{\cal X})\mathcal{B}=\de\left\langle \frac{1}D\frac{\delta h}{\delta {\psi}},\de\psi\right\rangle
\]
Thus,
\[
\frac{\de}{\de t}\oint_{\gamma(t)}  (\mathcal{A}-\mathcal{A}_B) =0
\ \implies\ 
\frac{\de}{\de t}\eta(t)^*(\omega-{\cal B}(t))=0
\ \implies\ 
\eta(t)^*(\omega-{\cal B}(t))=\omega-{\cal B}(0)
\,.
\]
Recall that we had set $\mathcal{A}-\mathcal{A}_B=\de S$ and indeed we have $(\partial_t+\pounds_{\cal X})(\mathcal{A}-\mathcal{A}_B)=\de F$. Thus, 
\[
\omega-{\cal B}(0)=\de^2S(0)=0
\ \implies\ 
\omega-{\cal B}(t)=0\]
Not sure this is a good approach...
\end{framed}
}%%%%%%%%%%%%%%%%%%%%%%%%%%
where we recall the notation $\langle \widehat{A}\rangle :=D^{-1}\big\langle\mathcal{P},\widehat{A}\big\rangle$. As a first consequence, we obtain the following  dynamics for the Poincar\'e integral around an arbitrary loop $\gamma (t)= \eta (t, \gamma _0)$:
\[
\frac{\de}{\de t}\oint_{\gamma(t)}  \mathcal{A} =-\oint_{\gamma(t)}\left\langle\de\frac{\delta h}{\delta \mathcal{P}}\right\rangle
\ \implies\ 
\frac{\de}{\de t}\eta^*\omega=\eta^*\de \left\langle\de\frac{\delta h}{\delta \mathcal{P}}\right\rangle
\,,
\]
where  $ \eta (t)$ is the flow of $ \mathcal{X} $ and the implication is a direct consequence of Stokes theorem. Notice that this relation differs   from \eqref{caterina}, since the latter involves a phase-space path $\eta(t)$ defined as the flow of a different vector field $\cal X$. We hope that no confusion arises by using the same notation for both $\cal X$ and $\eta$ as in Section \ref{sec:EF}.
From \eqref{anna}, Cartan's magic formula $ \pounds _ \mathcal{X} = {\rm d} \mathbf{i} _ \mathcal{X} + \mathbf{i} _ \mathcal{X} {\rm d} $ yields
\[
\mathbf{i} _{\cal X}\omega=\de \frac{\delta h}{\delta D}+\left\langle\de \frac{\delta h}{\delta \mathcal{P}}\right\rangle
=
\left\langle\de \bigg(\frac{\delta h}{\delta D}+\frac{\delta h}{\delta \mathcal{P}}\bigg)\right\rangle,
\]
where we have used $\de {\cal A}=-\omega$ and $\operatorname{Tr}\mathcal{P}=D$. Since $ \omega $ is non-degenerate, we can write
\begin{equation}\label{chi_expression} 
{\cal X}=\left\langle X_{\,\delta h/\delta D+\delta h/\delta \mathcal{P}}\right\rangle=X_{\,\delta h/\delta D}+\left\langle X_{\,\delta h/\delta \mathcal{P}}\right\rangle
.
\end{equation} 
A direct calculation yields
\[
\frac{\delta h}{\delta D}=-\frac1{2D}\big\langle i\hbar[\de\mathcal{P},X_{\widehat{H}}]\big\rangle=-
\frac1{2D}\big\langle i\hbar\{\mathcal{P},{\widehat{H}}\}+i\hbar\{\widehat{H},\mathcal{P}\}\big\rangle
\]
and
\begin{align*}
\frac{\delta h}{\delta \mathcal{P}}=&\, \widehat{H}+\frac{i\hbar}{D}\,[\de\mathcal{P},X_{\widehat{H}}]-\frac{i\hbar}{2D^2}\,[\mathcal{P},\{D,\widehat{H}\}]
\\
=&\, \widehat{H}+\frac{i\hbar}{D}\,\big(\{\mathcal{P},{\widehat{H}}\}+\{\widehat{H},\mathcal{P}\}\big)+\frac{i\hbar}{2D^2}\,[\{D,\widehat{H}\},\mathcal{P}].
%\de\frac{\delta h}{\delta D}+\left\langle\de\frac{\delta h}{\delta \mathcal{P}}\right\rangle=\left\langle\mathcal{P},
%\de \widehat{H}
%+\de\left(\frac{i\hbar}{2D^2} {\rm d}  D \cdot [\mathcal{P},X_{\widehat{H}}]\right)\right\rangle-\de
%\Big\langle\mathcal{P},\frac{i\hbar}{2D^2} \big[{\rm d} \mathcal{P}, X_{\widehat{H}}\big]\Big\rangle
\end{align*}
On the one hand, the advection equation $ \partial _t D + \operatorname{div}( D \mathcal{X} )=0$, with $ \mathcal{X} $ given in \eqref{chi_expression}, yields
\begin{equation}\label{adv_D} 
%\partial_t\rho_c+ \operatorname{div}\Big( \rho_c\left\langle X_{\,\delta h/\delta D+\delta h/\delta \mathcal{P}}\right\rangle\!\Big)=0
%\,\qquad
\frac{\partial D}{\partial t} + \left\{ D,\frac{\delta h}{\delta {D}}\right\}+\operatorname{Tr}\left\{\mathcal{P},\frac{\delta h}{\delta \mathcal{P}}\right\}=0\,,
\end{equation} 
where we recall $D=\rho  _c$. On the other hand, the variations in \eqref{VP7} proportional to $ \Sigma $ yield
\begin{equation}\label{tilde_rho_adv} 
i\hbar\frac{\partial \mathcal{P}}{\partial t}+i\hbar\left\{ \mathcal{P},\frac{\delta h}{\delta D}\right\}+i\hbar\operatorname{div}\! \Big( \mathcal{P}\left\langle X_{\,\delta h/\delta \mathcal{P}}\right\rangle\!\Big)=\bigg[\frac{\delta h}{\delta \mathcal{P}},\mathcal{P}\bigg]
\,.
\end{equation}
In addition, the advection equation $\partial _t\mathcal{P}  + \operatorname{div}(\mathcal{P} {\cal X}) = [ \xi , \mathcal{P}  ]$, together with \eqref{tilde_rho_adv}, yields  the condition 
$
\left[\mathcal{P}, i \hbar \xi - {\delta h}/{\delta \mathcal{P}  }  \right] =0
$.

As we can see, if ${\delta h}/\delta \mathcal{P}=0$, then  equation \eqref{adv_D} recovers a classical Liouville evolution of the type ${\partial_tD} + \left\{ D,{\delta h}/{\delta {D}}\right\}$. Instead, if ${\delta h}/{\delta {D}}=0$ and $\de ( {\delta h}/\delta \mathcal{P})=0$, the integral of equation \eqref{tilde_rho_adv} returns  quantum Liouville dynamics. These special cases will be discussed in more detail in Section \ref{sec:relevcases}, which will also cover the Ehrenfest model in \eqref{Ehrenfest1}.

\begin{remark}[Role of the Berry connection]\label{rem:berry}
Notice that equation \eqref{tilde_rho_adv}  may be combined with the chain rule ${\delta h}/{\delta \mathcal{P}  }=D^{-1}{\delta h}/{\delta  \rho  }$ to express \eqref{rhoeq} in the form
$
i\hbar D(\partial_t +  {\cal X}  \cdot{\rm d})\rho=[{\delta h}/{\delta  \rho  }, \rho]
$,
with $\cal X$ given as in \eqref{chi_expression}. In turn, the definition $\rho=\psi\psi^\dagger$ leads to the Schr\"odinger equation
\[
i\hbar (\partial_t +  {\cal X}  \cdot{\rm d})\psi=
\frac1{2D}\frac{\delta h}{\delta \psi  }
\]
so that, upon noticing that the chain rule relation $\delta h/\delta\psi=2(\delta h/\delta\rho)\psi$ yields $\operatorname{Im}\langle\psi|\delta h/\delta\psi\rangle=2\operatorname{Im}\langle\rho|\delta h/\delta\rho\rangle=0$, the dynamics of the Berry connection reads 
$
(\partial_t+\pounds_{\cal X})\mathcal{A}_B= D^{-1}\langle {\delta h}/{\delta {\psi}},\de\psi\rangle\linebreak=-\langle \de({\delta h}/{\delta {\mathcal{P}}})\rangle.
$
Thus, \eqref{anna} can be equivalently written as
\beq\label{giada}
(\partial_t+\pounds_{\cal X})(\mathcal{A}-\mathcal{A}_B)=\de \!\left({\cal X}\cdot \mathcal{A} -\frac{\delta h}{\delta D}\right)
\,,
\eeq
%which reflects the constraint  \eqref{sara}, upon recalling the second relation in \eqref{gina} and $\sigma=D\de S$. Indeed, 
Since $\omega=- {\rm d} {\cal A}$ and ${\cal B}=\de{\cal A}_B$, equation \eqref{giada} implies
$\eta(t)^*(\omega+{\cal B}(t))=\omega+{\cal B}(0)$. This relation is also obtained by differentiating the constraint \eqref{constraint1}, in which case one has $\omega+{\cal B}(0)=0$. In  more generality, if $\omega+{\cal B}$ is symplectic initially, then it stays so indefinitely.

%and if we enforce \eqref{constraint1} at the initial time then we have $\omega+{\cal B}(0)=0$, thereby recovering ${\cal B}(t)+\omega=0$ at all times. Alternatively, if $D^{-1}\sigma$ in \eqref{constraint1} is allowed to have a nontrivial differential as in Remark \ref{VNops}, then \eqref{giada} would imply $\eta(t)^*(\omega+{\cal B}(t))=\omega+{\cal B}(0)\neq0$.  
\end{remark}

In conclusion, from \eqref{VP7} we obtain a closure model for the hybrid quantum--classical wave equation \eqref{hybrid_KvH}. Since we recall $D=\operatorname{Tr}\mathcal{P}$, this closure model may be entirely expressed in terms of $\mathcal{P}$ and we can state the following:
\begin{proposition}[Variational closure]\label{propclosure} The closure model \eqref{adv_D}-\eqref{tilde_rho_adv}, with the Hamiltonian \eqref{h_D_tilde_rho},  obtained by the replacement $m \rightarrow D \mathcal{A}$ in the  variational principle \eqref{VP5} is equivalent to 
\beq
i\hbar\partial_t \mathcal{P}+i\hbar\operatorname{div} \!\big( \mathcal{P}\left\langle X_{\,\delta {\cal H}/\delta \mathcal{P}}\right\rangle\!\big)=\bigg[\frac{\delta {\cal H}}{\delta \mathcal{P}},\mathcal{P}\bigg]\,,
\label{andrea}
\eeq
where
\beq
{\cal H}( \mathcal{P}  )=\int \bigg\langle\mathcal{P},\widehat{H}+\frac{i\hbar}{2\operatorname{Tr}\mathcal{P}}\, \big[{\rm d} \mathcal{P}, X_{\widehat{H}}\big]\bigg\rangle\, \Lambda 
\label{h_tilde_rho}
\eeq
satisfies ${\cal H}( \mathcal{P}  )=h(D, \mathcal{P}  )$.
\end{proposition}
For completeness, we present the explicit expression of the  functional derivative:
\begin{align*}
\frac{\delta {\cal H}}{\delta \mathcal{P}}=&\,  \widehat{H}+\frac1{2\operatorname{Tr}\mathcal{P}}\bigg(2{i\hbar}\big(\{\mathcal{P},{\widehat{H}}\}+\{\widehat{H},\mathcal{P}\}\big)
\\
&\, +\frac{i\hbar}{\operatorname{Tr}\mathcal{P}}\,[\{\operatorname{Tr}\mathcal{P},\widehat{H}\},\mathcal{P}] 
-
\big\langle i\hbar\{\mathcal{P},{\widehat{H}}\}+i\hbar\{\widehat{H},\mathcal{P}\}\big\rangle \operatorname{Id}\bigg),
\end{align*}
which needs to be substituted in  \eqref{andrea}.

We have obtained a variational closure model for the dynamics of coupled quantum and classical systems. The former are identified by a quantum density matrix $\hat\rho_q=\int\!\mathcal{P}\,\Lambda$, while the latter are given in terms the Liouville distribution $\rho_c=D= \operatorname{Tr}\mathcal{P}$. We notice that in the present construction this phase-space distribution satisfies the evolution law given by the first relation in \eqref{action_summary}, where $\dot\eta(t)={\cal X}(t, \eta(t))$ and the vector field ${\cal X}$ is given as in \eqref{chi_expression}. In other words,  the equation \eqref{adv_D} is a transport equation of the type $\partial_t D+\operatorname{div}(D{\cal X})=0$, which preserves the initial sign of $D$ whenever the vector field ${\cal X}$ is sufficiently smooth. The fact that a classical Liouville density remains positive over time represents the main advantage of the present closure model presented here. Since $\hat\rho_q$ is also positive by construction, the present model satisfies both properties 1 and 2 from the list of desirable self-consistency criteria in the Introduction.

\begin{remark}[Unobservable classical phases\label{classphases}]
While we have shown that the closure relation \eqref{sara} yields a positive-definite classical density, one may wonder what makes this closure so special. To see this, it is convenient to write the closure relation as in \eqref{constraint1}, where  $\sigma=D\de S$. Then, one easily checks that enforcing \eqref{constraint1} makes the Hamiltonian \eqref{covHam}  invariant under phase transformations $S\mapsto S+\varphi$. Indeed, we recognize that the  closure model for the measure-valued projection $\mathcal{P}=D\psi\psi^\dagger$  obtained upon enforcing \eqref{constraint1} is entirely independent of the classical phase. This particular aspect has a deep physical meaning. As discussed in \cite{boucher,Sudarshan}, the argument is that classical phases are \emph{unobservable} in the sense that they do not affect the  expectation values of observable quantities. While Sudarshan proposed enforcing the principle of unobservable phases by adopting superselection rules  \cite{Sudarshan}, here the same principle was applied by invoking a symmetry argument which makes the model completely phase-independent. Notice that the same principle may be applied to the classical Koopman-van Hove theory, thereby recovering the standard Koopman-von Neumann construction. This was recently shown  in \cite{GBTr22}, where the present model is discussed from a more physical viewpoint.
\end{remark}

So far, however, we have not yet explained how the current model can be used to compute expectation values of quantum-classical observables. For example, if $q$ and $\hat{x}$ are the classical and quantum position, respectively, how do we compute the expectation value of the hybrid observable $\widehat{A}(q,p)=q\hat{x}$? This question will be addressed in Section \ref{sec:clasquantdens}. Instead, the next sections provide a geometric presentation of the  Hamiltonian structure underlying the present closure model.

\subsection{Hamiltonian structure of the closure\label{sec:HamStr}}

In this section we will prove that the closure \eqref{adv_D}-\eqref{tilde_rho_adv} is a Hamiltonian system with respect to a noncanonical Poisson bracket. In order to prove this result we first compute the expression of the bracket by evaluating the time derivative of an arbitrary function $f( D,\mathcal{P} )$ along the solutions $D(t),  \mathcal{P} (t) $ of \eqref{adv_D}-\eqref{tilde_rho_adv}, for any Hamiltonian $h$. The bracket expression is found from the relation $\dot f=\{\!\!\{f,h\}\!\!\}$. 
A direct computation of $\frac{d}{dt} f( D(t), \mathcal{P}  (t))$ along the solutions of \eqref{adv_D}-\eqref{tilde_rho_adv} gives the following bracket expression:
\begin{equation}\label{bracket_candidate}
\begin{aligned}
\{\!\!\{f,h\}\!\!\}(D,\mathcal{P})=&\ \int \! D \omega \big( X_{{\delta f}/{\delta D} },X_{{\delta h}/{\delta D} } \big) \Lambda -\int  \!D\left\langle \frac{i}\hbar\!\left[\frac{\delta f}{\delta \mathcal{P}},\frac{\delta h}{\delta \mathcal{P}}\right]\right\rangle\Lambda
\\
&\ +
\int \!D \Big( \omega \big(  \left\langle  X_ {{\delta f}/{\delta \mathcal{P}} } \right\rangle ,X_{{\delta h}/{\delta D} }\big) -  \omega \big(  \left\langle X_{ {\delta h}/{\delta \mathcal{P}}} \right\rangle ,X_{{\delta f}/{\delta D} }\big) \Big) \, \Lambda\\
&\
+\int \!D\omega \big( \langle X_{  {\delta f}/{\delta \mathcal{P}}  }\rangle,\langle X_{  {\delta h}/{\delta \mathcal{P}} }\rangle\big)  \,\Lambda
\,,
\end{aligned}
\end{equation} 
where  $ \omega (X_f, X_g)
%= \left\langle {\rm d} f, \omega ^\sharp ( {\rm d} h) \right\rangle 
=\{f, g\}$ is the canonical Poisson bracket on $T^*Q$ and we recall the notation $ \left\langle A \right\rangle = \left\langle \rho  , A \right\rangle = D ^{-1} \left\langle \mathcal{P}  , A \right\rangle $. Also, $D$ is a strictly positive density, that is   $D \in \operatorname{Den}_+(T^*Q)$, while ${\cal P}\in\mathcal{F}( T^* Q , \operatorname{He}(\mathscr{H}_{\scriptscriptstyle Q}))^*$ and we identify  $\mathcal{F}( T^* Q , \operatorname{He}(\mathscr{H}_{\scriptscriptstyle Q}))^*\simeq \operatorname{Den}( T^* Q ) \otimes \operatorname{He}(\mathscr{H}_{\scriptscriptstyle Q})$.  If we  introduce the convenient notation
\beq
A:B= \operatorname{Tr}(AB),
\label{easynotn}
\eeq
then the bracket structure \eqref{bracket_candidate} can be equivalently expressed as 
\begin{align}\nonumber
\{\!\!\{f,h\}\!\!\}(D,\mathcal{P})=&\ \int \! D \left\{\frac{\delta f}{\delta D} ,\frac{\delta h}{\delta D}\right\}  \Lambda -\int  \!D\left\langle \frac{i}\hbar\!\left[\frac{\delta f}{\delta \mathcal{P}},\frac{\delta h}{\delta \mathcal{P}}\right]\right\rangle\Lambda
\\\nonumber
&\ +
\int \!D\left\langle\left\{\frac{\delta f}{\delta \mathcal{P}},\frac{\delta h}{\delta D}\right\}-\left\{\frac{\delta h}{\delta \mathcal{P}},\frac{\delta f}{\delta D}\right\}\right\rangle\Lambda\\
&\
+\int \!\frac1D \bigg(\mathcal{P}  : \left\{\frac{\delta f}{\delta \mathcal{P}},\frac{\delta h}{\delta \mathcal{P}}\right\}: \mathcal{P}   \bigg) \Lambda 
\,.
\label{bracket_candidate2}
\end{align}
We have the following result:
\begin{proposition}\label{It_is_Poisson}  The closure \eqref{adv_D}-\eqref{tilde_rho_adv} is a Hamiltonian system with respect to the Poisson bracket \eqref{bracket_candidate2} on $\operatorname{Den}_+(T^*Q)\times \mathcal{F} (T^*Q, \operatorname{He}(\mathscr{H}_{\scriptscriptstyle Q})) ^*$  and with the Hamiltonian \eqref{h_D_tilde_rho}.
The bracket admits the functional
\[
C( D, \mathcal{P} )=\operatorname{Tr} \int \!D \Upphi ( \mathcal{P} /D) \Lambda 
\]
as a Casimir function, for  any analytic matrix function $ \Upphi $.
\end{proposition} 
%\textbf{Proof.} 
The first part of this statement will be proved later. As we will see, this follows as a corollary on a more general result concerning a class of Poisson structures that arise by Poisson reduction, see Corollary \ref{corollary}. 
In addition, upon denoting $\upphi(A)=\operatorname{Tr}(\Upphi(A))$, the statement about the Casimir functions follows from a direct computation using
$
{\delta C}/{\delta D} = \upphi ( \mathcal{P} /D)  - \left\langle \nabla \upphi ( \mathcal{P} /D), \mathcal{P} /D \right\rangle$, where   $\nabla \upphi$ denotes the standard derivative of $\upphi$ with respect to its argument. 
%\qquad\blacksquare

The Poisson bracket in  \eqref{bracket_candidate2} depends on the two variables $D$ and $\mathcal{P}  $ which have no specific relation between them. 
It is however clear from the Hamiltonian dynamics that if $D_0= \operatorname{Tr}\mathcal{P}_0  $ at initial time, then the relation $D=\operatorname{Tr}\mathcal{P} $ holds at all times, showing that the dynamics of $D$ is determined from that of $\mathcal{P}  $. In terms of the Poisson bracket, this suggests that the expression
\begin{equation}\label{bracket_candidate_rho}
\{\!\!\{{\cal K},{\cal H}\}\!\!\}_1(\mathcal{P})=-\int  \!\left\langle \mathcal{P}  ,\frac{i}\hbar\!\left[\frac{\delta {\cal K}}{\delta \mathcal{P}},\frac{\delta {\cal H}}{\delta \mathcal{P}}\right] \right\rangle\Lambda +\int \!\frac1{\operatorname{Tr}\mathcal{P}  }\bigg(\mathcal{P}  : \left\{\frac{\delta {\cal K}}{\delta \mathcal{P}},\frac{\delta {\cal H}}{\delta \mathcal{P}}\right\}: \mathcal{P}   \bigg) \Lambda 
\end{equation} 
found from the Poisson bracket \eqref{bracket_candidate2} by dropping the terms involving the $D$-derivatives and using the notation \eqref{easynotn}, should define a Poisson bracket structure.  Indeed, we have the following result:

\begin{corollary}\label{giulia} The expression \eqref{bracket_candidate_rho} 
defines a Poisson bracket formally arising by restriction of \eqref{bracket_candidate2} to the Poisson submanifold of $ \operatorname{Den}_+(T^*Q)\times \mathcal{F} (T^*Q, \operatorname{He}(\mathscr{H}_{\scriptscriptstyle Q})) ^*$ that is determined by the relation $D= \operatorname{Tr}\mathcal{P}$.
This bracket admits the Casimir functionals $C( \mathcal{P} )=\int (\operatorname{Tr}{\cal P})\operatorname{Tr}\Upphi({\cal P}/\operatorname{Tr}{\cal P})\,\Lambda$ for any analytic matrix function $ \Upphi $.
\end{corollary} 
\textbf{Proof.} 
We start by writing the Hamiltonian vector field
\[
\mathsf{X}_h( D, \mathcal{P}  )= \left(  - \operatorname{div}( D \mathcal{X} ), - \operatorname{div}(\mathcal{P}  \mathcal{X} ) - i \hbar ^{-1} \left[ \frac{\delta h}{\delta \mathcal{P}  }, \mathcal{P}  \right] \right)    
\]
associated to the Poisson structure \eqref{bracket_candidate}  and we consider the linear inclusion $ \mathcal{P}  \mapsto \iota (\mathcal{P}  )=  (D= \operatorname{Tr}\mathcal{P}, \mathcal{P}  )$. We observe that, for each $\mathcal{P}  $, the restriction of the Hamiltonian vector field to $D= \operatorname{Tr}\mathcal{P}$ is tangent to the range of $\iota$, thereby showing that the subspace given by the range of $\iota$ is a quasi Poisson submanifold \cite{OrRa2004}. Hence there is a unique Poisson bracket $\{\!\!\{F,H\}\!\!\}_1(\mathcal{P})$ such that $\{\!\!\{f \circ \iota,h \circ \iota\}\!\!\}_1=\{\!\!\{ f, g\}\!\!\} \circ \iota$. It is directly checked that this relation is satisfied with \eqref{bracket_candidate_rho} and \eqref{bracket_candidate2} by inserting $F=f \circ \iota$ and $H=h\circ \iota$ in \eqref{bracket_candidate_rho} and noting the relation
${\delta (f \circ \iota)}/{\delta \mathcal{P}  } = {\delta f}/{\delta \mathcal{P}  } + ({\delta f}/{\delta D}) \operatorname{Id}$.
Hence the desired result is proved. 
The Casimir property of the functional $C( \mathcal{P} )$  follows either from a direct computation, or as a restriction of the Casimir functional $C( \mathcal{P} , D)$ to the Poisson submanifold. $\qquad\blacksquare$ 

\medskip
As anticipated, the explicit proof of the Poisson structure comprised by the hybrid bracket \eqref{bracket_candidate2} requires the use of Poisson reduction. This is the topic of the next section, where Proposition \ref{It_is_Poisson}  is proved explicitly.

\subsection{Poisson reduction and quantum-classical brackets\label{sec:HamStr2}} 

This section is devoted to the proof of Proposition \ref{It_is_Poisson} and of some of its corollaries. Before embarking on the explicit proof, however, it is helpful to set up the geometric framework that will be used throughout the various steps. This is the focus of the first part of this section.

Assume that the group $\operatorname{Diff}(T^*Q)$ acts from the right on a manifold ${\cal O}$ as $n\mapsto\Phi_\eta n$ with  cotangent-lift momentum map ${\cal J}:T^*{\cal O}\to\mathfrak{X}(T^*Q)^*$ given by $ \left\langle \mathcal{J} (\alpha _{n}), \mathcal{X}  \right\rangle = \left\langle  \alpha _n, \mathcal{X} _ \mathcal{O} ({n}) \right\rangle $, for all $ \alpha _{n} \in T^*_{n} \mathcal{O} $ and $ \mathcal{X} \in \mathfrak{X} (T^*Q)$.
Also, let $\mathscr{F}:{\cal O}\to\operatorname{Den}(T^*Q)$ be an equivariant map, i.e., $\eta^*\left( \mathscr{F}(n) \right) =\mathscr{F}(\Phi_\eta n)$, for all $ \eta \in \operatorname{Diff}(T^*Q)$, and assume $\mathscr{F}(n)>0$ for all $n \in \mathcal{O} $. In addition, denote
\[
G:= \mathcal{F} (T^*Q, \mathcal{U} (\mathscr{H}_{\scriptscriptstyle Q}))
\,.
\] 

For each $n \in \mathcal{O} $, consider the $n$-dependent one-form on $\operatorname{Diff}(T^*Q)\times T^*G$ given by
\begin{equation}\label{1form} 
\Theta _n( \eta ,   U,   P)( \dot \eta , \dot U, \dot P)=  \int \mathscr{F}(n) ( \mathcal{A} \circ \eta ) \cdot \dot \eta\,  \Lambda+\Theta _{\rm can}(U,P)( \dot  U, \dot  P),
\end{equation} 
where the second term is the canonical one-form $\Theta _{\rm can}(U,P)( \dot  U, \dot  P)= \int  \langle P, \dot  U  \rangle \Lambda $ on $T^*G$. Then, for each $n \in \mathcal{O} $, we move on to constructing the (weak) symplectic form $ \Omega _n = - {\rm d} \Theta _n$ on $\operatorname{Diff}(T^*Q) \times T^*G$ as follows:
\[
\Omega _n( \eta , U, P)\big(( \dot \eta , \dot U, \dot P), ( \delta  \eta , \delta  U, \delta  P) \big) = \int\mathscr{F}(n)( \omega \circ \eta )( \dot  \eta , \delta  \eta ) \,\Lambda +\Omega  _{\rm can}(U,P)\big(( \dot  U, \dot  P),  ( \delta   U, \delta   P)\big),
\]
with $ \Omega _{\rm can}$ the canonical symplectic form on $T^*G$.
The symplectic form $ \Omega _n$ formally leads to defining a $n$-dependent Poisson bracket $\{ \cdot , \cdot \}_n$ on $\operatorname{Diff}(T^*Q) \times T^*G$. This  bracket can be trivially extended to define a Poisson structure on $\operatorname{Diff}(T^*Q) \times T^*G \times \mathcal{O} $ as follows:
\begin{align}\nonumber
\{f,g\}( \eta , U,P,n):=&\ \{f( \cdot , n), g( \cdot , n)\}_n( \eta ,U,P)
\\
=& \int\!\frac1{\mathscr{F}(n)}\,(\eta^*\omega^{-1})\bigg(\frac{\delta f}{\delta \eta},\frac{\delta g}{\delta \eta}\bigg)\Lambda+\{f,g\}_{\rm can}(U,P),
\label{PB_1} 
\end{align} 
where $\eta^*$ denotes pullback and $\omega^{-1}$ is the Poisson bivector associated to the canonical symplectic structure on $ T^*Q$, i.e. $
 \omega ^{-1} ( \theta_1, \theta_2 )=  \theta_1 \cdot \,\Bbb{J}\theta_2$,
for any two one-forms $\theta_1$ and $\theta_2 $. The Poisson structure \eqref{PB_1} generates trivial dynamics for $n$, that is $\dot{n}=0$.

Finally, we endow the semidirect-product group in \eqref{sdpgroup}, that is
\[
\operatorname{Diff}(T^*Q) \,\circledS\, G
= \operatorname{Diff}(T^*Q) \,\circledS\, \mathcal{F} (T^*Q, \mathcal{U} (\mathscr{H}_{\scriptscriptstyle Q})),
\]
with the usual  multiplication $(\eta _1,U_1)(\eta _2, U_2)=(\eta _1 \circ \eta _2, (U_1 \circ \eta _2)U_2)$ and we notice that there is a natural right action of $\operatorname{Diff}(T^*Q) \,\circledS\, G$ on $ \operatorname{Diff}(T^*Q) \times T^*G \times \mathcal{O} $. Indeed, this action is given by
\begin{equation}\label{SDP_action} 
( \eta , U, P, n) 
%( \varphi , V) = 
\mapsto
( \eta \circ \varphi , ( U \circ \varphi )V, ( \varphi ^* P)V, \Phi _ \varphi n),
\end{equation} 
where $ \varphi ^*$ denotes the pull-back of tensor densities.

With the above setting we can state the following result:
\begin{proposition}\label{Big_prop} The action \eqref{SDP_action} of $\operatorname{Diff}(T^*Q) \,\circledS\, G$ on $\operatorname{Diff}(T^*Q) \times T^*G \times \mathcal{O}$ is Poisson relative to the Poisson structure \eqref{PB_1}. By Poisson reduction relative to this group action, the reduced space 
\[
 \big(\operatorname{Diff}(T^*Q) \times T^*G \times \mathcal{O}\big)/( \operatorname{Diff}(T^*Q) \,\circledS\, G) \simeq \mathfrak{g} ^* \times \mathcal{O} 
 \] 
 is endowed with the Poisson bracket
\begin{equation}\label{Big_PB}
\begin{aligned} 
\{f,g \}( \mu , \mathsf{n})
= & \int \!\frac{1}{\mathscr{F}(\mathsf{n})}\, \omega ^{-1} \bigg( \mathcal{J} \bigg( \mathsf{n}, \frac{\delta f}{\delta \mathsf{n} } \bigg) , \mathcal{J} \bigg(\mathsf{n},  \frac{\delta g}{\delta \mathsf{n} } \bigg) \bigg) \Lambda 
+ 
\int\!\left\langle \mu  ,\left[ \frac{\delta f}{\delta \mu }, \frac{\delta g}{\delta \mu }\right]  \right\rangle \Lambda  \\
 + &\!\int\! \frac{1}{\mathscr{F}(\mathsf{n})} \,\omega ^{-1} \bigg( \mathcal{J} \bigg(\mathsf{n}, \frac{\delta f}{\delta \mathsf{n} } \bigg) , \bigg\langle \mu , {\rm d} \frac{\delta g}{\delta \mu } \bigg\rangle  \bigg) \Lambda 
-
\int\! \frac{1}{\mathscr{F}(\mathsf{n})} \,\omega ^{-1}\! \bigg( \mathcal{J} \bigg( \mathsf{n}, \frac{\delta g}{\delta \mathsf{n} } \bigg) , \bigg\langle \mu , {\rm d} \frac{\delta f}{\delta \mu } \bigg\rangle  \bigg) \Lambda \\
+ &\! \int\! \frac{1}{\mathscr{F}(\mathsf{n})} \,\omega ^{-1} \bigg( \bigg\langle \mu , {\rm d} \frac{\delta f}{\delta \mu } \bigg\rangle, \bigg\langle \mu , {\rm d} \frac{\delta g}{\delta \mu } \bigg\rangle  \bigg) \Lambda .
\end{aligned}
\end{equation} 
%Here, $\omega^{-1}$ is the Poisson bivector associated to the canonical symplectic structure on $ T^*Q$, i.e. {\color{blue}$ \omega ^{-1} ( \theta_1, \theta_2 )=  \theta_1 \cdot \,\Bbb{J}\theta_2$, for any two one-forms $\theta_1$ and $\theta_2 $}.
\end{proposition} 
\textbf{Proof.} We note that, for each $n \in \mathcal{O} $, the one-form in \eqref{1form} has the following invariance with respect to $\operatorname{Diff}(T^*Q) \,\circledS\, G$:
\[
\Theta _{ \Phi _ \varphi n}( \eta \circ \varphi ,( U \circ \varphi )V, ( \varphi ^* P)V) \big(  \dot  \eta \circ \varphi , ( \dot  U \circ \varphi )V, ( \varphi ^* \dot   P)V \big) = \Theta _n( \eta , U, P)\big( \dot  \eta ,\dot  U, \dot  P\big),
\]
for all $( \varphi , V) \in \operatorname{Diff}(T^*Q) \,\circledS\, G$. As a consequence, the symplectic form $ \Omega _n$ on $ \operatorname{Diff}(T^*Q) \times T^*G$ inherits the same invariance property. From this, one directly concludes that the action \eqref{SDP_action} on $\operatorname{Diff}(T^*Q) \times T^*G \times \mathcal{O} $ is Poisson relative to the Poisson structure \eqref{PB_1}.  Thus, one can implement Poisson reduction with respect to the action of the group $ \operatorname{Diff}(T^*Q) \,\circledS\, G$, thereby obtaining a Poisson bracket structure on the quotient $\big(\operatorname{Diff}(T^*Q) \times T^*G \times \mathcal{O}\big)/( \operatorname{Diff}(T^*Q) \,\circledS\, G)$. The latter is identified with $\mathfrak{g} ^* \times \mathcal{O} $ via the map $[ \eta , U,P, n] \mapsto ( \eta _*(P U^\dagger), \Phi _{ \eta } ^{-1}n):= ( \mu , \mathsf{n})$, with $[ \eta , U,P, n] $ denoting the equivalence class in the quotient space.
We will find this reduced bracket on $\mathfrak{g} ^* \times \mathcal{O}$ by exploiting the variational approach, thereby deducing the dynamics of the Poisson-reduced variables $( \mu , \mathsf{n})$.

As noticed earlier,  the Poisson bracket \eqref{PB_1} on $\operatorname{Diff}(T^*Q) \times T^*G \times \mathcal{O} $ yields the trivial dynamics  $ \dot n=0$, i.e., $n=n_0$. For such a fixed $n_0$, the remaining content of the Hamilton equations arising from \eqref{PB_1} are associated to the symplectic form $ \Omega _{n_0}$ on $\operatorname{Diff}(T^*Q) \times T^*G$. 
In particular, given a Hamiltonian $H: \operatorname{Diff}(T^*Q) \times T^*G \times \mathcal{O} \rightarrow \mathbb{R} $, these equations follow from the phase-space variational principle for the canonical Hamiltonian motion associated to $ \Omega _{n_0}= - {\rm d} \Theta _{n_0}$ on $\operatorname{Diff}(T^*Q) \times T^*G $, that is
$
\delta \!\int_{t_1}^{t_2}\! \big[  \Theta _{n_0}( \eta , U, P) \cdot ( \dot  \eta , \dot  U , \dot  P)    - H( \eta , U,P, n_0) \big]  {\rm d} t=0$.
More explicitly,
\begin{equation}\label{general_VP} 
\delta \!\int_{t_1}^{t_2} \left[ \int \big(\mathscr{F}(n_0)  (\mathcal{A} \circ \eta ) \cdot \dot \eta\, +  \langle P, \dot U  \rangle \big)\Lambda   - H( \eta , U,P, n_0) \right]  {\rm d} t=0.
\end{equation} 

Assuming that $H$ is invariant with respect to $\operatorname{Diff}(T^*Q) \,\circledS\, G$, denote its reduced Hamiltonian  as $h:\mathfrak{g}  ^* \times  \mathcal{O} \rightarrow \mathbb{R} $. Thus, by reduction, the variational principle \eqref{general_VP} yields
\begin{equation}\label{general_VP_reduced} 
\delta \!\int_{t_1}^{t_2}  \left[ \int  \big(\mathscr{F}(\mathsf{n}) \mathcal{A}  \cdot  {\cal X}  + \left\langle \mu , \xi \right\rangle \big)\,\Lambda  - h (\mu , \mathsf{n}) \right]  {\rm d} t=0,
\end{equation}
where we defined $ \mathcal{X} = \dot \eta \circ \eta ^{-1} $, $ \xi = \eta _*(P U ^\dagger)$, $\mathsf{n}= \Phi  ^{-1}_ {\eta}n$. Since the  variations $ \delta \eta $ and $ \delta U$ are free, we obtain  the constrained variations
\begin{equation}\label{EP_variations_2}
\delta {\cal X}= \partial_t{\cal Y}+{\cal X}\cdot\nabla{\cal Y}-{\cal Y}\cdot\nabla{\cal X}
,\qquad\quad 
\delta \xi =\partial _t \Sigma + [ \Sigma , \xi  ] + {\rm d}  \Sigma  \cdot {\cal X} - {\rm d}   \xi \cdot {\cal Y}
,\qquad\quad
\delta \mathsf{n}  = -{\cal Y}_ \mathcal{O} (\mathsf{n}).
\end{equation}
Also, since the variations $ \delta P$ are free, so are the variations $ \delta \mu $.

Thus, upon collecting the terms proportional to $ \mathcal{Y} $, $ \Sigma $, and $ \delta \mu $, respectively, the reduced variational principle \eqref{EP_variations_2} yields the conditions
\begin{equation*} 
{\cal X}=\frac{1}{\mathscr{F}(\mathsf{n})}\, \Bbb{J} \bigg(\langle  \mu ,  {\rm d} \xi \rangle - \mathcal{J} \bigg( \mathsf{n},\frac{\delta h}{\delta \mathsf{n}} \bigg) \bigg)
\,,\qquad\quad 
 \partial_t \mu + \operatorname{ad}^*_ \xi \mu + \operatorname{div}( \mu \mathcal{X} )=0
 \,,\qquad\quad
\frac{\delta h}{\delta \mu }= \xi .
\end{equation*} 
The reduced equation of motion for $\mathsf{n}(t)\in\mathcal{O}$ is found by inserting the first condition above in the advection equation $ \partial _t \mathsf{n} + \mathcal{X} _ \mathcal{O} (\mathsf{n})=0$, which follows from $\mathsf{n}(t)=\Phi_{\eta(t)}^{-1}n_0 $. Then, the final equations are 
\begin{equation}\label{reduced_equ} 
\partial_t \mu + \operatorname{ad}^*_ { \frac{\delta h}{\delta \mu } } \mu + \operatorname{div}( \mu \mathcal{X} )=0
\,, \quad  
\partial _t \mathsf{n} + \mathcal{X} _ \mathcal{O} (\mathsf{n})=0
\,, \quad\ 
  {\cal X}=\frac{1}{\mathscr{F}(\mathsf{n})} \,\Bbb{J} \bigg(\bigg\langle  \mu ,  {\rm d} \frac{\delta h}{\delta \mu }  \bigg\rangle - \mathcal{J} \bigg( \mathsf{n},\frac{\delta h}{\delta \mathsf{n}} \bigg) \bigg).
\end{equation} 
By Poisson reduction, these equations are Hamiltonian with respect to the reduced Poisson bracket. The explicit form of the bracket is obtained by computing $ \frac{d}{dt} f( \mu (t), \mathsf{n}(t))$ along a solution of \eqref{reduced_equ}, for an arbitrary functional $f$. This leads to the Poisson structure \eqref{PBgenclas}. $\qquad\blacksquare$

\begin{remark} Using the same (weak) symplectic form on $ \operatorname{Diff}(T^* {Q} )$, one can consider the more general situation of a Poisson bracket constructed in the same way on $\operatorname{Diff}(T^* {Q} ) \times \mathcal{M} \times \mathcal{O} $, where $ \mathcal{M} $ is an other symplectic manifold, acted on symplectically by $ \operatorname{Diff}(T^* {Q} )$. In this case, Poisson reduction by $ \operatorname{Diff}(T^* {Q} )$ can be used to endow $ \mathcal{M} \times \mathcal{O} $ with a reduced Poisson bracket. This more general case will be pursued elsewhere.
\end{remark} 

\begin{corollary}\label{corollary} The bracket \eqref{bracket_candidate2} associated to the closure \eqref{adv_D}-\eqref{tilde_rho_adv} is a Poisson bracket on  
\[
\operatorname{Den}_+(T^*Q) \times \big( \operatorname{Den}(T^* Q ) \otimes  \operatorname{He}( \mathscr{H}_{\scriptscriptstyle Q})\big)   .
\]
\end{corollary}
\textbf{Proof.} We apply Proposition \ref{Big_prop} to the manifold $ \mathcal{O} = \operatorname{Den}_+( T^*Q)$ and with $\mathscr{F}=id: \mathcal{O} \rightarrow \operatorname{Den}( T^*Q)$, which is equivariant with respect to the action by pull-back of densities. Using the notation $\mathsf{n}=D$ and $ \mathcal{P} = i \hbar ^{-1} \mu  $, and noting that $ \frac{1}{\mathscr{F}(\mathsf{n})} \,\mathcal{J} \big( \frac{\delta h}{\delta \mathsf{n}} \big) $ becomes $- {\rm d} \frac{\delta h}{\delta D}$, the expression \eqref{Big_PB} reduces to  \eqref{bracket_candidate2}. $\qquad\blacksquare$ 

\medskip 
\noindent The various relations among the Poisson brackets appeared so far are summarized by the following diagram:
{
%\begin{figure}[h!]
\noindent\footnotesize
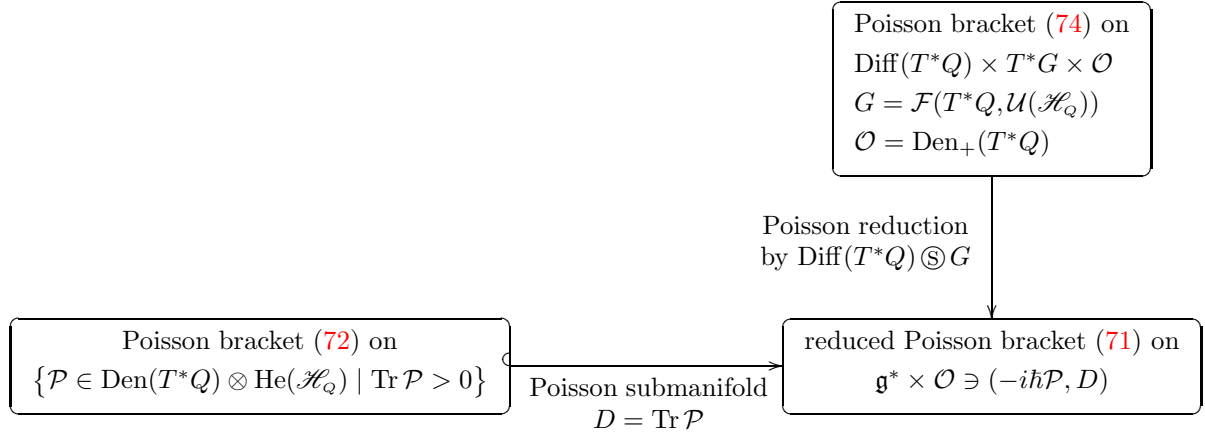
\begin{center}
\hspace{2cm}\begin{xy}
\hspace{-1cm}
\xymatrix{
& & \hspace{1cm} &
*+[F-:<3pt>]{
\begin{array}{l}
\vspace{0.1cm}\text{Poisson bracket \eqref{PB_1} on}\\
\vspace{0.1cm}\text{$\operatorname{Diff}(T^*Q) \times T^*G \times \mathcal{O}$}\\
\vspace{0.1cm} G=\mathcal{F} (T^*Q, \mathcal{U} (\mathscr{H}_{\scriptscriptstyle Q}))\\
\vspace{0.1cm}\mathcal{O} = \operatorname{Den}_+( T^*Q)
\end{array}
}\ar[dd]_{ \begin{array}{c} \\\text{Poisson reduction}\\ \text{by $\operatorname{Diff}(T^*Q) \,\circledS\,  G$} \end{array} }\\
& & & \\
&*+[F-:<3pt>]{
\begin{array}{c}
\vspace{0.1cm}\text{Poisson bracket \eqref{bracket_candidate_rho} on}\\
\vspace{0.1cm}\text{$\big\{ \mathcal{P} \in \operatorname{Den}(T^* {Q})\otimes \operatorname{He}(\mathscr{H}_{\scriptscriptstyle Q}) \mid \operatorname{Tr} \mathcal{P} >0\big\} $}
\end{array}
}\ar@{^{(}->}[rr]_{  \hspace{0.6cm} \begin{array}{c} \text{Poisson submanifold}\\ \text{$D= \operatorname{Tr}\mathcal{P}  $} \end{array} }
& \hspace{1.7cm} &
*+[F-:<3pt>]{
\begin{array}{c}
\vspace{0.1cm}\text{reduced Poisson bracket \eqref{bracket_candidate2} on}\\
\vspace{0.1cm}\text{$\mathfrak{g} ^* \times \mathcal{O} \ni ( - i \hbar \mathcal{P}  ,  D)$}
\end{array}
}
}
\end{xy}
%\end{figure}
\end{center}
}

While the first term in the hybrid Poisson bracket \eqref{bracket_candidate_rho} essentially coincides with the familiar Lie-Poisson bracket underlying the quantum Liouville equation, the second term looks somewhat mysterious and requires a special discussion. This term corresponds to the last term in the bracket \eqref{Big_PB}. As we will see, the first and the last term in the bracket \eqref{Big_PB} possess a similar nature, which will be discussed in the next section. In particular, each of these terms extends the usual Lie-Poisson structure \eqref{LiouvBracket} underlying the classical Liouville equation.  
\rem{%%%%%%%%%%%%%%%%%%%%%%%%%%%%%%
It is interesting to also note that in absence of the $ \mathcal{F} (T^*Q, \mathcal{U} (\mathscr{H}_{\scriptscriptstyle Q})) $-component, the variational principle \eqref{unreduced_1} reduces to
\begin{equation}\label{unreduced_1_simpler}  
\delta \int_{t_1}^{t_2} \left[   \int_{T^*Q}  \operatorname{Tr}(\mathcal{P}  _0)  (\mathcal{A} \circ \eta ) \cdot \dot \eta  \,\Lambda   - h \left(  \eta _*   \mathcal{P}  _0 \right)  \right]  {\rm d} t=0,
\end{equation} 
where we have chosen $D_0= \operatorname{Tr}(\mathcal{P}  _0)>0$. It is of the form \eqref{PS_principle} with the one-form
\[
\Theta ( \eta ) \cdot \dot \eta =   \int_{T^*Q} \operatorname{Tr}(\mathcal{P}  _0) (\mathcal{A} \circ \eta ) \cdot \dot \eta  \, \Lambda.
\]
As opposed to \eqref{unreduced_1}, $ \Omega = - {\rm d} \Theta $ is a (weak) symplectic form, namely,
\begin{equation}\label{2_form_1_simpler} 
\Omega ( \eta ) \left(  \dot \eta ,  \delta \eta  \right) =  \int_{T^*Q} \operatorname{Tr}(\mathcal{P}  _0) ( \omega  \circ \eta ) ( \dot \eta, \delta \eta ) \Lambda.
\end{equation} 
In \eqref{unreduced_1_simpler} both the one-form and the Hamiltonian are $G_{\mathcal{P}  _0}$ invariant, therefore the reduced dynamics on $G/G_{\mathcal{P}  _0}$ is Poisson. As before, we obtain the following result as a particular instance of the preceding.
}     %%%%%%%%%%%%%%%%%%%%%%%%%%%%%%

\subsection{Poisson structures of classical Liouville dynamics\label{sec:HamStr3}}

This section shows how some terms in the bracket \eqref{Big_PB} provide an entire class of noncanonical Poisson structures extending the usual formulation of classical Liouville dynamics. Since here we identify classical dynamics with the Liouville equation, its fundamental Poisson structure  is given by \eqref{LiouvBracket}, which is a Lie-Poisson bracket as discussed in Section \ref{sec:KvHgeom}. Depending on the representation adopted for the classical Liouville density, the original Lie-Poisson structure \eqref{LiouvBracket} is transformed into an alternative noncanonical bracket.  For example, the KvN prescription $\rho=|\Psi|^2$ was recently shown to provide a new type of noncanonical Poisson bracket underlying the KvN equation \cite{TrJo21}. In more generality, this process leads to a new class of Poisson structures comprising the last term in \eqref{bracket_candidate_rho} as one particular example.
\begin{corollary} The last term of \eqref{Big_PB}, namely
\begin{equation}\label{PBgenclas} 
\{\!\!\{ f, g\}\!\!\}(\mathsf{n})%=\int_{T^*Q}\frac1{\mathscr{F}(n)}\,\omega^{-1} \bigg(  n\diamond\frac{\delta f}{\delta n}  , n\diamond\frac{\delta f}{\delta n}\bigg)  \,\Lambda\,,
=\int\frac1{\mathscr{F}(\mathsf{n})}\ \omega^{-1\!} \bigg(  \mathcal{J} \bigg(\mathsf{n},\frac{\delta f}{\delta \mathsf{n}}\bigg)  , \mathcal{J} \bigg(\mathsf{n},\frac{\delta g}{\delta \mathsf{n}}\bigg)\!\bigg)  \,\Lambda\,.
\end{equation} 
is a Poisson bracket on $ \mathcal{O} $. Also, let the group $\operatorname{Aut}_ \mathcal{A} (T^*Q\times S^1) $ defined in \eqref{stricts} act from the left on $ \mathcal{O} $ as $\mathsf{n} \mapsto\Phi _ {\eta}^{-1} \mathsf{n}$. Then, this action is canonical with respect to the Poisson structure \eqref{PBgenclas} and admits $\mathscr{F}:{\cal O}\to\operatorname{Den}(T^*Q)$ as an equivariant momentum map.
\end{corollary} 
\noindent\textbf{Proof.} 
The first part of the statement follows by Poisson reduction of the Poisson bracket
$
\{f,g\}( \eta ,n):=\{f( \cdot , n), g( \cdot , n)\}_n( \eta )
$
on $ \operatorname{Diff}(T^*Q) \times \mathcal{O} $, where $\{ \cdot , \cdot \}_n$ is the $n$-dependent Poisson bracket on $\operatorname{Diff}(T^*Q)$ associated to the $n$-dependent (weak) symplectic form on $\operatorname{Diff}(T^*Q)$
\beq\label{bigsympform}
\Omega _n( \eta )(u_ \eta , v_ \eta ) = \int\mathscr{F}(n)( \omega \circ \eta )(u_\eta , v_ \eta ) \,\Lambda\,.
\eeq
Indeed,  
it suffices to apply Proposition \ref{Big_prop} to the case when the group $G$ is absent.
To show that the action $n\mapsto\Phi_{\eta} n$ is Poisson, it is enough to prove that $\{\!\!\{ f \circ \Phi _ \eta ,  g \circ \Phi _ \eta \}\!\!\}(\mathsf{n})=\{\!\!\{ f  ,  g   \}\!\!\}( \Phi _ \eta(\mathsf{n}))$ for all $f,g \in \mathcal{F} ( \mathcal{O} )$ and $ \eta \in \operatorname{Diff}_ \omega (T^*Q)$. Indeed, we have the equality 
\[
\mathcal{J} \bigg( \frac{\delta (f \circ \Phi _ \eta )}{\delta \mathsf{n}} \bigg) = \eta _* \bigg( \mathcal{J} \bigg( \frac{\delta f }{\delta \mathsf{n}} \circ \Phi _ \eta\bigg) \bigg) .
\]
Hence, for all $f,g \in \mathcal{F} ( \mathcal{O} )$ and $ \eta \in \operatorname{Diff}_ \omega (T^*Q)$, we compute
\begin{align*}
\{\!\!\{ f \circ \Phi _ \eta ,  g \circ \Phi _ \eta \}\!\!\}(\mathsf{n}) =&\int\frac1{\mathscr{F}(\mathsf{n})}\ \omega^{-1\!} \bigg( \eta _* \bigg( \mathcal{J} \bigg( \frac{\delta f }{\delta \mathsf{n}} \circ \Phi _ \eta\bigg) \bigg)  , \eta _* \bigg( \mathcal{J} \bigg( \frac{\delta g }{\delta \mathsf{n}} \circ \Phi _ \eta\bigg) \bigg) \!\bigg)  \,\Lambda\\
=&\int\frac1{\eta _* (( \mathscr{F}\circ \Phi _ \eta) (\mathsf{n}) ) } \ \omega^{-1\!} \bigg(  \mathcal{J} \bigg( \frac{\delta f }{\delta \mathsf{n}} \circ \Phi _ \eta\bigg)  ,  \mathcal{J} \bigg( \frac{\delta g }{\delta \mathsf{n}} \circ \Phi _ \eta\bigg) \!\bigg) \circ \eta ^{-1}   \,\Lambda\\
=&\ \{\!\!\{ f  ,  g   \}\!\!\}( \Phi _ \eta\mathsf{n}),
\end{align*} 
where we used $ \omega ^{-1} ( \eta _* \alpha , \eta _* \beta )= \omega ^{-1} ( \alpha , \beta ) \circ \eta ^{-1} $. 
The final part of the theorem is proved by direct verification upon recalling that the equivariance  $\eta_*\left( \mathscr{F}(\mathsf{n}) \right) =\mathscr{F}(\Phi_\eta^{-1} \mathsf{n})$ 
implies the infinitesimal  property $ \operatorname{div} (\mathscr{F}(n) \mathcal{X} )=\langle\de_n\mathscr{F},{\cal X}_{\cal O}(n)\rangle$, where $\de_n$ denotes the differential on $\cal O$. 
Then, for all $ f \in \mathcal{F} ( \mathcal{O} )$ and all $ \xi \in \mathcal{F} (T^*Q) $, we can write
\begin{align*}
\{\!\!\{ f, \langle\mathscr{F},\xi \rangle\}\!\!\}(\mathsf{n})=&- \int \frac{1}{\mathscr{F}(\mathsf{n})}\  \mathcal{J}\bigg( \mathsf{n}, \frac{\delta  \left\langle \mathscr{F}, \xi \right\rangle }{\delta \mathsf{n}} \bigg) \cdot \Bbb{J} \mathcal{J} \bigg( \mathsf{n}, \frac{\delta f}{\delta \mathsf{n}} \bigg) \,\Lambda \\
=&  -\left\langle \frac{\delta  \left\langle \mathscr{F}, \xi \right\rangle }{\delta \mathsf{n}} , \bigg(   \frac1{\mathscr{F}(\mathsf{n})}\ \Bbb{J}   \mathcal{J} \bigg(\mathsf{n},\frac{\delta f}{\delta \mathsf{n}}\bigg) \bigg) _{\! \mathcal{O}} (\mathsf{n}) \right\rangle 
\\
=&- \int\!\xi\,\de_\mathsf{n}\mathscr{F}\cdot\bigg(\frac1{\mathscr{F}(\mathsf{n})}\ \Bbb{J}   \mathcal{J} \bigg(\mathsf{n},\frac{\delta f}{\delta \mathsf{n}}\bigg)  \bigg)_{\!\cal O}(\mathsf{n}) \ \Lambda
\\
= &-  \int \!\xi \operatorname{div} \bigg( \Bbb{J}    \mathcal{J} \bigg(\mathsf{n},\frac{\delta f}{\delta \mathsf{n}}\bigg) \bigg) \, \Lambda\\
=& -\int\!    \mathcal{J}  \bigg(\mathsf{n},\frac{\delta f}{\delta \mathsf{n}}\bigg)\cdot X_\xi \,\Lambda
\\
=& - \left\langle \frac{\delta f}{\delta n},  (X_\xi)_{\cal O}(\mathsf{n}) \right\rangle 
\,.
\end{align*}
Here, we have used $X_\xi=\Bbb{J}\de\xi$ in the last equality. Thus, since the left action $\mathsf{n} \mapsto
\Phi_\eta^{-1}\mathsf{n}$ of $\operatorname{Aut}_ \mathcal{A} (T^*Q\times S^1) $ leads to the infinitesimal generator $\xi _ \mathcal{O} (\mathsf{n})=-(X_\xi)_{\cal O}(\mathsf{n})$, this verifies the defining relation $\{\!\!\{ f, \langle\mathscr{F},\xi \rangle\}\!\!\}(\mathsf{n})= \langle{\delta f}/{\delta \mathsf{n}},\xi _ \mathcal{O} (\mathsf{n})\rangle$ for momentum maps on Poisson manifolds.
$\qquad\blacksquare$

\medskip
\noindent 
Given a Hamiltonian functional $h({\sf n})$, the Poisson bracket \eqref{PBgenclas}  yields the dynamics
\[
\dot{\sf n}+{\cal X}_{\cal O}({\sf n})=0
\,,\qquad\text{ with }\qquad
{\cal X}=-\frac1{\mathscr{F}({\sf n})}\,\Bbb{J}{\cal J}\bigg({\sf n},\frac{\delta h}{\delta {\sf n}}\bigg).
\]
Importantly, this form of dynamics yields a family of alternative representations of the classical density evolution. Indeed, since $\mathscr{F}$ is an equivariant momentum map, it is also a Poisson map taking the Poisson bracket \eqref{PBgenclas}  into \eqref{LiouvBracket}. Then, classical Liouville dynamics may be obtained by the Guillemin-Sternberg collectivization of momentum maps \cite{GuSt80}. Indeed, if the Hamiltonian is of the \emph{collective} form $h({\sf n})=\int H\mathscr{F}({\sf n})\,\Lambda$, then the collectivization ensures that the momentum map $\rho_c=\mathscr{F}({\sf n})$ obeys the classical Liouville equation $\partial_t\rho_c=\{H,\rho_c\}$. 

As an immediate example, the usual Poisson bracket \eqref{LiouvBracket} for the classical Liouville equation arises from \eqref{PBgenclas} in the special case of (strictly positive) densities  ${\cal O}=\operatorname{Den}_+(T^*{{Q}})$ with $\mathscr{F}(\mathsf{n})=\mathsf{n}$.
Another interesting case  is given by the space ${\cal O}=L^2_0(T^*{{Q}})$ of 
(nonzero) Koopman wavefunctions with $\mathscr{F}(\mathsf{n})=|\mathsf{n}|^2$. Upon using the identification $L^2(T^*{{Q}})\simeq\operatorname{Den}^{1/2}(T^* {Q})$ of Koopman wavefunctions with half-densities on $T^* Q$, the natural action of $\operatorname{Diff}(T^*Q)$ on half-densities yields a noncanonical Poisson structure for Koopman-von Neumann theory as recently presented in \cite{TrJo21}. Alternatively, one can consider the  manifold ${\cal O}= \{{\cal P}\in\mathcal{F} (T^*Q, \operatorname{He}(\mathscr{H}_{\scriptscriptstyle Q})) ^*\,|\,\operatorname{Tr}\mathcal{P} >0\}$ of (strictly positive) densities with values in the space of  Hermitian operators on $\mathscr{H}_{\scriptscriptstyle Q}$. Here, we recall that $\mathcal{F} (T^*Q, \operatorname{He}(\mathscr{H}_{\scriptscriptstyle Q})) ^*\simeq \operatorname{Den}(T^* \mathcal{Q})\otimes \operatorname{He}(\mathscr{H}_{\scriptscriptstyle Q}) $.
If one sets $\mathscr{F}(\mathsf{n})=\operatorname{Tr} \mathsf{n}$ and uses the pullback action on $\cal O$, then the Poisson bracket \eqref{PBgenclas} yields the last term of \eqref{bracket_candidate} (or, equivalently, the last term in \eqref{bracket_candidate_rho}) for ${\sf n}={\cal P}$. Thus,  the  expression 
\beq\label{gino}
\{\!\!\{ f, g\}\!\!\}_0(\mathcal{P}   )= \int\! \operatorname{Tr}( \mathcal{P}   )\, \omega \Big( \big\langle X_{  {\delta F}/{\delta \mathcal{P} }  }\big\rangle,\big\langle X_{  {\delta H}/{\delta \mathcal{P} } }\big\rangle\Big)  \,\Lambda,
\eeq 
%It arise by Poisson reduction of the weak symplectic form \eqref{2_form_1_simpler}.
defines a Poisson bracket structure  on  $\cal O$.

In more generality, \eqref{PBgenclas} identifies a new general class of Poisson structures for different representations of the  Liouville density in classical mechanics.
To our knowledge, this general class  has not appeared before in the literature and one is led to ask if certain bracket structures in  this family may be advantageous beyond the particular context of hybrid quantum-classical systems. For example, following \cite{Joseph20}, the authors of \cite{TrJo21} proposed a possible role of the bracket \eqref{PBgenclas}  for applications in plasma physics. In this case, one would set ${\cal O}=L^2(T^*Q)$ so that  \eqref{PBgenclas}  defines a noncanonical Poisson structure for Koopman-von Neumann wavefunctions in classical mechanics. We leave the investigation of other cases of interest for future work.

After this digression into noncanonical Hamiltonian structures for classical mechanics, we will now focus on the construction of a hybrid density operator mimicking the analogous quantity \eqref{hybridDenOp} from the original theory. In particular, we  will apply the geometric methods treated in the present section to  study the equivariance properties of this hybrid density, along with  their implications on the dynamics of the classical and quantum subsystems.

\subsection{Hybrid density and equivariance\label{sec:clasquantdens}}

As we have seen in Section \ref{sec:closure}, the entire information in the present closure model is contained in the quantity $\mathcal{P}=D\psi\psi^\dagger$. However, the total energy identified by the Hamiltonian functional \eqref{h_D_tilde_rho} clearly differs from the usual expression of the expectation value of the Hamiltonian operator, that is the first term $\int \langle\widehat{\cal P}, \widehat{H}\rangle\Lambda$ in \eqref{h_D_tilde_rho}. This leaves us with two possible interpretations. On the one hand, we could  accept that the usual expectation of the Hamiltonian is not conserved during quantum-classical interaction, and the energy balance involves extra work -- here given by the second term in \eqref{h_D_tilde_rho} -- that is produced by correlation effects. This seems to be the possibility suggested by the authors of \cite{Carroll}. On the other hand, we may simply use integration by parts to isolate a  form of the hybrid quantum--classical density that is alternative to that presented in \eqref{hybridDenOp}. As we show in this section, this second possibility leads, once again, to a sign-indefinite hybrid density, although now both the quantum and the classical densities alone remain always positive.

We observe that combining the factorization \eqref{EFDef} with the constraint \eqref{constraint1} changes the original expression \eqref{hybridDenOp} into the following:
\begin{align}
\nonumber
\widehat{\cal D}
%=&\, 
%\color{red}D{\rho}+\frac{i\hbar}2\operatorname{div}\!\left(D\big[\rho,\Bbb{J}\nabla\rho\big]\right)
%\\
%=&\, 
%D{\rho}+\frac{i\hbar}2\left(\{D\rho,\rho\}+\{\rho,D\rho\}\right)
%\\
%=&\, 
%D\left({\rho}+{i\hbar}\{\rho,\rho\}\right)+\frac{i\hbar}2[\rho,\{D,\rho\}]
%\\
=&\, 
\mathcal{P}+\frac{i\hbar}2\operatorname{div}\!\left(D^{-1}\big[\mathcal{P}, {\Bbb{J}} {\rm d} \mathcal{P}\big]\right)
\\
=&\, 
\mathcal{P}+\frac{i\hbar}2\left(\{D^{-1}\mathcal{P},\mathcal{P}\}+\{\mathcal{P},D^{-1}\mathcal{P}\}\right)
\nonumber
\\
=&\, 
\mathcal{P}+\frac{i\hbar}D\Big(\{\mathcal{P},\mathcal{P}\}+\frac1{2D}[\mathcal{P},\{D,\mathcal{P}\}]\Big)
\,,
\label{hybdens}
\end{align}
so that the Hamiltonian functional \eqref{h_D_tilde_rho} reads $h(D, \mathcal{P}  )=\operatorname{Tr}\int \!\widehat{H}(z)\widehat{\cal D}(z)\,\Lambda$, in analogy with \eqref{Dintro}. We notice that the expression \eqref{hybdens} is the sum of  $\mathcal{P}=D\psi\psi^\dagger$ and another term whose trace and integral vanish identically. The latter term is responsible for the dynamics of quantum--classical correlation effects such as quantum decoherence. Notice that this term contributes to the expectation of neither  purely classical nor purely quantum observables, since $D= \operatorname{Tr}\widehat{\cal D}=\operatorname{Tr}\mathcal{P}$  and $\hat \rho  _q=\int \!\widehat{\cal D}  \Lambda =\int \!\mathcal{P}  \Lambda$.

The hybrid operator \eqref{hybdens} may now be used to compute quantum-classical expectation values. For example, if $\widehat{A}(z)$ denotes a hybrid observable, as discussed at the end of Section \ref{sec:closure}, its expectation value will be given by $\int\langle\widehat{\cal D},\widehat{A}\rangle\,\Lambda$. Then, the dynamics of expectation values and the associated Heisenberg picture of hybrid dynamics deserve a dedicated discussion that is left for future work. Instead, here we will show that the hybrid density operator \eqref{hybdens} enjoys two equivariance properties that may be used in writing the dynamics of the classical and quantum state. In particular, we have the following statement:

\begin{proposition}[Equivariance of the hybrid density operator]\label{equivprop}
The quantum--classical density operator $\widehat{\cal D}=\widehat{\cal D}(D,\mathcal{P})$ in \eqref{hybdens} satisfies the following equivariance properties
\beq
\widehat{\cal D}(D,{\sf U}\mathcal{P}{\sf U}^\dagger)={\sf U}\widehat{\cal D}(D,\mathcal{P}){\sf U}^\dagger
\,,\qquad\quad
\widehat{\cal D}({\upeta}^*D,{\upeta}^*\mathcal{P})={\upeta}^*\big(\widehat{\cal D}(D,\mathcal{P})\big)
\label{equivD}
\,,
\eeq
%\todo{\color{magenta} FGB: Maybe ${\upeta}^*\widehat{\cal D}(D,\mathcal{P})$ is more clear as ${\upeta}^* \left( \widehat{\cal D}(D,\mathcal{P}) \right) $}
for any unitary operator ${\sf U}\in{\cal U}(\mathscr{H}_{\scriptscriptstyle Q})$ and any symplectic diffeomorphism ${\upeta}\in\operatorname{Diff}_\omega(T^*{Q})$.
\end{proposition}
\textbf{Proof.} Since ${\sf U}$ is not a function on $T^*{Q}$, the first relation is verified immediately. The second relation follows from
\[
\int\langle{\upeta}^*\widehat{\cal D}(D,\mathcal{P}),\widehat{H}\rangle\,\Lambda=\int\langle\widehat{\cal D},{\upeta}_*\widehat{H}\rangle\,\Lambda=\int\langle\widehat{\cal D}({\upeta}^*D,{\upeta}^*\mathcal{P}),\widehat{H}\rangle\,\Lambda
\,.
\]
Here, the second equality is verified as follows. First, it is convenient to write $\widehat{\cal D}$ in the form
$
\widehat{\cal D}
=
D{\rho}+({i\hbar}/2)\operatorname{div}(D[\rho,\Bbb{J}\de\rho])
$, 
so that
\begin{align*}
\int\langle\widehat{\cal D},{\upeta}_*\widehat{H}\rangle\,\Lambda
=& 
\int \!D \,\Big\langle\rho,{\upeta}_*\widehat{H}+\frac{i\hbar}{2} \big[{\rm d} \rho, X_{{\upeta}_*\widehat{H}}\big]\Big\rangle\, \Lambda 
\\
=&
\int \! \,\Big\langle{\upeta}^*(D\rho),\widehat{H}+\frac{i\hbar}{2} \big[{\rm d}( {\upeta}^*\rho), X_{\widehat{H}}\big]\Big\rangle\, \Lambda 
\\
=&
\int \! \,\Big\langle({\upeta}^*(D\rho)+\frac{i\hbar}2\operatorname{div}\!\left(\big[{\upeta}^*(D\rho),\Bbb{J}\de({\upeta}^*\rho)\big]\right),\widehat{H}\Big\rangle\, \Lambda 
\,,
\end{align*}
where we have used $X_{{\upeta}_*\widehat{H}}={\upeta}_*X_{\widehat{H}}$ and $X_{\widehat{H}}=\Bbb{J}\de{\widehat{H}}$. Also, notice the relation $({\upeta}^*X_{\widehat{H}}){\cdot({\upeta}^*\de\rho)}={\upeta}^*(X_{\widehat{H}}\cdot\de\rho)$ and similarly for ${\upeta}^*(\de\rho\cdot X_{\widehat{H}})$.
 Then, the proof is completed by using ${\upeta}^*\mathcal{P}=({\upeta}^*D)({\upeta}^*\rho)$, so that ${\upeta}^*\rho=({\upeta}^*D)^{-1\,}{\upeta}^*\mathcal{P}. \qquad\blacksquare$ 

\medskip
\noindent
Notice that here $\upeta$ denotes a symplectomorphism, unlike $\eta$ which instead was used in the previous sections to denote a generic diffeomorphism. Analogously, here ${\sf U}$ denotes a standard unitary operator on $\mathscr{H}_{\scriptscriptstyle Q}$, while we recall that $U=U(z)$ retains parametric dependence on the classical coordinates.

The equivariance properties \eqref{equivD} of the hybrid density operator \eqref{hybdens} under both quantum and classical transformations mimic analogous relations already applying to the original quantity \eqref{hybridDenOp} from the full hybrid Koopman theory \cite{GBTr20}.   Proposition \ref{equivprop} ensures that  the present closure model satisfies the  property 3 from the list of desirable consistency relations appearing in the Introduction. This property
has long been sought in the theory of hybrid quantum--classical systems \cite{boucher} and stands as one of the key geometric features of the present construction. 
Importantly, the relations \eqref{equivD} lead to  writing the dynamics of the classical Liouville density $D$ and the quantum density matrix $\hat{\rho}_q=\int\mathcal{P}\,\Lambda$ as
\beq\label{erika}
\frac{\partial D}{\partial t}=\operatorname{Tr}\{\widehat{H},\widehat{\cal D}\}
\,,\qquad\qquad
i\hbar\frac{\de \hat{\rho}_q}{\de t}=\int[\widehat{H},\widehat{\cal D}]\,\Lambda
\,,
\eeq
respectively. Again, these dynamical equations are formally the same as those appearing in the original theory of hybrid wavefunctions \cite{BoGBTr19} from Section \ref{sec:QChybrids}, although in that case the hybrid density $\widehat{\cal D}$ is given by \eqref{hybridDenOp}. In particular, the equations \eqref{erika}  imply that the model in Proposition \ref{propclosure} satisfies the properties 4 and 5 from the list of desirable self-consistency criteria in the Introduction. In order to show how \eqref{erika} follow from \eqref{equivD}, we present the following statement:

\begin{proposition}[Collectivization by equivariant maps]\label{collectivization}
Let $ \mathcal{H} :P\to\Bbb{R}$ a Hamiltonian on the Poisson manifold $P$ and let $J:P\to\mathfrak{g}^*$ be a momentum map corresponding to a left canonical $G-$action on  $P$. Assume that $G$ also acts on the left on a manifold $N$ and let ${\cal D}:P\to N$ be a $G-$equivariant map such that there exists a function $\widetilde{ \mathcal{H} }:N\to\Bbb{R}$ with $ \mathcal{H} =\widetilde{ \mathcal{H} }\circ{\cal D}$. Then, if the curve $p(t)\in P$ satisfies Hamilton's equations, the momentum map $J(t):=J(p(t))$ evolves according by
\begin{equation}\label{dot_J} 
\frac{\de J}{\de t}=-{\cal J}\bigg({\cal D},\frac{\delta \widetilde{ \mathcal{H} }}{\delta {\cal D}}\bigg)
\end{equation} 
where $ \mathcal{J} :T^* N \rightarrow  \mathfrak{g} ^* $ is the cotangent-lift momentum map associated to the $G-$action on $N$.
\end{proposition}
\textbf{Proof.} The result follows from a direct computation using the definition of the momentum maps. For convenience, in this proof we will denote the duality pairing on an arbitrary space by $\langle \cdot,\cdot\rangle$ and we hope that no confusion arises with the pairing on $\mathscr{H}_{\scriptscriptstyle Q}$ introduced previously. If $\dot p(t)= X_\mathcal{H} (p(t))$ are Hamilton's equations on $P$, for each $ \xi \in \mathfrak{g} $ we have
\begin{align*}
\left\langle \frac{\de J}{\de t} , \xi \right\rangle &= {\rm d} J_ \xi (p) \cdot X_{ \mathcal{H} }(p) = - \{ \mathcal{H} , J_ \xi \}(p) = - \left\langle \frac{\delta \mathcal{H} }{\delta p} , \xi _P(p) \right\rangle \\
&=- \bigg\langle \frac{\delta \widetilde{ \mathcal{H} }}{\delta \mathcal{D} } ,  T_p \mathcal{D} ( \xi _P(p)) \bigg\rangle = - \bigg\langle \frac{\delta \widetilde{ \mathcal{H} }}{\delta \mathcal{D} } ,  \xi _N( \mathcal{D} (p)) \bigg\rangle =- \bigg\langle \mathcal{J} \bigg( \mathcal{D} , \frac{\delta \widetilde{ \mathcal{H} }}{\delta \mathcal{D} }  \bigg) , \xi \bigg\rangle \,,
\end{align*} 
which proves the desired result. $\qquad\blacksquare$ 

\medskip\noindent
In analogy with the Guillemin-Sternberg collectivization by momentum maps \cite{GuSt80}, here we call the Hamiltonian $\widetilde{ \mathcal{H} }\circ{\cal D}$  ``{collective Hamiltonian}''.
At this point, we are ready to show how the equations \eqref{erika} follow directly from \eqref{equivD}. Indeed, we have the result below, which combines Proposition \ref{collectivization} with Corollary \ref{giulia}:

\begin{corollary}\label{marta} If $\mathcal{P}  $ is a solution of \eqref{andrea} then $D=\operatorname{Tr}\mathcal{P}$ and $\hat \rho  _q= \int \!\mathcal{P}\,  \Lambda $ satisfy the first and the second in \eqref{erika}, respectively.
%\beq\label{erika}
%\frac{\partial D}{\partial t}=\operatorname{Tr}\{\widehat{H},\widehat{\cal D}\}
%\,,\qquad\qquad
%i\hbar\frac{\de  \hat{\rho}_q}{\de  t}=\int[\widehat{H},\widehat{\cal D}]\,\Lambda\,.
%\eeq
\end{corollary}
\textbf{Proof.} For both identities, we apply Proposition \ref{collectivization} with
\[
P= \big\{( \mathcal{P}   , D) \in \operatorname{Den}(T^*Q)\times \big(\operatorname{Den}(T^*Q) \otimes \operatorname{He}( \mathscr{H}_Q)\big)\, \big|\, D= \operatorname{Tr}\mathcal{P}\big\}
\]
endowed with the Poisson bracket \eqref{bracket_candidate_rho}. In addition, we let 
\[
N= \operatorname{Den}(T^*Q) \otimes \operatorname{He}( \mathscr{H}_Q)
\]
with $\widehat{\cal D}:P \rightarrow N$ given by \eqref{hybdens}, $D= \operatorname{Tr}\mathcal{P}$, and  $ \mathcal{H} $ given in \eqref{h_tilde_rho}. Hence, the collective Hamiltonian $\widetilde{\mathcal{H} }: N \rightarrow \mathbb{R}$ reads $\widetilde{\mathcal{H} }(\widehat{\cal D})= \operatorname{Tr}\int \widehat{H}\widehat{\cal D} \Lambda $.

Now, if $\mathcal{P}\in P$, one verifies that $J^{(1)}:\mathcal{P} \mapsto D\in \operatorname{Den}(T^*Q)$,  and $J^{(2)}:\mathcal{P} \mapsto -i \hbar   \hat{\rho}_q\in \mathfrak{u}( \mathscr{H}_Q)$ are momentum maps for the standard left actions
\[
\Phi^{(1)}_{(\upeta,e^{{\rm i}\varphi})}\mathcal{P}=\upeta_*\mathcal{P}
\,,\qquad
\Phi^{(2)}_{\sf U}\mathcal{P}={\sf U}\mathcal{P}{\sf U}^\dagger
\]
of the groups $\operatorname{Aut}_{\cal A}(T^*{Q}\times S^1)$ and $\mathcal{U}(\mathscr{H}_{\scriptscriptstyle Q})$, respectively.
The corresponding cotangent-lift momentum maps for the same actions on $N$ are $ \mathcal{J} ^{(1)}:T^* N \rightarrow \operatorname{Den}(T^*Q)$ and $ \mathcal{J}^2:T^* N \rightarrow \mathfrak{u}(\mathscr{H}_Q)$ given by $ \int F \mathcal{J} ^{(1)}( \widehat{\cal D}  , \alpha  )\,\Lambda = \int \langle \alpha,   F_P(\widehat{\cal D}  )\rangle \Lambda $ and $ \langle \mathcal{J} ^{(2)}( \widehat{\cal D}  , \alpha  ), \xi  \rangle =  \int \langle \alpha, \xi _P(\widehat{\cal D}  )\rangle \Lambda $, with $F \in \mathcal{F} (T^*Q)$, $ \xi \in \mathfrak{u}( \mathscr{H}_Q)$. The group actions above lead to the infinitesimal generators  $ F_P( \widehat{\cal D}  )= \{F, \widehat{\cal D}  \}$ and $ \xi _P( \widehat{\cal D}  )= [ \xi , \widehat{\cal D} ]$, so that one obtains
\[
\mathcal{J} ^{(1)}(  \widehat{\cal D} , \alpha ) = \operatorname{Tr} \{  \widehat{\cal D} , \alpha  \} \qquad\text{and}\qquad \mathcal{J} ^{(2)}( \widehat{\cal D}  , \alpha )= \int [ \alpha , \widehat{\cal D} ] \Lambda\, .
\]
Since $ \delta \widetilde{ \mathcal{H} }/\delta \widehat{\cal D} =\widehat{H}$, equation \eqref{dot_J} with $J^{(1)}(\mathcal{P}  )=D$ and $J^{(2)}(\mathcal{P}  )=- i \hbar \hat \rho  _q$ gives the first and the second in \eqref{erika}, respectively. $\qquad\blacksquare$
%\[
%\frac{\partial D}{\partial t}=  - \mathcal{J} ^{(1)}( \widehat{\cal D}, \widehat{H}) = \operatorname{Tr}\{\widehat{H}, \widehat{\cal D}\} \qquad\text{and}\qquad  i\hbar\frac{\partial  \hat{\rho}_q}{\partial  t}= \color{blue}\mathcal{J} ^{(2)}( \widehat{\cal D}, \widehat{H}) = \int [ \widehat{H} , \widehat{\cal D} ] \Lambda\,. \qquad\blacksquare
%\]

\medskip
Having characterized the hybrid density operator \eqref{hybdens} in terms of its equivariance properties and its role in the evolution of the quantum and classical subsystems, we move on to present a specialization of our model to the case of quantum two-level systems. In particular, we will  exploit the Lie algebra isomorphism $\mathfrak{su}(2)\simeq\mathfrak{so}(3)$ by using the Bloch vector representation.

\subsection{Quantum two-level systems\label{sec:twolevel}}

In order to present a special case of interest, here we consider the interaction of a classical system with a quantum two-level system whose observables are expressed in terms of Pauli matrices. When the classical system is a harmonic oscillator, this problem is of paramount importance in several contexts, from chemistry to quantum control, and is commonly referred to as ``spin-boson problem''. For a discussion of its role in the context of quantum control and controllability, here we refer the reader to \cite{RaBl05}.

The state of the quantum--classical system is described by a hybrid wave function $ \Upsilon (t)  \in L^2(T^*Q, \mathbb{C} ^2 )$ with $ \int \| \Upsilon (t,z)\| ^2 \Lambda =1$, where $\|\cdot\|$ denotes the norm in $\mathscr{H}_{\scriptscriptstyle Q}=\Bbb{C}^2$. The exact factorization reads
\[
\Upsilon (t,z)= \chi (t,z) \psi (t,z), \qquad\  \chi(t) \in L^2(T^*Q), \qquad\  \psi(t) \in \mathcal{F} (T^*Q, \mathbb{C} ^2 ), \qquad\  \| \psi (t,z)\|=1, 
%\;\forall\; z \in T^* Q
\]
and we define as before $D= | \chi |^2 $, $ \rho  = \psi \psi ^\dagger$, and $\mathcal{P}  = D \rho $. Upon denoting by $\widehat{\bsigma}=(\widehat{\sigma}_1,\widehat{\sigma}_2,\widehat{\sigma}_3)$ the array of Pauli matrices,  the Bloch vector field $ \bn(t,z)$, with $|\bn(t,z)|=1$, is defined by  $ \psi \psi ^\dagger = \frac12 \big(1+\bn\cdot\widehat{\bsigma}\big)$. Then, upon introducing the spin vector field $\bs=\hbar\bn/2$ and vector field density $\tilde\bs=D\bs$, we write
\begin{equation}\label{tilde_rho_s}
\begin{aligned} 
\mathcal{P}(t,z)&=\frac12D(t,z)\big(1+\bn(t,z)\cdot\widehat{\bsigma}\big)\\
%&= \frac{1}{2} D(t,z) + \hbar ^{-1} D(t,z) \bs(t,z) \cdot \widehat{\bsigma}\\
&= \frac{1}{2} D(t,z) + \hbar ^{-1} \tilde \bs(t,z) \cdot \widehat{\bsigma}.
\end{aligned}
\end{equation}
The hybrid Hamiltonian $\widehat{H} \in \mathcal{F} (T^*Q, \operatorname{He}( \mathbb{C} ^2 ))$ is chosen as
\begin{equation}\label{hybrid_H} 
\widehat{H}(z)=H_0(z)+\frac{\hbar}2\bH(z)\cdot\widehat{\bsigma},
\end{equation} 
for $H_0 \in \mathcal{F} (T^*Q, \mathbb{R} )$ and $ \bH \in \mathcal{F} (T^*Q, \mathbb{R} ^3  )$. While $H_0(z)$ is the Hamiltonian of the purely classical system, the interaction Hamiltonian $\bH(z)\cdot\widehat{\bsigma}$ typically involves a magnetic field so that $\bH(z)=\mathbf{H}(q)$.

As customary in dealing with two-level quantum systems, we formulate the dynamics in terms of the variables $\tilde\bs:T^*Q\rightarrow \mathbb{R} ^3 $ instead of $\mathcal{P}:T^*Q\rightarrow \operatorname{He}( \mathbb{C} ^2 )$. The usual isomorphism $\Bbb{R}^3\simeq \mathfrak{su}(2)$ reads
\[
\mathbb{R} ^3  \ni\bxi \longmapsto \xi=-\frac{i}2\bxi\cdot\widehat{\bsigma} \in \mathfrak{su}(2),
\]
so that the relation \eqref{tilde_rho_s} leads to  $\left\langle \mathcal{P}  , i \hbar \xi \right\rangle =\tilde \bs \cdot \boldsymbol{\xi}$.
Also, from \eqref{h_D_tilde_rho} and \eqref{hybrid_H}, we can write the Hamiltonian as follows
\[
\begin{aligned} 
h(D, \mathcal{P}  )&=\int\!\Big\langle\mathcal{P},\widehat{H}+\frac{i\hbar}{2D} \big[{\rm d} \mathcal{P}, X_{\widehat{H}}\big]\Big\rangle \Lambda=\int \!\Big(\langle\mathcal{P},\widehat{H}\rangle+\Big\langle\frac{i\hbar}{2D} X_{\widehat{H}}, \big[{\rm d} \mathcal{P}, \mathcal{P}  \big]\Big\rangle\Big) \,\Lambda \\
&=\int \!\left( DH_0 + \tilde\bs \cdot \bH + D^{-1} X_{\bH}\cdot\tilde\bs\times\nabla \tilde\bs \right) \,\Lambda=: h(D,\tilde\bs).
\end{aligned} 
\]
Consequently, the variational principle \eqref{VP7} becomes
\beq
\delta\int^{t_2}_{t_1}\left[\int\! \left( D{\cal A}\cdot{\cal X}+\tilde\bs\cdot\bxi
\right) \,\Lambda-h(D,\tilde\bs)\right] {\rm d} t
=0,
\label{VP8}
\eeq
with respect variations given by the first and the third relation in \eqref{EP_variations_2} as well as
\[
\delta \bxi = \ \partial _t \bSigma +  \bSigma \times \bxi   + {\rm d}  \bSigma  \cdot {\cal X} - {\rm d}   \bxi \cdot {\cal Y}\,\qquad\qquad
\delta \tilde\bs =\  \bSigma\times \tilde\bs - \operatorname{div}( {\cal Y}\tilde\bs)\,.
\]
In addition, we have the two accompanying equations
\beq
\partial_t D+ \operatorname{div}( {\cal X}D)=0
\,,\qquad\qquad
\partial_t \tilde\bs+ \operatorname{div}( {\cal X}\tilde\bs)=\bxi\times \tilde\bs
\,.
\label{aux-eqs}
\eeq
Then, upon following the same steps as in Section \ref{sec:closure}, we obtain
\[
\pounds_{{\cal X}} \mathcal{A} 
%=\de\left({\cal X}\cdot\boldsymbol{\cal A}-\frac{\delta h}{\delta D}\right)+\frac1D\left\langle\mathcal{P},\de\bigg(i\hbar\xi-\frac{\delta h}{\delta \mathcal{P}}\bigg)\right\rangle+\frac1D\left\langle i\hbar\mathcal{P},\de \xi\right\rangle
=\de \!\left({\cal X}\cdot \mathcal{A} -\frac{\delta h}{\delta D}\right)-D^{-1}\tilde\bs\cdot\de \frac{\delta h}{\delta \tilde{\bs}}
\ \implies\ {\cal X}=X_{\,\delta h/\delta D}+ D^{-1}  X_{\,\delta h/\delta \tilde{\bs}}\cdot\tilde\bs
\,,
\]
along with $\tilde\bs\times\bxi=\tilde\bs\times \delta h/\delta \tilde{\bs}$. 

Eventually, the equations \eqref{aux-eqs} become
\begin{equation}\label{aux-eqs_1} 
\partial_t {D} +\left\{  {D},\frac{\delta h}{\delta D}\right\}
+\left\{ \tilde{s}_k,\frac{\delta h}{\delta \tilde{s_k}}\right\}=0
\,,\qquad 
\partial_t \tilde\bs+\left\{ \tilde\bs,\frac{\delta h}{\delta D}\right\}
+\left\{D^{-1} \tilde\bs \tilde{s}_k,\frac{\delta h}{\delta \tilde{s_k}}\right\}=\frac{\delta h}{\delta \tilde{\bs}}\times\tilde\bs
\,,
\end{equation} 
where the second is a Landau-Lifshitz equation in the phase-space frame moving with velocity ${\cal X}=X_{\,\delta h/\delta D}+ D^{-1}  X_{\,\delta h/\delta \tilde{\bs}}\cdot\tilde\bs$.
The explicit form of the equations above is obtained by replacing
\[
\frac{\delta h}{\delta D}=H_0-D^{-2}
   \tilde\bs\cdot(\de \tilde\bs\times X_{\bH})
\,,\qquad
 \frac{\delta h}{\delta \tilde{\bs}}=\bH+2D^{-1}\de\tilde\bs\times   X_{\bH}+D^{-2}\{D,\bH\}\times\tilde\bs.
 %\bH+D^{-1}\nabla\tilde\bs\times   X_{\bH}+\operatorname{div}(D^{-1}\tilde\bs\times X_{\bH})
\]
Notice that the hybrid density \eqref{hybdens} is now expressed as
\begin{align*}
\widehat{\cal D}
%=&\, 
%\frac12\big(D+D\bn\cdot\widehat{\bsigma}\big)+\frac{i\hbar}8\operatorname{div}\!\left(D\big[\big(\bn\cdot\widehat{\bsigma}\big),\big(\Bbb{J}\nabla\bn\cdot\widehat{\bsigma}\big)\big]\right)
%\\
%=&\, 
%\frac12\big(D+D\bn\cdot\widehat{\bsigma}\big)+\frac{i\hbar}8\operatorname{div}\!\left(Dn_k\Bbb{J}\nabla n_j\big[\widehat{\sigma}_k,\widehat{\sigma}_j\big]\right)
%\\
%=&\, 
%\frac12\big(D+D\bn\cdot\widehat{\bsigma}\big)+\frac{i\hbar}8\operatorname{div}\!\left(2iDn_k\Bbb{J}\nabla n_j\epsilon_{kj\ell}\widehat\sigma_\ell\right)
%\\
=&\, 
\frac12D+\frac1\hbar\widehat{\bsigma}\cdot\left(\tilde\bs-\operatorname{div}\!\left(D^{-1}\tilde\bs\times \Bbb{J} \de \tilde\bs\right)\right)
\\
=&\, 
\frac12D+\frac1\hbar\widehat{\bsigma}\cdot\left(\tilde\bs-\left\{D^{-1\,}\widehat{ s}, \tilde\bs\right\}\right)
\,,
\end{align*}
where we have used the hat map $\widehat{ s}_{jk}=-\epsilon_{jk\ell}\tilde\bs_\ell$ in the last equality.

One last point concerns the Poisson bracket \eqref{bracket_candidate2}, which now reads
\begin{multline}\label{bracket_candidate2-spin}
\{\!\!\{f,h\}\!\!\}(D,\tilde\bs)=
\int\! D\left\{\frac{\delta f}{\delta D},\frac{\delta h}{\delta D}\right\}\Lambda
\\+
\int \!\tilde\bs \cdot\left(\frac{\delta f}{\delta \tilde\bs}\times\frac{\delta h}{\delta \tilde\bs}+
\left\{\frac{\delta f}{\delta \tilde\bs},\frac{\delta h}{\delta D}\right\}-\left\{\frac{\delta h}{\delta \tilde\bs},\frac{\delta f}{\delta D}\right\}+\frac1{D}   \left\{\frac{\delta f}{\delta \tilde\bs},\frac{\delta h}{\delta \tilde\bs}\right\}\cdot\tilde\bs\right) \,\Lambda
\,.
\end{multline} 
As this is again written  in terms of canonical brackets, we hope it helps clarifying the structure of the operators involved in the original version \eqref{bracket_candidate2}. 
From Proposition \ref{It_is_Poisson}, it follows that $C( D, \tilde\bs)= \int D\Upphi(|\tilde\bs|/D)\,\Lambda$ are Casimir functionals for \eqref{bracket_candidate2-spin} for any function $\Upphi$.

\section{Conclusion  and outlook\label{conclusions}}

After reviewing the Koopman-van Hove theory of classical and quantum--classical systems in Sections \ref{sec:KvH} and \ref{sec:QChybrids}, Section \ref{sec:closure1} has presented a variant of this construction that ensures a positive classical density at all times, along with a positive-definite quantum density matrix. In this way, it is possible to unambiguously identify a classical and a quantum state  throughout the entire hybrid evolution. 
The question  whether the quantum--classical wave equation \eqref{hybrid_KvH} leads to a classical distribution \eqref{rhoc} that is left positive in time remains open. 
In this paper we chose to overcome this question by directly enforcing a constraint to ensure that the classical density obeys a transport equation   preserving the sign of the initial condition. The possibility of enforcing this constraint is made available by the variational structure  underlying the exact factorization \eqref{EFDef}, which was recently discussed in \cite{FoHoTr19}. Then, a closure of KvH hybrids was obtained by enforcing the relation $\eta(t)^*(\omega+{\cal B}(t))=\omega+{\cal B}(0)$, as pointed out in Remark   \ref{rem:berry}. The resulting dynamical model involves a Hamiltonian  given in \eqref{h_D_tilde_rho} as the sum of the usual expression $\int\! D\langle\widehat{H}\rangle\Lambda$ from mean-field theory and another term $\int \langle i\hbar[{\rm d} \mathcal{P},  X_{\widehat{H}}]\rangle \Lambda/2$ that is responsible for generating quantum--classical correlations in time. 

The expression \eqref{hybdens} identifies a quantum--classical density operator $\widehat{\cal D}(t, z)$ such that the total energy of the system is written as $\int\langle\widehat{\cal D},\widehat{H}\rangle\Lambda$. This hybrid density was shown to be equivariant with respect to both quantum and classical transformations, thereby enjoying a long-sought property in quantum--classical dynamics \cite{boucher}. Also, we notice that, similarly to its original form \eqref{hybridDenOp}, the hybrid density operator \eqref{hybdens} is sign-indefinite and therefore it may not be directly used to make  considerations in statistical physics. Thus, while the  closure in Proposition \ref{propclosure} emerges as the only deterministic hybrid model retaining the classical and quantum character of each associated  subsystem, the development of a statistical theory of quantum--classical systems needs further progress.

The closure model was shown to possess an intricate noncanonical bracket structure that is Poisson and it arises from standard Poisson reduction.
Different variants of the Poisson structure \eqref{bracket_candidate2} were considered along with their correspondents in purely classical dynamics. Finally, the hybrid closure was specialized to the case of quantum two-level systems, so that the formalism allows for a vector algebra formulation of quantum spin dynamics.

\subsection{Relevant special cases\label{sec:relevcases}}
In this section, we show how the closure model recovers both purely quantum and classical cases, as well as other simple hybrid models appearing in the literature. 

\medskip
\noindent
{\bf Classical dynamics.} In the purely classical case, one has $\widehat{H}(z)={H}(z)\operatorname{Id}$, so that the Hamiltonian functional \eqref{h_D_tilde_rho} reduces to $h(D)=\int DH\,\Lambda$. Then, the Poisson structure \eqref{bracket_candidate2} collapses to the usual Lie-Poisson bracket $\int\! D\{{\delta f}/{\delta D},{\delta h}/{\delta D}\}\,\Lambda$ for the classical Liouville equation
$
{\partial_t D} + \{ D,{H}\}=0
$.

\medskip
\noindent
{\bf Quantum dynamics.}  Analogously, in the purely quantum case $\de\widehat{H}(z)=0$, the Hamiltonian functional \eqref{h_D_tilde_rho} reduces to $h(D)=\langle\widehat{H}\rangle=\langle \widehat{H},\hat\rho_q\rangle$, with $\hat\rho_q=\int\mathcal{P}\,\Lambda$. Then, the Poisson structure \eqref{bracket_candidate2} collapses to the usual Lie-Poisson bracket  $-\left\langle {i}\hbar^{-1}[{\delta f}/{\delta \hat\rho_q},{\delta h}/{\delta \hat\rho_q}]\right\rangle$ for the quantum Liouville equation
$
 i\hbar\,{\de \hat\rho_q}/{\de t} =[\widehat{H},\hat\rho_q]
$.
 
\medskip
\noindent
{\bf Mean-field dynamics.} Additionally, if we enforce $\mathcal{P}=D\hat\rho_q$, then the Hamiltonian \eqref{h_D_tilde_rho} reduces to $h(D,\hat\rho_q)=\int\!D\langle\hat\rho_q,\widehat{H}\rangle\,\Lambda$. Since $\hat\rho_q=\int\mathcal{P}\,\Lambda$, the chain rule yields ${\delta h}/{\delta \mathcal{P}}={\delta h}/{\delta \hat\rho_q}$, so that the Poisson structure \eqref{bracket_candidate2} collapses to the direct-sum bracket $\int\! D\{{\delta f}/{\delta D},{\delta h}/{\delta D}\}\,\Lambda-\left\langle {i}\hbar^{-1}[{\delta f}/{\delta \hat\rho_q},{\delta h}/{\delta \hat\rho_q}]\right\rangle$, thereby returning the mean-field model
\[
\frac{\partial D}{\partial t} + \big\{ D,\langle\widehat{H}\rangle\big\}
\,,\qquad\quad
i\hbar\frac{\de \hat\rho_q}{\de   t} =\bigg[\int\! D\widehat{H}\,\Lambda,\hat\rho_q\bigg]
\,.
\]
We realize that in this case the hybrid density \eqref{hybdens} reduces to $\widehat{\cal D}=D\hat\rho_q$.

\medskip
\noindent
{\bf Ehrenfest dynamics.} Another case of interest may be obtained by dropping the commutator term in the Hamiltonian \eqref{h_D_tilde_rho}, which then reduces to $h(D,\mathcal{P})=\int \langle\mathcal{P},\widehat{H}\rangle\,\Lambda$ and similarly $\widehat{\cal D}=\mathcal{P}$. In this particular case, the Poisson bracket \eqref{bracket_candidate2} yields the equations
\[
\frac{\partial D}{\partial t} + \operatorname{div}(D\langle X_{\widehat{H}}\rangle)=0
\,,\qquad\quad
i\hbar\frac{\partial \mathcal{P}}{\partial t}+ i\hbar\operatorname{div}(\mathcal{P}\langle X_{\widehat{H}}\rangle)=\big[\widehat{H},\mathcal{P}\big]
\,.
\]
We recognize that the second equation coincides with the {Ehrenfest model} in \eqref{Ehrenfest1}. The fact that the latter possesses the same Poisson bracket as in \eqref{bracket_candidate_rho} shows explicitly that the Ehrenfest model is indeed Hamiltonian.
Notice that this  model is often confused with the mean-field system presented above.

\subsection{Future perspectives\label{sec:outlook}}
The present work raises several questions and we list some of them below:
\begin{itemize}

\item {\bf Noncanonical classical coordinates.} In several cases, the classical subsystem possesses itself a noncanonical structure even in the absence of coupling to quantum degrees of freedom. For example, this is the case for classical angular momentum, or simply classical spin dynamics. We intend to develop a quantum--classical theory for noncanonical (e.g. Lie-Poisson) Hamiltonian systems that can possibly be applied, for example, to the dynamics of hybrid  quantum--classical spin systems. 

%\item {\bf Lyapunov stability.} The Poisson tensor of noncanonical systems possesses a nontrivial kernel that gives rise to Casimir invariants. In turn, the existence of such conserved quantities allows for the interesting possibility of performing a Lyapunov stability analysis going beyond the standard linear studies  often adopted in physics. This method is known as \emph{energy-Casimir method} and goes back to the 80's \cite{HoMaRaWe85}. We intend to pursue this future direction in the context of the present noncanonical structure: the identification of Casimir invariants for the Poisson bracket \eqref{bracket_candidate2} would likely allow for an efficient way of studying the stability properties of hybrid quantum--classical systems.

\item {\bf Finite-dimensional closures.} The nonlinear character of the hybrid quantum--classical equations \eqref{adv_D}-\eqref{tilde_rho_adv} makes their numerical implementation particularly challenging. This leads naturally to the need for new finite-dimensional closures that can be suitable for the design of numerical algorithms. In this context, the use of symplectic integrators can be particularly advantageous for examining the long-time  hybrid dynamics. In more generality, we intend to design new trajectory-based closures that are suitable for the construction of geometric integrators for quantum--classical hybrid dynamics.

\item {\bf Hybrid wavefunctions.} While the original quantum--classical wave equation evidently involves a hybrid wavefunction, the theory presented in the last section only involves density variables. Then, it is natural to ask if a wavefunction formulation is still available. Indeed, since the fundamental quantity $\mathcal{P}$ may be written as $\mathcal{P}(z)=\Upsilon(z)\Upsilon(z)^\dagger$, one is tempted to write the reduced quantum--classical hybrid dynamics in terms of hybrid Koopman wavefunctions. This direction was recently pursued in \cite{GBTr22}, where we  unfolded several differences between the nonlinear construction in Section \ref{sec:closure1} and the quantum--classical wave equation in Section \ref{sec:QChybrids}.

\item {\bf Hybrid density operator.} While the hybrid density operator $\widehat{\cal D}$ in \eqref{hybridDenOp} was recently studied in \cite{BoGBTr19,GBTr20}, little is yet known about the variant in \eqref{hybdens}. In particular, we intend to understand how its evolution law is related to the hybrid wavefunctions introduced in the previous point above, thereby leading to a comparison between the present closure model and the original hybrid theory.
This will be important to draw conclusion on the expectation value dynamics comprising quantum--classical correlations, a topic that was not treated in this work.

\item {\bf Poisson/symplectic reduction.} An interesting question for the model \eqref{adv_D}-\eqref{tilde_rho_adv} is the determination of the symplectic leaves of the Poisson brackets and their physical meaning. For instance the symplectic leaves of the Poisson bracket \eqref{gino} are formally found by implementing symplectic reduction associated to the momentum map of the right action of $ \operatorname{Diff}(T^*Q)_{ \mathcal{P}  _0}$  on $ \operatorname{Diff}(T^*Q)$ endowed with the symplectic form $\Omega ( \eta )(u_ \eta , v_ \eta ) = \int D_0 ( \omega \circ \eta )(u_\eta , v_ \eta ) \Lambda$.
\end{itemize}

The points raised above are only some of the several questions emerging in the study of hybrid systems. Other questions involve, for example, the inclusion of field theories. A whole new research theme seems to be opening in front of us!

\medskip
\noindent
{\bf Acknowledgments.}  We wish to express our gratitude to Tony Bloch for several inspiring discussions over the years, and for encouraging us to work in this direction. Also, we would like to thank Darryl Holm for his valuable feedback on these and related topics. This work was made possible through the support of Grant 62210 from the John Templeton Foundation. The opinions expressed in this publication are those of the authors and do not necessarily reflect the views of the John Templeton Foundation. CT also acknowledges partial support by the Institute of Mathematics and its Applications and by the Royal Society Grant IES\textbackslash R3\textbackslash203005.

\end{document}